\documentclass[pra,aps,twocolumn,sshowpacs,floatfix]{revtex4-1}
\usepackage{graphicx}
\usepackage{amssymb}
\usepackage{amsmath}
\usepackage{xspace} 
\usepackage{color}

\renewcommand{\vec}[1]{\mathbf{#1}}

\begin{document}
\title{Projector-based renormalization approach to electron-hole-photon systems in their nonequlibrium steady state}
\author{Klaus W. Becker}
\affiliation{
Institut f{\"u}r Theoretische Physik, Technische Universit{\"a}t Dresden, D-01062 Dresden, Germany}
\author{Holger Fehske}
\affiliation{
Institut f{\"u}r Physik, Universit{\"a}t Greifswald, D-17489 Greifswald, Germany}
\author{Van-Nham Phan}
\thanks{Corresponding author: phanvannham@duytan.edu.vn}
\affiliation{Institute of Research and Development, Duy Tan University, 3 Quang Trung, Danang, Vietnam}


\begin{abstract}
We  present an extended version of the projector-based renormalization method  that can be used to address not only equilibrium but also non-equilibrium situations in coupled fermion-boson systems. The theory is applied to interacting electrons, holes, and photons in a semiconductor microcavity, where the loss of cavity photons into vacuum is of particular importance. The method incorporates correlation and fluctuation processes beyond mean-field theory in a wide parameter range of detuning, Coulomb interaction, light-matter coupling, and damping, even in the case when the number of quasiparticle excitations is large. This enables the description of exciton and polariton formation, and their possible condensation through spontaneous phase symmetry breaking by analyzing the ground-state, steady-state, and spectral properties of a rather generic electron-hole-photon Hamiltonian, which also includes the coupling to two fermionic baths and a free-space photon reservoir. Thereby, the steady-state behavior of the system is obtained by evaluating expectation values in the long-time limit by means of the Mori-Zwanzig  projection technique. Tracking and tracing different order parameters, the fully renormalized single-particle spectra and the steady-state luminescence,  we demonstrate  the Bose-Einstein condensation of excitons and polaritons  and its smooth transition when the excitation density is increased.
\end{abstract}
\date{\today}+
\maketitle

\section{Introduction}
\label{I}
Semiconductor microcavity systems with quantum well potentials  have created fascinating  possibilities with regard to the formation of diverse condensed phases~\cite{DWSBY02,EM04}. These condensates constitute a macroscopic, long-range quantum phase-coherent state that exhibits unconventional transport and luminescence properties in particular. Coupled electron-hole-photon (e-h-p) systems have led to very early speculations about a Bose-Einstein condensation of excitons, i.e., of electron-hole pairs formed by the  attractive Coulomb interaction, at low but sufficient large particle densities~\cite{KK64}.

While the short life time of optically generated excitons seems to be a serious problem establishing a Bose-Einstein condensate (BEC) in bulk semiconductors, such as $\rm Cu_2O$, even in potential traps~\cite{SSKSKSSNKF12}, quantum wells realized in layered semiconductors significantly reduce the rate at which electrons and holes recombine into photons (albeit there is not yet compelling evidence for an exciton BEC in these systems). Increasing the excitation density, phase-space (Pauli-blocking) and Fermi-surface effects become important and, as  a result, the exciton BEC may cross over into an e-h BCS phase~\cite{CN82,NS85}. In response to a specific electronic band structure, such as those near a semiconductor-semimetal transition, the exciton condensate can also exist in equilibrium whereby it typifies an excitonic insulator phase~\cite{Mo61,Kno63,ZIBF12}.

Of course the e-h-p system is also influenced by its interaction with the surroundings.  In the case of  a semiconductor microcavity the loss of cavity photons into the vacuum space is of particular importance. This means that the microcavity system is essentially in a non-equilibrium state. To maintain the system in a stationary quasi-equilibrium state one has to supply continuously electrons and holes to the e-h-p system which compensates the decay of photons into the environment.  Unfortunately, however, only for low excitation densities, when photon effects are still irrelevant, the properties of the e-h-p system  reduce to the equilibrium physics. At large excitation density, the photonic effects play a predominant role, and the condensate turns from excitonic to polaritonic. Polaritons in semiconductor microcavities have also been observed to exhibit BEC~\cite{DWSBY02,Kaea06}.  At even higher excitation densities, the excitonic component saturates, whereas the photonic order parameter continues to increase.  Here, the relationship between a polariton BEC  and photon lasing has to be clarified~\cite{KO11,BKY14}.

The main objective of this paper is to describe both the equilibrium and the non-equilibrium properties of the e-h-p system on an equal footing. To this end, we employ a minimal model for the e-h-p gas  that includes attractive interactions between electrons and holes as well as between cavity photons and electron-hole excitations~\cite{SKL06,SKL07,YKNOY13,YO16,HLO16}. Moreover the decay of cavity photons to an external vacuum and the pumping from two fermionic  baths to the electrons and holes of the e-h-p system are taken into account. The major difficulty  results from  the lack of reliable techniques to tackle such a model in the whole parameter regime.  So far most theoretical approaches~\cite{KESL04,BHIY10,LEKMSS04,KESL05} have addressed  the equilibrium properties separately from those of the steady state~\cite{La98,CC05,CC14}.  Only recently a steady-state framework~\cite{YKOY12,YNKOY15} based on a non-equilibrium Green's function approach\cite{SKL06,SKL07}  was formulated which allows us to treat equally the equilibrium BEC and BCS phases at low excitation densities just as the non-equilibrium state at high excitation densities. 

In this work, we utilize an alternative theoretical tool, the projector-based renormalization method (PRM)~\cite{BHS02,PBF10,PFB11}. 
The PRM was applied before exclusively to  equilibrium phenomena, and also  to describe the equilibrium properties of e-h-p systems~\cite{PBF16} at small--to--moderate excitation densities, where the leakage of photons to the vacuum is not important.  We show that the PRM can be extended to non-equlibrium situations, and applied to the model under consideration even in the case when the number of excitations is large. Here,  the steady-state properties can be found from time-dependent expectation values for long times which will be  evaluated by means of the  Mori-Zwanzig projection technique. Thereby, in contrast to the work~\cite{YKOY12,YNKOY15}, the PRM  incorporates fluctuation processes beyond mean field theory for all excitation densities.  This allows us to address the great variety of e-h-p condensation phenomena mentioned above. 
 
The paper is organized as follows. In Sec.~\ref{II} we introduce our theoretical model for a pumped-decaying exciton-polariton system and briefly discuss its adaption to a steady-state situation.   Since the present theoretical study is based on the PRM, we outline this technique and its improvements in Sec. \ref{III}. More details of the PRM approach can be found in the Appendices~\ref{A}--\ref{C}. The steady-state expectation values are evaluated in Sec.~\ref{IV}, the single-particle spectral function in Sec.~\ref{V} and the steady-state luminescence in  Sec.~\ref{VI}. Finally, in Sec.~\ref{VII} some characteristic numerical results will be presented and discussed. Section \ref{VIII} contains a brief summary and our main conclusions.  

\section{Modeling of pumped-decaying exciton-polariton systems}
\label{II} 

As a typical example of an  e-h-p system we will consider electrons and holes, confined in a semiconductor quantum
well structure, are exposed to photons, entrapped in a microcavity.  In such a setup Bose-Einstein condensates of bound electron-hole pairs (excitons) and polaritons may appear, which possibly can cross over into a BCS-like coherent state under quasi-equilibrium conditions at high particle densities, in case the quasiparticle lifetime is larger than the thermalization time~\cite{YNKOY15}. In general, however, these systems are driven out of equilibrium by coupling to multiple baths, and such nonequilibrium electron-hole condensates in the solid state are subject to dissipation, dephasing, and decay.  Therefore pump and loss channels have to be taken into account. In the following we introduce appropriate microscopic models for the system and for the reservoirs to which it is coupled in order to include these effects.

\subsection{System Hamiltonian}
\label{II.A}
Our starting point is the e-h-p Hamiltonian~\cite{KO11,PBF16} of an isolated semiconductor 
quantum-well/microcavity system,
\begin{eqnarray}
\label{1}
\check{\mathcal{H}}_{\rm S}=  \check{ \mathcal{H}}_{0, \rm S}+ \check{\mathcal{H}}_{\rm el-ph}  
+\check{\mathcal{H}}_{\rm el-el}\,
\end{eqnarray}
with
\begin{eqnarray}
\label{2}
&&  \check{\mathcal{H}}_{0, \rm S} =  \sum_{\vec{k}} \check\varepsilon^e_{\vec{k}} \check e_{\vec{k}}^{\dagger}
\check e_{\vec{k}}^ {}+\sum_{\vec{k}}\check \varepsilon^h_{\vec{k}} \check h_{\vec{k}}^{\dagger}\check h_{\vec{k}}^ {} + \sum_{\vec{q}}\check \omega_{\vec{q}}\check \psi_{\vec{q}}^{\dagger}\check \psi_{\vec{q}}^{}\,, \\
\label{3}
&& \check{\mathcal{H}}_{\rm el-ph}   = -\frac{g}{\sqrt{N}}\sum_{\vec{q}\vec{k}}
[\check e_{\vec{k}+\vec{q}}^{\dagger} \check h_{-\vec{k}}^{\dagger}\check \psi_{\vec{q}}^ {}+\textrm{H.c.}]\,,
 \\
\label{4} 
&& \check{\mathcal{H}}_{\rm el-el} = -\frac{U}{N} \sum_{\vec k} \check \rho^e_{\vec k} \check\rho^h_{-\vec k}\,,
\end{eqnarray}
describing  free particles (electrons created by $\check e_{\vec k}^\dag$, holes by $\check h_{\vec k}^\dag$, and photons by $\check \psi_{\vec q}^\dag$), the coupling ($\propto g$) of electron-hole pairs  to the radiation field, and the local  Coulomb interaction ($\propto U$) between electrons (density operators $\check \rho^e_{\vec k}=\sum_{\vec k_1} \check e^\dag_{\vec k + \vec k_1} \check e_{\vec k_1}^{}$) and holes  ($\check \rho^h_{\vec k}= \sum_{\vec k_1} \check h^\dag_{\vec k + \vec k_1} \check h_{\vec k_1}^{}$), respectively. In $\check{\mathcal{H}}_{0, \rm S}$,   $\check \varepsilon^e_{\vec{k}}$ ($\check \varepsilon^h_{\vec{k}}$) denotes the dispersion of electrons (holes),  
\begin{equation}
\label{5}
\check \varepsilon^e_{\vec{k}} = -2t \sum_i^{D}  \cos k_i+\frac{E_g+4tD}{2} = \check \varepsilon^h_{\vec{k}}\,, \\
\end{equation}
where  $D$ is the dimension of the hypercubic lattice, $t$ is the particle transfer amplitude between neighboring sites, $E_g$ gives the minimum distance (gap) between the bare electron and hole bands, and $\check \varepsilon_{\vec{k}}^{e}  = \check \varepsilon_{\vec{k}}^{h} $ is set for simplicity. The photon field is characterized by 
\begin{equation}
\label{6}
\check \omega_{\vec{q}}=\sqrt{(c{\vec q})^{2}+ \omega_{c}^{2} }
\end{equation}
with the  zero-point cavity frequency $\omega_c$.

\subsection{Coupling to reservoirs}
\label{II.B}

Next we model the coupling of the e-h-p system, being an open quantum system in reality,  to its environment.  
In the first place, two pumping baths for electrons and holes made possible the injection of free fermions into the system.
In addition, the cavity photons are connected to a free-space photon reservoir, allowing for a leakage of photons into the surroundings. 
To maintain a steady state, the loss of cavity photons to the external reservoir must be compensated by bringing in fermionic carriers.  
Then for the total system the following Hamiltonian seems to be adequate
\begin{eqnarray}
\label{9}
&&\check{ \mathcal H} =   \check{\mathcal H}_{\rm S} +  \check{\mathcal H}_{\rm R} +  
\check{\mathcal H}_{\rm SR}  \, ,
 \end{eqnarray}
where $ \check{ \mathcal H}_{\rm S}$ is given by Eq.~\eqref{1}, and $\check{\mathcal H}_{\rm R}$ 
and $\check{\mathcal H}_{\rm SR}$  are defined as:
\begin{eqnarray}
\label{10}
\check{ \mathcal H}_{\rm R} &=&  \sum_{\vec p}\check  \omega^e_{\vec p} \, \check b^\dag_{e, \vec p} \check b_{e, \vec p}  +
 \sum_{\vec p} \check \omega^h_{\vec p} \, \check b^\dag_{h, \vec p} \check b_{h, \vec p}  \nonumber \\
 &+& \sum_{\vec p} \check \omega^\varphi_{\vec p} \, \check \varphi^\dag_{\vec p} \check \varphi_{\vec p}\,,  
 \\
 && \nonumber \\
 \label{11}
\check{ \mathcal H}_{\rm SR} &=& \frac{1}{N}\sum_{\vec k \vec p} (\Gamma^e_{\vec k\vec p} 
 \check e^\dag_{\vec k} \check b_{e,\vec p} + {\rm H.c.}) \nonumber \\
&+& \frac{1}{N} \sum_{\vec k \vec p} (\Gamma^h_{\vec k\vec p} \check h^\dag_{-\vec k} \check b_{h,-\vec p} + {\rm H.c.})  \nonumber \\
&+&\frac{1}{N} \sum_{\vec q \vec p}( \Gamma^{\psi}_{\vec q\vec p}  \check \psi_{\vec q}^\dag \check \varphi_{\vec p} + {\rm H.c.}) \, .
\end{eqnarray}
$\check{\mathcal H}_{\rm R}$ is the Hamiltonian for the two fermionic baths and the free-space photon reservoir which are interacting with the e-h-p system via $\check{\mathcal H}_{\rm SR}$. 
 The quantities $\check b_{e,\vec p}^{(\dag)}$ and $\check b_{h,\vec p}^{(\dag)}$ are the fermion creation/annihilation  operators of the two pumping baths, and  $\check \varphi_{\vec p}^{(\dag)}$
are the boson creation  and annihilation operators 
of the free-space  photons. Finally, $\Gamma^{e,h}_{\vec k\vec p}$ and $\Gamma^\psi_{\vec q\vec p}$
in Eq.~\eqref{11} are the coupling constants between the system and the respective reservoirs. 

Let us also define the particle number of the total system by
\begin{eqnarray}
\label{14} 
\mathcal N &=& \frac{1}{2} \sum_{\vec k} (\check e^{\dag}_{\vec k} \check e_{\vec k} 
+ \check h^{\dag}_{\vec k} \check h_{\vec k} )
 + \sum_{\vec q} \check \psi_{\vec q}^\dag  \check \psi_{\vec q}  \nonumber \\
 &+&  \frac{1}{2} \sum_{\vec p} (\check b^{\dag}_{e, \vec p} \check b_{e, \vec p} + \check b^{\dag}_{h, \vec p}
 \check b_{h,\vec p} )
 + \sum_{\vec p}  \check \varphi_{\vec p}^\dag \check \varphi_{\vec p} \, ,
 \end{eqnarray}
which is a constant of motion  $[\check{\mathcal H}, \mathcal N] =0$.

We maintain that the total system in a non-equilibrium situation evolves under Hamiltonian  $\check{\mathcal H} = {\mathcal H}_{\rm S} + \check{\mathcal H}_{\rm R} 
+ \check{\mathcal H}_{\rm SR}$. Thereby  $\check{\mathcal H}_{\rm S}$ is 
``simple'' in the sense that it can be diagonalized, even though many-body aspects due to the presence of 
$\check{\mathcal H}_{\rm el-el}$ and $\check{\mathcal H}_{\rm el-ph}$  require a special treatment.  
$\check{\mathcal H}_{\rm SR}$ is responsible for the non-equilibrium situation since it 
governs the pumping and damping of electrons and holes and the leakage of photons into the free space. 
Note that $\mathcal H_{\rm SR}$ is not translationally invariant.

We now assume that 
$\check{\mathcal H}_{\rm SR}$ vanishes for times $t < t_0$, where
$t_0 \rightarrow -\infty$ might be used as a suitable starting point. That is, before at $t_0$ the interaction 
$\check{\mathcal H}_{\rm SR}$ 
is turned on,  the reservoirs and the e-h-p system are in separate thermal equilibrium states. 
Then the state of the total system is described by a product  of the e-h-p system density operator 
$\check \rho_{\rm S}$ and the reservoir density operator $\check \rho_{\rm R}$
\begin{eqnarray}
\label{12} 
\check \rho_0= \check \rho_{t_0 \rightarrow -\infty} = \check \rho_{\rm S} \, \check \rho_{\rm R} \, ,
 \end{eqnarray}
where $\check \rho_{\rm S}$ commutes with $\check{\mathcal H}_{\rm S}$. To simplify the considerations we suppose  
the electronic baths and the external photon reservoir to be huge compared to 
 $\check{\mathcal H}_{\rm S}$. As a result, in the steady state the two electronic baths  remain
in thermal equilibrium,  even when they are coupled to the e-h-p system. Similarly 
the free-space photons act as a reservoir for cavity photons escaped from the e-h-p system.

Below, the task is to evaluate time-dependent expectation values of  observables $\check{\mathcal A}$ 
for times $t \gg t_0$,
 \begin{eqnarray}
\label{13} 
\langle \check{\mathcal A}(t)\rangle = {\rm Tr \, }[\check \rho_0 \, \check{\mathcal A}(t)]  \, ,
\end{eqnarray}
when the system has  approached  a steady state. 
Therefore we use the Heisenberg picture, in which the time-dependence of $\check{\mathcal A}$ is governed by the full Hamiltonian $\check{\mathcal H}$, and $\check \rho_0$ is time independent. Note that  $\check \rho_0$ and $\check{\mathcal H}$ do not commute.
This property causes the genuine time dependence of expectation values \eqref{13}.
Being $\tau_R$ some internal relaxation time, for times $t \gg \tau_R$ 
the system is expected to merge into a periodically driven steady state 
and remembers no longer its initial state.

\subsection{Steady-state description}

\label{II.C}
Now let us consider a steady-state situation in which both loss and pump processes are spatially homogenous with 
a coherent photon field that is only formed for $\vec q=0$. For large times,
the steady state will evolve according to 
\begin{eqnarray}
\label{15}
\langle \check \psi^\dag_{\vec q}(t) \rangle& =&    \delta_{\vec q 0} \, \langle \psi^\dag_{0 } \rangle \,  e^{i\mu t}\,,\\
\label{16a}
\langle (\check e^\dag_{\vec k} \check h^\dag_{-\vec k})(t)\rangle 
 &=&   d_{\vec k }^* \,  e^{i \mu t} \, ,\\ 
 \label{16b}
\langle (\check e^\dag_{\vec k} \check e_{\vec k})(t)\rangle &=&  n^e_{\vec k } \,,\\  
 \label{16c}
 \langle (\check h^\dag_{-\vec k} \check h_{-\vec k})(t)\rangle &=&   n^h_{-\vec k } \,,
\end{eqnarray}
where the quantities $\langle \psi^\dag_{0} \rangle$,  $d_{\vec k }^*$, $n^e_{\vec k }$ and $n^h_{-\vec k }$ 
are time-independent and---together with $\mu$---are subject to the evaluation below. 
{\it Ansatz} \eqref{15}--\eqref{16c} implies that the dynamics of certain variables is captured on a rotating frame with a frequency
$\mu$, where in the steady state $\langle \psi^\dag_{0} \rangle$,  $d_{\vec k }^*$, $n^e_{\vec k }$ and $n^h_{-\vec k }$ 
become time-independent~\cite{YO16}. 

 In the first evaluation step the explicit time dependence in
 $\langle \check \psi^\dag_{\vec k}(t) \rangle$ and 
 $\langle (\check e^\dag_{\vec k} \check h^\dag_{-\vec k})(t)\rangle$
 will be eliminated.  This is achieved by performing a time-dependent gauge transformation: 
\begin{align}
\label{17}
\big(e_{\vec k}^\dag,  h_{-\vec k}^\dag,  \psi^\dag_{\vec q} \big) & =
e^{-i\mu\mathcal N t } \, \big(\check e_{\vec k}^\dag,  \check h_{-\vec k}^\dag, \check \psi^\dag_{\vec q} \big)\, 
e^{i\mu \mathcal N t }  \\
& =\big(e^{-i(\mu/2)t} \check e_{\vec k}^\dag, e^{- i(\mu/2)t} \check h_{-\vec k}^\dag, e^{- i\mu t} 
\check\psi_{\vec q}^\dag \big)\,, \nonumber  \\
&\nonumber\\
\label{18}
\big(b_{e,\vec p}^\dag, b_{h, \vec p}^\dag, \varphi^\dag_{\vec p} \big) & =
e^{-i\mu \mathcal N t } \, \big(\check b_{e,\vec p}^\dag,  \check b_{h, \vec p}^\dag, \check \varphi^\dag_{\vec p} \big)\,
e^{i\mu \mathcal N t } \\
& =\big(e^{-i(\mu/2)t} \check b_{e, \vec p}^\dag, e^{-i(\mu/2)t} \check b_{h, \vec p}^\dag, e^{- i\mu t} \check \varphi_{\vec p}^\dag \big) \, .\nonumber
\end{align}
Let us look at an example: The equation of motion for 
the operator $\check \psi_{\vec k}^\dag (t)$ reads $({\rm d/dt})   \check \psi_{\vec k}^\dag(t)  
=  (i/\hbar) [\check{\mathcal H}, \check \psi_{\vec k}^\dag](t)$. Going over from 
$\check \psi^\dag_{\vec k}$ to   
 the new variable ${\psi_{\vec k}^{\dag}}= { \check \psi_{\vec k}^\dag}  
 \, e^{-i\mu t}$,  the equation for 
 $  {\psi_{\vec k}^\dag}(t) $ becomes $({\rm d/dt})   \psi_{\vec k}^\dag (t) 
 =  (i/\hbar) [\check{\mathcal H} -\mu \mathcal N, { \psi_{\vec k}^\dag}](t)$. Thus, 
using the following replacements 
 \begin{align}
\label{19a}
 \varepsilon^\alpha_{\vec k}  &= \check \varepsilon^\alpha_{\vec k} - \frac{1}{2} \mu \, , 
\quad
\omega_{\vec q} = \check \omega_{\vec q} -\mu\,,   \\\label{19b}
 \omega^\alpha_{\vec p} &= \check \omega^\alpha_{ \vec p} - \frac{1}{2} \mu\, , 
  \quad \omega^\varphi_{\vec p} = \check \omega^\varphi_{ \vec p} -  \mu \, ,
 \end{align} 
($\alpha= e,h$), the explicit time dependences in Eqs.~\eqref{15} and \eqref{16a} disappears. 
Following the equations of motion of the variables $e^\dag_{\vec k}, h^\dag_{\vec k}, \psi^\dag_{\vec q}$, we therefore introduce a Hamiltonian,
\begin{equation}
\label{19c}
\mathcal H = \check{\mathcal H} -\mu \mathcal N \, ,
\end{equation}
where both parts on the right hand side keep their operator form when 
written in the new variables.  Note that replacements \eqref{19a} and \eqref{19b}
only apply to the time dependence of $\check{\mathcal A}(t)$  in Eq.~\eqref{13} (Heisenberg picture) 
but not to the density operator $\check \rho_0$, which keeps its operator form expressed by the variables without $\check{}$ symbols, 
and will be called $\rho_0$. The total particle number $\mathcal N$, written in the variables
$e^\dag_{\vec k}, h^\dag_{\vec k}, \psi^\dag_{\vec q}$  has the same operator form as in Eq.~\eqref{14} and obeys  $[\mathcal H, \mathcal N]=0$. 
Thus the total particle flux ${\rm d} \langle \mathcal N \rangle/{\rm dt} =0$ disappears, which means 
that a  change of the particle numbers of the e-h-p subsystem and the electronic reservoirs 
must be balanced by a change of the free space photons.  

We wish to stress that only in thermal equilibrium the quantity $\mu$ will turn out to act as a chemical potential. 
For time-dependent problems, such as the considered open e-h-p system, the dynamics is captured 
on a rotating frame with the frequency $\mu$.  Thereby $\mu$  is a given parameter which has to be fixed  in a steady state~\cite{YO16}.

\subsection{Total Hamiltonian}

\label{II.D} 
With the above transformations and replacements the total Hamiltonian ${\mathcal H}$  takes the form:
\begin{eqnarray}
\label{20}
\mathcal H &=& \mathcal H_{\rm S}
 +\mathcal H_{\rm R} + \mathcal H_{\rm SR} \, ,
 \end{eqnarray}
where $\mathcal H_{\rm S}$ describes the interacting e-h-p subsystem 
 \begin{equation}
 \label{21}
\mathcal H_{\rm S} = \mathcal H_{0} + \mathcal H_{c}+ \mathcal H_{g} + \mathcal H_{U}\,,\\
\end{equation}
 with
\begin{align}
\label{22}
& \mathcal{H}_0 = \sum_{\vec{k}}{\varepsilon}_{\vec{k}}^{e}
e_{\vec{k}}^{\dagger}e_{\vec{k}}^ {}
+\sum_{\vec{k}}{\varepsilon}_{\vec{k}}^{h} h_{-\vec{k}}^{\dagger}h_{-\vec{k}}^ {}    
+ \sum_{\vec{q}}\omega_{\vec{q}}
\psi_{\vec{q}}^{\dagger}\psi_{\vec{q}}   \,,    \\
\label{23}
&
\mathcal H_{c}=  
 \sum_{\vec{k}}(  \Delta\, e_{\vec{k}}^{\dagger}h_{-\vec{k}}^{\dagger}+\textrm{H.c.})
+ \sqrt{N} ( \Gamma \psi_{0}^{\dagger}+\textrm{H.c.}) \,,
 \\
\label{24}
& \mathcal H_{g} = 
-\frac{g}{\sqrt{N}}\sum_{\vec{k}\vec{q}}(e_{\vec{q}+\vec{k}}^{\dagger}
h_{-\vec{k}}^{\dagger}\psi^{}_{\vec{q}}+\textrm{H.c.}) \,,\\
\label{25}
& \mathcal H_{U} =  -\frac{U}{N}\sum_{\vec{k}_{1}\vec{k}_{2}\vec{k}}
e_{\vec{k}_{1}+\vec k}^{\dagger}e^{}_{\vec{k}_{1}}\, h_{\vec{k}_{2}-\vec k}^{\dagger}h^{}_{\vec{k}_2} \,.
 \end{align}
Here, the first term $\mathcal H_0$ of  $\mathcal H_{\rm S}$ is diagonal,
whereas the second part $\mathcal H_c$ is non-diagonal and contains infinitesimally 
small external fields $  \Delta=0^+$ and $ \Gamma =0^+$, which are  
introduced  to account for possible ground-state phases with broken gauge symmetry.
As shown below,  in the course of the renormalization procedure, the fields $  \Delta$ and $ \Gamma$ 
take over the role of order parameters for the exciton and photon condensates.  
 Finally, the terms $ \mathcal H_{g}$ and  $ \mathcal H_{U}$  in Eqs.~\eqref{24} and \eqref{25} 
  stand for the interactions between excitons  and photons  and  
 for the Coulomb attraction between electrons and holes.

 The remaining terms  in Eq.~\eqref{20} are the reservoir Hamiltonian  $\mathcal H_{\rm R}$ and the interaction Hamiltonian $\mathcal H_{\rm SR}$ between the reservoirs and the e-h-p system. Written in the variables introduced in Eqs.~\eqref{17} and~\eqref{18}, they have the same operator structure as Eqs.~\eqref{10} and \eqref{11}:
\begin{align}
\label{26}
&\mathcal H_{\rm R} =  \sum_{\vec p} \omega^e_{\vec p} \, b^\dag_{e, \vec p} b_{e, \vec p}  +
 \sum_{\vec p} \omega^h_{\vec p} \, b^\dag_{h, \vec p} b_{h, \vec p} 
  + \sum_{\vec p} \omega^\varphi_{\vec p} \, \varphi^\dag_{\vec p} \varphi_{\vec p} \, , \\ 
\label{27}
& \mathcal H_{\rm SR}=\sum_{\vec k \vec p}( \Gamma^e_{\vec k \vec p} \, e^\dag_{\vec k} b_{e,\vec p} + {\rm H.c.})+  \sum_{\vec k \vec p}( \Gamma^h_{\vec k \vec p} \,  h^\dag_{-\vec k} b_{h,-\vec p} + {\rm H.c.}) \nonumber \\ 
&\qquad\quad\qquad\qquad + \sum_{\vec q \vec p}( \Gamma^{\psi}_{\vec q \vec p} \,  \psi_{\vec q}^\dag \varphi_{\vec p} + {\rm H.c.})
 \, . 
\end{align}

In order to separate the mean-field contributions 
from $\mathcal H_{g}$ and $\mathcal H_{U}$, we introduce time ordered operators: 
\begin{align}
\label{28a}
&:e^\dag_{\vec k + \vec q}h^\dag_{-\vec k} \psi_{\vec q}:  = \, : e^\dag_{\vec k + \vec q} h^\dag_{-\vec k}: \, 
:\psi_{\vec q}: =  e^\dag_{\vec k + \vec q} h^\dag_{-\vec k} \psi_{\vec q} \\
&\qquad\qquad\qquad\qquad- \delta_{\vec q,0} \big( d^*_{\vec k}\,  :\psi_0: + \langle \psi_0\rangle\, :e^\dag_{\vec k} h_{-\vec k}^\dag: 
\big)\,, \nonumber 
\end{align}
\begin{align}
\label{28b}
&:e_{\vec{k}_{1}+\vec k}^{\dagger}e^{}_{\vec{k}_{1}}\,  h_{\vec{k}_{2}-\vec k}^{\dagger}h^{}_{\vec{k}_2}: \, =
e_{\vec{k}_{1}+\vec k}^{\dagger}e^{}_{\vec{k}_{1}} \,  h_{\vec{k}_{2}-\vec k}^{\dagger}h^{}_{\vec{k}_2} \\
& \quad - \delta_{\vec k,0} \, \big(n^e_{\vec k_1} h^\dag_{\vec k_2} h_{\vec k_2} + 
n^h_{\vec k_2} e^\dag_{\vec k_1} e_{\vec k_1} -  n^e_{\vec k_1}\, n^h_{\vec k_2} \big)
\nonumber \\
& \quad - \delta_{\vec k_1, -\vec k_2} \big( d^*_{\vec k+ \vec k_1} :h_{-\vec k_1} e_{\vec k_1}: +
d_{\vec k_1}\, :e^\dag_{\vec k + \vec  k_1} h^\dag_{-\vec k - \vec k_1}:
\big) \, .\nonumber 
\end{align}
Here, $:\mathcal A: \,=  \,\mathcal A - \langle \mathcal A \rangle $, and
 $n^e_{\vec k_1}$ and $n^h_{\vec k_2} $ are occupation numbers evaluated with 
 the density operator $\rho_0$: 
\begin{equation}
\label{29}
n^e_{\vec k_1} = \langle e^\dag_{\vec k_1} e_{\vec k_1}\rangle \, , \quad 
 n^h_{\vec k_2} = \langle  h^\dag_{\vec k_2} h_{\vec k_2}\rangle \, .
\end{equation} 
Obviously, a finite  $d_{\vec k}^*$ indicates a particle-hole (exciton) condensate:
\begin{equation}
\label{30}
d^*_{\vec k} = \langle e^\dag_{\vec k} h^\dag_{-\vec k}\rangle \, .
\end{equation}
With Eqs.~\eqref{28a}--\eqref{28b} Hamiltonian $\mathcal H_{\rm S}$ is rewritten as 
\begin{equation}
\label{31}
 \mathcal H_{\rm S} =  \hat {\mathcal H}_{0} + \hat {\mathcal H}_{c} + \hat {\mathcal H}_{g} 
+ \hat {\mathcal H}_{U} 
 \, ,
\end{equation}
where $\hat { \mathcal H}_0$ and $\hat {\mathcal H}_{c}$ have acquired one-particle contributions from 
separations \eqref{28a} and \eqref{28b}:
 \begin{align} 
\label{32}
 &\hat{\mathcal{H}}_0 = \sum_{\vec{k}} \hat{\varepsilon}_{\vec{k}}^{e}
e_{\vec{k}}^{\dagger}e_{\vec{k}}^ {}
+\sum_{\vec{k}}\hat {\varepsilon}_{\vec{k}}^{h} h_{-\vec{k}}^{\dagger}h_{-\vec{k}}^ {}    
+ \sum_{\vec{q}}\omega_{\vec{q}}
\psi_{\vec{q}}^{\dagger}\psi_{\vec{q}}\,, \\\label{33}
&\hat {\mathcal H}_{c}=  \sum_{\vec{k}}(\hat \Delta \, e_{\vec{k}}^{\dagger}h_{-\vec{k}}^{\dagger}+\textrm{H.c.})
 + \sqrt N ( \hat \Gamma \psi^\dag_0 + {\rm  H.c.} )\, .
\end{align} 
Thereby, the field parameters $  \Delta$ and $ \Gamma$  have changed into 
\begin{align}
\label{34a}
& \hat\Delta=  \Delta  - \frac{g}{\sqrt N} \langle \psi_0 \rangle  - \frac{U}{N} \sum_{\vec k} d_{\vec k}\,,   \\\label{34b}
&  \hat\Gamma = \Gamma - \frac{g}{N} \sum_{\vec k} d_{\vec k} \, ,
 \end{align}
and  the electronic one-particle  energies contain the Hartree shifts:
\begin{align} \label{35a}
  &\hat {\varepsilon}_{\vec{k}}^{e}=\varepsilon_{\vec{k}}^{e}
    - \frac{U}{N}\sum_{\vec{q}} n^h_{-\vec q} \,, \\ \label{35b}
   &  \hat{\varepsilon}_{\vec{k}}^{h}=\varepsilon_{\vec{k}}^{h}-\frac{U}{N}\sum_{\vec{q}} n^e_{\vec q} \,.
\end{align}
Finally, the former interactions \eqref{24}
and \eqref{25} have changed  into $\hat {\mathcal H}_{g}$ and $\hat {\mathcal  H}_{U}$, which now
consist of  fluctuation operators only:
 \begin{align}
\label{36}
\hat {\mathcal H}_{g} &=
-\frac{g}{\sqrt{N}}\sum_{\vec{k}\vec{q}}( :e_{\vec{q}+\vec{k}}^{\dagger}
h_{-\vec{k}}^{\dagger} \, \psi^{}_{\vec{q}}: +\textrm{H.c.})\,, \\
\label{37}
\hat {\mathcal H}_{U}&=  -\frac{U}{N}\sum_{\vec{k}_{1}\vec{k}_{2}\vec{k}}
:e_{\vec{k}_{1}+\vec k}^{\dagger}e^{}_{\vec{k}_{1}}\,  h_{\vec{k}_{2}-\vec k}^{\dagger}h^{}_{\vec{k}_2}:  \,.
 \end{align}


\section{PRM for an open electron-hole-photon system}
\label{III}

Applying the projector-based renormalization approach~\cite{BHS02,PBF16} to the open exciton-polariton system, one starts, as usual,  
from an appropriate separation of the total Hamiltonian  $\mathcal H$ into an ``unperturbed" part $\mathcal H_0$ and a ``perturbation'' $\mathcal H_1$. In a many-particle system, $\mathcal H_1$ is usually the interaction, which prevents a straightforward solution of $\mathcal H$ since it leads to transitions between the eigenstates of $\mathcal H_0$. However, integrating out the interaction by a sequence of small unitary transformations, the Hamiltonian can be transformed into a diagonal operator. Thereby, transitions from $\mathcal H_1$ between eigenstates of $\mathcal H_0$  will be stepwise eliminated. For the actual evaluation one starts from the largest transition energy of $\mathcal H_0$, called $\Lambda$, and proceeds in small steps $\Delta \lambda$ to lower  transition energies $\lambda$. Suppose all transitions between $\Lambda$ and $\lambda$ have already been eliminated, the resulting Hamiltonian, which contains only transitions with energies smaller than $\lambda$,  will be called $\mathcal H_\lambda$. An additional elimination step from $\mathcal H_\lambda$ to  a new Hamiltonian $\mathcal H_{\lambda - \Delta \lambda}$  with a somewhat reduced maximum transition energy $\lambda - \Delta \lambda$ is performed by means of a small unitary transformation, 
\begin{equation}
 \label{38} 
 \mathcal H_{\lambda -\Delta \lambda} = e^{X_{\lambda, \Delta \lambda}} \, \mathcal H_\lambda \,
 e^{-X_{\lambda, \Delta \lambda}} \, ,
\end{equation}
by which  all excitations in $\mathcal H_\lambda$ between 
$\lambda$ and $\lambda - \Delta \lambda$ will be eliminated.
Here, $X_{\lambda, \Delta \lambda} =- X_{\lambda, \Delta \lambda}^\dag$ 
is the generator of the unitary transformation. Its lowest-order expression is given by\cite{BHS02} 
\begin{equation}
\label{39}
X_{\lambda, \Delta \lambda} = \frac{1}{\mathbf L_{0, \lambda}} {\mathbf Q}_{\lambda - \Delta \lambda} {\mathcal H_{1, \lambda}} \, ,
\end{equation}
 Here, the quantities $\mathbf Q_{\lambda - \Delta \lambda}$ and $\mathbf L_{0,\lambda}$ 
are so-called superoperators which act on usual operators of the unitary space. Thereby
${\mathbf Q}_{\lambda - \Delta \lambda}= 1 - {\mathbf P}_{\lambda - \Delta \lambda}$ 
is a generalized projector that projects on all transition operators
(with respect to the unperturbed Hamiltonian $\mathcal H_0$) with energies larger than $\lambda -\Delta \lambda$, whereas ${\mathbf P}_{\lambda - \Delta \lambda}$ 
is the orthogonal projector, which  project on all transition operators with energies smaller than $\lambda -\Delta \lambda$.  Examples for the action of 
$\mathbf P_\lambda$ and $\mathbf Q_{\lambda}$ are found in the Secs. \ref{III.A} and \ref{III.B} below. 
Moreover, 
$\mathbf L_{0,\lambda}$ is the Liouville operator, which is defined by the commutator with $\mathcal H_{0,\lambda}$
applied to any operator variable $\mathcal A$, 
i.e., $\mathbf L_{0,\lambda} \mathcal A = [\mathcal H_{0,\lambda}, \mathcal A]$. The explicit form of the generator $X_{\lambda, \Delta \lambda}$ 
is given in Eqs.~\eqref{57} to \eqref{61}. 

We note that after each elimination step the unperturbed Hamiltonian as well as the perturbation become renormalized 
and therefore depend on $\lambda$.  Continuing  the renormalization scheme stepwise up to zero transition energy 
$\lambda=0$ all transitions with energies larger than  zero will be integrated out: In this way one arrives 
at a  fully renormalized Hamiltonian $\mathcal H_{\lambda=0}$,  which is diagonal (or quasi-diagonal) and therefore solvable. 
We finally like to point out that for  sufficiently small $\Delta \lambda$, the evaluation of the transformation step \eqref{38}
can be restricted to low orders in $\mathcal H_1$ which, in general, limits the validity of the 
approach to parameter values of $\mathcal H_1$ of the same magnitude as those of $\mathcal H_0$.

\subsection{Ansatz for the system Hamiltonian }
\label{III.A}
As  above mentioned the reservoirs are considered to be very large. Thus $\mathcal H_{\rm R}$
and $\mathcal H_{\rm SR}$ will not be renormalized by the PRM procedure. We 
therefore may restrict the renormalization to the e-h-p system only, and employ 
the following $\lambda$-dependent {\it ansatz} for $\mathcal H_{\rm S,\lambda}$, 
\begin{equation}
\label{40}
\mathcal H_{\rm S} \rightarrow \mathcal H_{\rm S, \lambda}= { \mathcal H}_{0,\lambda} + 
\hat {\mathcal H}_{c, \lambda}  
+ \hat {\mathcal H}_{g,\lambda} + \hat {\mathcal H}_{U, \lambda}\, ,
\end{equation}
where the operator structure of \eqref{40} is found from Eq.~\eqref{38}
by an expansion around $\lambda= \Lambda$ for small interactions
$\hat{\mathcal H}_{g}  + \hat{\mathcal H}_{U}$.  
As above-mentioned the parameters in $ \mathcal H_{0,\lambda}$ and $\hat {\mathcal H}_{c, \lambda}$ 
 depend on $\lambda$:
 \begin{align}
 \label{41}
 {\mathcal H}_{0,\lambda} &= \sum_{\vec{k}}{\varepsilon}_{\vec{k},\lambda}^{e}
e_{\vec{k}}^{\dagger}e_{\vec{k}}^ {}
+\sum_{\vec{k}}{\varepsilon}_{\vec{k},\lambda}^{h} h_{-\vec{k}}^{\dagger}h_{-\vec{k}}^ {}  
 + \sum_{\vec{q}}\omega_{\vec{q},\lambda}
\psi_{\vec{q}}^{\dagger}  \psi_{\vec{q}}\,,\\
 \label{42}
 \hat{\mathcal{H}}_{c,\lambda}&=\sum_{\vec{k}}
( \hat \Delta_{\vec k, \lambda} \, e_{\vec{k}}^{\dagger}h_{-\vec{k}}^{\dagger}+\textrm{H.c.}) 
+ \sqrt N ( \hat \Gamma_\lambda \psi^\dag_0 + {\rm H.c.} )
\, . 
\end{align}
Moreover, the quantity $\hat \Delta_{\vec k, \lambda}$  
has acquired an additional $\vec k$ dependence. 
The interactions take the form
\begin{align}
\label{43}
&\hat{\mathcal{H}}_{{g},\lambda} 
= -\frac{g}{\sqrt{N}} \sum_{\vec{k}\vec{q}}
 \mathbf P_\lambda \big(: e_{\vec{k}+\vec{q}}^{\dagger}h_{-\vec{k}}^{\dagger}\, 
\psi_{\vec{q}}:  +\textrm{H.c.} \big)\,, \\
\label{44}
&\hat{\mathcal{H}}_{{U},\lambda} = -\frac{U}{N} \sum_{\vec{k}_{1}\vec{k}_{2}\vec{k}_{3}}
\mathbf P_\lambda\big(:e_{\vec{k}_{1}}^{\dagger}e_{\vec{k}_{2}} \, 
h_{\vec{k}_{3}}^{\dagger}h_{\vec{k}_{1}+\vec{k}_{3}-\vec{k}_{2}}:  \big)  \, .   
\end{align}
As aforementioned,  $\mathbf P_\lambda=  1 - \mathbf Q_{\lambda}$ is a generalized projection operator, 
complementary to $\mathbf Q_{\lambda}$, which projects on all transition operators with energies smaller than $\lambda$. 
The coupling parameters $g$ and $U$ will remain
$\lambda$-independent in the renormalization procedure if one restricts oneself to renormalization
contributions up to order $g^2$ and $U^2$. 

Obviously the Hamiltonian $\mathcal H_{\rm S, \lambda =\Lambda}$ reduces to $\mathcal H_{\rm S}$ by construction,
provided the parameter values at the initial cutoff $\lambda = \Lambda$ fulfill 
\begin{align}
\label{45a}
&{\varepsilon}_{\vec{k},\Lambda}^{e}=\hat {\varepsilon}_{\vec{k}}^{e}\,, \quad{\varepsilon}_{\vec{k},\Lambda}^{h}=\hat{\varepsilon}_{\vec{k}}^{h}\,,\quad  \omega_{\vec q,\Lambda}= \omega_{\vec q} \,,\\
\label{45b}
 &  {\hat {\Delta}}_{\vec k,\Lambda}= \hat \Delta\,,\quad {\hat\Gamma}_{\Lambda}= \hat{\Gamma} \,.
\end{align} 

In order to study the action of $\mathbf P_\lambda$ in Eqs.~\eqref{43}  and \eqref{44} we start from the  decomposition of  
$\hat{\mathcal H}_{g, \lambda}$   into dynamical eigenmodes of $\mathcal H_{0,\lambda}$,  
 \begin{align}
\label{46}
\hat{ \mathcal{H}}_{{g},\lambda} 
&=-\frac{g}{\sqrt{N}} \sum_{\vec{k}\vec{q}} \Theta_{\vec k \vec q, \lambda} 
\big(e_{\vec{k}+\vec{q}}^{\dagger}h_{-\vec{k}}^{\dagger} \psi_{\vec{q}} +\textrm{H.c.} \big)
 \nonumber \\
 & + \frac{g}{\sqrt N} \sum_{\vec k} \Theta_{\vec k, \lambda} (  \langle \psi_{0}
 \rangle \, e^\dag_{\vec  k} h^\dag_{-\vec k}
 + {\rm H.c}) \nonumber \\
 & +\frac{g}{\sqrt N} \Theta_\lambda \sum_{\vec k} (d^*_{\vec k} \psi_{0} + {\rm H.c.}) \, ,
 \end{align}
 where Eq.~\eqref{36} was used. In Eq.~\eqref{46}, we have introduced the $\Theta$-functions
 \begin{align}
\label{47a}
 \Theta_{\vec{k}\vec{q},\lambda}&=\Theta(\lambda-{|\varepsilon_{{\bf k+{\bf q,\lambda}}}^{e}
 +\varepsilon_{-{\bf k,\lambda}}^{h}-\omega_{{\bf q,\lambda}}|}) \,,  \\\label{47b}
 \Theta_{\vec k,\lambda} &=  \Theta( \lambda -|\varepsilon_{\vec k, \lambda}^e 
 + \varepsilon_{-\vec k, \lambda}^h |)  \,,\\\label{47c}
 \Theta_\lambda &= \Theta(\lambda -|\omega_{\vec q=0, \lambda}|)\, ,
 \end{align}
which restrict transitions to those with excitation energies 
smaller than $\lambda$. Similarly one finds for $\hat{\mathcal H}_{U,\lambda}$:
  \begin{align}
 \label{48}
 \hat{\mathcal{H}}_{{U},\lambda} 
=& - \frac{U}{N} \sum_{\vec{k}_{1}\vec{k}_{2}\vec{k}_{3}}
\Theta_{\vec k_1 \vec  k_2 \vec k_3, \lambda}
:e_{\vec{k}_{1}}^{\dagger}e_{\vec{k}_{2}}: \, :h_{\vec{k}_{3}}^{\dagger}h_{\vec{k}_{1}+\vec{k}_{3}-\vec{k}_{2}}:  \nonumber \\
 & + \frac{U}{N} \sum_{\vec k} \Theta_{\vec k, \lambda}  \sum_{\vec k'} 
 (d_{\vec k'}  \, e^\dag_{\vec k} h^\dag_{-\vec k} + {\rm H.c.}  )
\end{align}
with 
 \begin{align}
\label{49}
  & \Theta_{\vec{k}_{1}\vec{k}_{2}\vec{k}_{3},\lambda}=\Theta(\lambda-|\varepsilon_{{\bf {\bf k_{1},\lambda}}}^{e}-\varepsilon_{{\bf k_{2},\lambda}}^{e}
 +\varepsilon_{{\bf {\bf k_{3},\lambda}}}^{h}-\varepsilon_{\vec{k}_{1}+{\bf k_{3}-k_{2},\lambda}}^{h}|).
  \end{align}

 In principle, the operator part  ${ \mathcal H}_{0,\lambda} + 
\hat {\mathcal H}_{c, \lambda}$ of  the {\it ansatz}  \eqref{40} for $\mathcal H_{\rm S}$
should take over the role of the unperturbed Hamiltonian and 
$\hat {\mathcal H}_{g,\lambda} + \hat {\mathcal H}_{U, \lambda}$ the role of the perturbation. 
This however would require a
 diagonalization of  ${ \mathcal H}_{0,\lambda} +  \hat {\mathcal H}_{c, \lambda}$ and an expansion of $\hat{\mathcal H}_{g, \lambda}$
and $\hat{\mathcal H}_{U, \lambda}$ into eigenmodes of this ``unperturbed" Hamiltonian. Since this procedure is 
rather complex, we prefer to use instead $\mathcal H_{0,\lambda}$ in the $\Theta$-functions of Eqs.~\eqref{46} and \eqref{48}.
Then the generator $X_{\lambda, \Delta \lambda}$ of the unitary transformation \eqref{38} has to be changed appropriately (see below).

One sees that  the last two terms in Eq.~\eqref{46} 
and the last term in \eqref{48} represent one-particle contributions. They should  best be included 
in the one-particle term  $\hat{\mathcal H}_{c,\lambda}$ of $\mathcal H_{\rm S, \lambda}$. 
That is only the first term  in Eq.~\eqref{46} and in Eq.~\eqref{48} should be considered as  ``true'' interactions. 
However, it has turned out that interactions formed by fluctuation operators  
should be  preferred in the unitary transformation Eq.~\eqref{38}. Therefore, instead of Eqs.~\eqref{46} and \eqref{48}, we   henceforth
use modified interactions $\mathcal H_{g,\lambda}$ and $\mathcal H_{U,\lambda}$ based on 
fluctuation operators,
 \begin{align}
\label{50}
&\mathcal H_{{g},\lambda} =-\frac{g}{\sqrt{N}} \sum_{\vec{k}\vec{q}} \Theta_{\vec k \vec q, \lambda} 
\big(:e_{\vec{k}+\vec{q}}^{\dagger}h_{-\vec{k}}^{\dagger} \psi_{\vec{q}}:  +\textrm{H.c.} \big)  \,,\\
 \label{51}
 &\mathcal{H}_{{U},\lambda} 
= - \frac{U}{N} \sum_{\vec{k}_{1}\vec{k}_{2}\vec{k}_{3}}
\Theta_{\vec k_1 \vec  k_2 \vec k_3, \lambda}
:e_{\vec{k}_{1}}^{\dagger}e_{\vec{k}_{2}} \, h_{\vec{k}_{3}}^{\dagger}h_{\vec{k}_{1}+\vec{k}_{3}-\vec{k}_{2}}:   \, ,
\end{align}
where the $\Theta$-functions in front apply to all parts of the respective fluctuation operators.   
Of course, we have to repair this ``mistake'' by including the 
corresponding ``counter-terms''  in  the one-particle part $\mathcal H_{c, \lambda}$ of $\mathcal H_{\rm S, \lambda}$. 
Thus, we finally arrive at the  following representation of $\mathcal H_{\rm S, \lambda}$: 
\begin{align}
\label{52a}
\mathcal H_{\rm S,\lambda} &= \mathcal H_{0,\lambda} + {\mathcal H}_{c, \lambda} 
+ \mathcal H_{1,\lambda} \,,  \\\label{52b}
\mathcal H_{1,\lambda} &= \mathcal{H}_{{g},\lambda} + \mathcal{H}_{{U},\lambda} \, .
\end{align}
Here, $ \mathcal{H}_{{g},\lambda}$ and $ \mathcal{H}_{{U},\lambda}$  
are given by Eqs.~\eqref{50} and \eqref{51},  whereas ${\mathcal H}_{c,\lambda}$ reads 
 \begin{equation}
 \label{53}
{\mathcal H}_{c,\lambda}=\sum_{\vec{k}}
( {\Delta}_{\vec k, \lambda} \, e_{\vec{k}}^{\dagger}h_{-\vec{k}}^{\dagger}+\textrm{H.c.}) 
+ \sqrt N ({\Gamma}_\lambda  \psi^\dag_0 + {\rm H.c.} )
\, ,
\end{equation}
with
\begin{align}
\label{54}
 {\Delta}_{\vec k, \lambda} & = \hat \Delta_{\vec k, \lambda}  + 
\frac{g}{\sqrt N} (\Theta_{\vec k, \lambda} - \Theta_{\vec k, \vec q=0, \lambda}) 
\langle  \psi_{0 } \rangle   \nonumber \\
&\quad+\frac{U}{N}\sum_{\vec k'} (\Theta_{\vec k,\lambda} - \Theta_{\vec k, \vec k', \vec k; \lambda}) \, d_{\vec k'}\,, \\
\label{55}
 {\Gamma}_{\lambda} & =\hat \Gamma_{\lambda} + 
\frac{g}{N} \sum_{\vec k}( \Theta_\lambda - \Theta_{\vec k, \vec q=0, \lambda})\,  d_{\vec k} \, .
\end{align}
As before, at the initial cutoff $\lambda = \Lambda$ Hamiltonian $\mathcal H_{\rm S, \lambda}$ must agree 
with $\mathcal H_{\rm S}$ (from Eq.~\eqref{21}), which is fulfilled by ensuring Eqs.~\eqref{45a} and~\eqref{45b}.

Let us add one remark: Carrying out the renormalization procedure 
 the additional contributions 
 in $\Delta_{\vec k, \lambda}$ and $\Gamma_\lambda$ in Eqs.~\eqref{54} and \eqref{55} are expected to have very  little influence on the results  
 since they vanish both at the beginning (cutoff $\Lambda$)  and  at the end ($\lambda=0$) of the PRM 
 procedure.

\subsection{Construction of the PRM generator}
\label{III.B}

Next, we establish the generator  $X_{\lambda, \Delta \lambda}$ of the unitary
transformation \eqref{38}. Following the lowest order expression \eqref{39}, we look for an   
$X_{\lambda, \Delta \lambda}$ having the same operator structure as $\mathcal H_{1,\lambda}$.  
For this we make the ansatz
\begin{eqnarray}
\label{57}
X_{\lambda, \Delta \lambda} &=& X^{g}_{\lambda,\Delta\lambda} + X^{U}_{\lambda,\Delta\lambda}
= - X_{\lambda, \Delta \lambda}^\dag
\end{eqnarray}
with
\begin{eqnarray}
\label{58}
 X^{g}_{\lambda,\Delta\lambda} &=&-\frac{g}{\sqrt{N}}\sum_{\vec{k}\vec{q}}
A_{{\bf k{\bf q}}}(\lambda,\Delta\lambda)
\big[:e_{\mathbf{k+q}}^{\dagger}h_{-\mathbf{k}}^{\dagger}\psi_{\mathbf{q}}:  -\textrm{H.c.}\big]\,,\nonumber\\
 && \\
\label{59}
 X^{U}_{\lambda,\Delta\lambda} &=&  -\frac{U}{N}
 \sum_{\vec{k}_{1}\vec{k}_{2}\vec{k}_{3}} B_{\vec k_1 \vec k_2; \vec k_3,\vec k_1 + \vec k_3 -\vec k_2 }(\lambda,\Delta\lambda) \nonumber \\
 && \times  :e_{\vec{k}_{1}}^{\dagger}e_{\vec{k}_{2}} \,  h_{\vec{k}_{3}}^{\dagger}h_{\vec{k}_{1}+\vec{k}_{3}-\vec{k}_{2}}:  \, ,
\end{eqnarray}
and
 \begin{align}
\label{60}
 & A_{{\bf k{\bf q}}}^ {}(\lambda,\Delta\lambda)=\frac{\Theta_{{\bf k{\bf q,\lambda}}}\big(1-\Theta_{{\bf k{\bf q,\lambda-\Delta\lambda}}}\big)}{{\varepsilon_{{\bf k+{\bf q,\lambda}}}^{e}+\varepsilon_{-{\bf k,\lambda}}^{h}-\omega_{{\bf q,\lambda}}^ {}}}  \,,\\
 & \nonumber \\
 \label{61}
 & B_{\vec k_1, \vec k_2;\vec k_3, \vec k_1 +\vec k_3 -\vec k_2}(\lambda,\Delta\lambda) \nonumber \\
 & \hspace*{1cm} =
 \frac{\Theta_{{\bf {\bf k{\bf _{1}k_{2}k_{3}},\lambda}}}\big(1-\Theta_{{\bf k{\bf _{1}k_{2}k_{3}},\lambda-\Delta\lambda}}\big)}{\varepsilon_{{\bf {\bf k_{1},\lambda}}}^{e}-\varepsilon_{{\bf k_{2},\lambda}}^{e}+\varepsilon_{{\bf {\bf k_{3},\lambda}}}^{h}-\varepsilon_{\vec{k}_{1}+{\bf k_{3}-k_{2},\lambda}}^h} \nonumber \\ 
&  \hspace*{1cm} = - B_{\bf k_2, k_1; \vec k_1 +\vec k_3 -\vec k_2,\vec k_3, }(\lambda,\Delta\lambda)\,.
\end{align}
 Here, the notation with four indices in $B_{\vec k_1,  \vec k_2, \vec k_3, \vec k_1+ \vec k_3 -\vec k_2}(\lambda, \Delta \lambda)$ 
 emphasizes the momentum conservation.  
It can be recognized that the products of $\Theta$-functions in Eqs.~\eqref{60} and \eqref{61} 
assure that excitations between $\lambda$
and $\lambda -\Delta \lambda$ are eliminated in each transformation step  $\Delta \lambda$.  
For small $\Delta \lambda$, the transformation \eqref{38} 
can be restricted to an expansion up to second order in $g$ and $U$, and to linear order in the order parameters 
$\Delta_{\vec k,\lambda}$ and $\Gamma_\lambda$. Then 
$\mathcal H_{{\rm S}, \lambda- \Delta \lambda}$ at the reduced cutoff $\lambda- \Delta \lambda$ reads 
\begin{eqnarray}
\label{62}
 \mathcal H_{\rm S, \lambda -\Delta \lambda} = && 
\mathcal H_{0,\lambda}  +  {\mathcal H}_{c, \lambda} 
+ \mathcal H_{1,\lambda} \nonumber \\
&&  +  [X_{\lambda, \Delta \lambda}, \mathcal H_{0,\lambda} + {\mathcal H}_{c, \lambda} + \mathcal H_{1, \lambda}   ] +  \cdots  
\end{eqnarray}
Relation \eqref{62} connects the parameter values of $\mathcal H_{\rm S, \lambda}$ at cutoff 
$\lambda$ with those at the reduced cutoff $\lambda -\Delta \lambda$.  That is, in order to find renormalization equations for the 
$\lambda$-dependent parameters
one has to evaluate the commutators. For instance,  from the first commutator $ [X_{\lambda, \Delta \lambda}, \mathcal H_{0,\lambda}]$, one finds the following 
renormalization contributions to $\Delta_{\vec k, \lambda}$ and $\Gamma_{\lambda}$:
\begin{align}
\label{63}
  \delta  {\Delta}_{\vec k, \lambda}^{(0)} &=   - \frac{g}{\sqrt N} A_{\vec k 0}(\lambda, \Delta \lambda) \,  
\omega_{0,\lambda} \langle \psi_{0}\rangle  \\
  & \quad - \frac{U}{N} \sum_{\vec k_1} B_{\vec k_1 \vec k, -\vec k_1, -\vec k}(\lambda, \Delta \lambda)  
(\varepsilon^e_{\vec k_1, \lambda} + 
\varepsilon^h_{-\vec k_1, \lambda}) \, d_{\vec k_1}\,, \nonumber \\
\label{64}
 \delta {\Gamma}_{\lambda}^{(0)}  &= \frac{g}{N} \sum_{\vec k} A_{\vec k 0}(\lambda, \Delta \lambda) 
(\varepsilon^e_{\vec k,\lambda} + \varepsilon^h_{-\vec k, \lambda}) \,  d_{\vec k} \, . 
\end{align}
Combining these relations with the remaining renormalization contributions from the last two commutators 
in \eqref{62},  one arrives at the following renormalization equations: 
 \begin{eqnarray}
\label{65}
\Delta_{\vec k, \lambda -\Delta \lambda} &=& \Delta_{\vec k, \lambda}
+  \delta  {\Delta}_{\vec k, \lambda}^{(0)} 
 +  \delta  {\Delta}_{\vec k, \lambda}^{(c)} 
+ \delta  {\Delta}_{\vec k, \lambda}^{(U)}\,, \\
\label{66}
\Gamma_{\lambda -\Delta \lambda} &=& \Gamma_{\lambda} + \delta \Gamma_\lambda^{(0)} 
+ \delta \Gamma_\lambda^{(c)}  + \delta \Gamma_\lambda^{(g)} 
\, .  
\end{eqnarray}
Here, $\delta  {\Delta}_{\vec k, \lambda}^{(c)}$ and $\delta  {\Delta}_{\vec k, \lambda}^{(U)}$ 
are defined in Eqs.~\eqref{A6ab} and \eqref{A15}, whereas  $\delta \Gamma_\lambda^{(c)}$ and 
$\delta \Gamma_\lambda^{(g)}$ are given in \eqref{A6a} and \eqref{A9d}, respectively.   The renormalization equations 
for the remaining parameters $\varepsilon^e_{\vec k, \lambda}$, $\varepsilon^h_{\vec k, \lambda}$, 
and $\omega_{\vec q, \lambda}$ of $\mathcal H_\lambda$ are derived in Appendix \ref{A} [Eqs.~\eqref{A9a}-\eqref{A9c}] as well. 

To solve the renormalization equations, one starts 
from the initial parameter values at cutoff $\Lambda$ [Eqs.~\eqref{45a} and~seems not\eqref{45b}] and proceeds in small steps $\Delta \lambda$
until $\lambda=0$ is reached. In doing so, all transitions from $\mathcal H_{1,\lambda}$ between $\Lambda$ and $\lambda=0$ will be eliminated.
We arrive at the fully renormalized Hamiltonian $\tilde{\mathcal H}_{\rm S}= \mathcal H_{{\rm S},\lambda=0}
= \mathcal H_{0, \lambda =0}+ \mathcal H_{c, \lambda =0}$: 
\begin{eqnarray}
\label{67}
\tilde{\mathcal H}_{\rm S}& = & \sum_{\vec{k}}\tilde{\varepsilon}_{\vec{k}}^{e}e_{\vec{k}}^{\dagger}e_{\vec{k}}^{}+\sum_{\vec{k}}\tilde{\varepsilon}_{\vec{k}}^{h}h_{\vec{k}}^{\dagger}h_{\vec{k}}^{}
+\sum_{\vec q}\tilde\omega_{\vec q}\psi_{\vec q}^\dagger  \psi_{\vec q}  \\
& &+  \sum_{\vec k}( \tilde \Delta_{\vec k} \, e_{\vec k}^\dagger h_{-\vec k}^\dagger+\textrm{H.c.}) 
+ \sqrt N (\tilde \Gamma \psi^\dag_0 +{\rm H.c.}) \nonumber \, . \end{eqnarray}
Accordingly, $\tilde{\varepsilon}^e_{\mathbf{k}}$, $\tilde{\varepsilon}^h_{\mathbf{k}}$, $\tilde{\omega}_{\bf q}$, 
$\tilde{\Delta}_{\vec k}$, and $\tilde \Gamma$ are the fully renormalized energy parameters at $\lambda =0$. 
They have to be determined self-consistently from the whole set of renormalization equations.

Since all transition operators from $\mathcal H_{1,\lambda}$ have been used up 
in the renormalization procedure, Hamiltonian $\tilde{\mathcal H}_{\rm S}$ is a one-particle 
operator which can be diagonalized. First, one defines ``displaced'' photon operators 
 \begin{equation}
\label{68}
\tilde \Psi^\dag_{\vec q} = \psi^\dag_{\vec q} 
+ \frac{\sqrt{N} \tilde \Gamma^*}{\tilde \omega_{\vec q=0}} \, \delta_{\vec q,0} \, ,
\end{equation}
which--up to a constant--leads to  
\begin{eqnarray}
\label{69}
\tilde{\mathcal H}_{\rm S}& = & \sum_{\vec{k}}\tilde{\varepsilon}_{\vec{k}}^{e}e_{\vec{k}}^{\dagger}e_{\vec{k}}^{}+\sum_{\vec{k}}\tilde{\varepsilon}_{\vec{k}}^{h}h_{\vec{k}}^{\dagger}h_{\vec{k}}^{}
+\sum_{\vec q}\tilde\omega_{\vec q} \tilde \Psi_{\vec q}^\dagger  \tilde\Psi_{\vec q} \nonumber \\
& + & \sum_{\vec k}( \tilde \Delta_{\vec k} \, e_{\vec k}^\dagger h_{-\vec k}^\dagger+\textrm{H.c.}) 
 \, . 
\end{eqnarray}
The electronic part of $\tilde{\mathcal H}_{\rm S}$  
is diagonalized by a subsequent Bogolyubov transformation 
\begin{eqnarray}
\label{70}
C_{1\vec{k}}^{\dagger} & = & \xi_{\vec{k}}e_{\vec{k}}^{\dagger}+\eta_{\vec{k}}^* h_{-\vec{k}}^ {}\,,  \\
\label{71}
C_{2\vec{k}}^{\dagger} & = & -\eta_{\vec{k}}e_{\vec{k}}^{\dagger}+ \xi_{\vec{k}}^* h_{-\vec{k}}^ {}
\end{eqnarray}
with coefficients  
\begin{eqnarray}
\label{72} 
 && 
|\xi_{\vec{k}} |^{2}=\frac{1}{2}\left[1+\textrm{sgn}(\tilde{\varepsilon}_{\vec{k}}^{e}+\tilde{\varepsilon}_{\vec{k}}^{h})\frac{\tilde{\varepsilon}_{\vec{k}}^{e}+\tilde{\varepsilon}_{\vec{k}}^{h}}{W_{\vec{k}}}\right]\,,   \\
  && 
  \label{73}
|\eta_{\vec{k}}|^{2}=\frac{1}{2}\left[1-\textrm{sgn}(\tilde{\varepsilon}_{\vec{k}}^{e}+\tilde{\varepsilon}_{\vec{k}}^{h})
 \frac{\tilde{\varepsilon}_{\vec{k}}^{e}+\tilde{\varepsilon}_{\vec{k}}^{h}}{W_{\vec{k}}}\right]\,,  \\
\label{74}
&&\eta_{\vec k} \xi_{\vec k} = {\rm sgn}(\tilde{\varepsilon}_{\vec{k}}^{e}+\tilde{\varepsilon}_{\vec{k}}^{h})
\frac{\tilde\Delta_{\vec k}}{W_{\vec k}}\,,
\end{eqnarray}
where
\begin{eqnarray}
&& \label{75}
W_{\vec{k}}=\sqrt{(\tilde{\varepsilon}_{\vec{k}}^{e}+\tilde{\varepsilon}_{\vec{k}}^{h})^{2}+4|\tilde \Delta_{\vec{k}}|^{2}}
\, . \\
&& \nonumber 
\end{eqnarray}
In terms of the new quasiparticle operators $C^{(\dag)}_{1\vec k}$ and $C^{(\dag)}_{2\vec k}$ 
Hamiltonian $\tilde{\mathcal H}$ becomes diagonal:
\begin{align}
\label{76}
\tilde{\mathcal{H}}_{\rm S}= & \sum_{\vec{k}} \tilde E_{1\vec{k}}C_{1\vec{k}}^{\dagger}C_{1\vec{k}}^ {}+\sum_{\vec{k}}
\tilde E_{2\vec{k}} C_{2\vec{k}}^{\dagger}C_{2\vec{k}}^ {}
+\sum_{\vec{q}}{\tilde \omega}_{\vec{q}} \tilde {\Psi}_{\vec{q}}^{\dagger}
\tilde {\Psi}_{\vec{q}}^ {}  \end{align}
with the quasiparticle energies 
\begin{equation}
\label{77}
\tilde E_{(1,2)\vec{k}}=\frac{\tilde{\varepsilon}_{\vec{k}}^{e}-\tilde{\varepsilon}_{\vec{k}}^{h}}{2}\pm\textrm{sgn}(\tilde{\varepsilon}_{\vec{k}}^{e}+\tilde{\varepsilon}_{\vec{k}}^{h})\frac{W_{\vec{k}}}{2}\,.
\end{equation}
 As usual, the order parameter $\tilde{\Delta}_{\vec k}$ also acts  
 as gap parameter for the quasiparticle bands.

\section{Steady state}
 \label{IV}
The fully transformed (renormalized) Hamiltonian $\tilde{\mathcal H}$
of the total system is \begin{equation}
\label{78}
\tilde{\mathcal H} = \tilde{\mathcal H}_{\rm S} + \mathcal H_{\rm R} + \mathcal H_{\rm SR} \, ,
\end{equation}
 where $\tilde{\mathcal H}_{\rm S} $, $ \mathcal H_{\rm R}$, and $\mathcal H_{\rm SR}$ are given by 
 Eqs.~\eqref{76}, \eqref{26}, and \eqref{27}, respectively.
As aforementioned, $\mathcal H_{\rm R}$ and $\mathcal H_{\rm SR}$ will not be
affected by the unitary transformations \eqref{38}.
We are now going to calculate the steady-state expectation values 
$\langle  \psi^\dag_{\vec q =0} \rangle$, $\langle  e^\dag_{\vec k} h_{-\vec k}^\dag \rangle = d_{\vec k}^*$,
$\, n^e_{\vec k}$, and $n^h_{-\vec k}$. 

\subsection{Density operator for the initial state}
\label{IV.A}
First, the initial density operator $\rho_0$
must be specified. According to Eq.~\eqref{13}, $\rho_0$ is a product of the density  operator $\rho_{\rm S}$  
for  the e-h-p subsystem and the  density $\rho_{\rm R}$ for the reservoirs:
\begin{equation}
\label{79}
\rho_0 = \rho_{\rm S} \rho_{\rm R} \, .
\end{equation} 
Moreover, $\rho_{\rm R}$ factorizes into the density matrices $ \rho_{{\rm R},e }$ and $\rho_{{\rm R},h}$ 
of the two electronic baths and into  the density    $ \rho_{{\rm R}, \psi}$ 
of the  free-space photons,
\begin{eqnarray}
\label{80}
\rho_{\rm R} =  \rho_{{\rm R},e }\,  \rho_{{\rm R},h} \, \rho_{{\rm R}, \psi} \, ,
\end{eqnarray}
where
\begin{eqnarray}
\label{81a}
&& \rho_{{\rm R},e } =\frac{e^{-\beta \sum_{\vec p} (\omega^e_{\vec p} - (\mu_e - \mu/2)) \, 
b^\dag_{e,\vec p} b_{e, \vec p}}}  {Z_{{\rm R},e}}\,,  \\
\label{81b} && \rho_{{\rm R},h } =\frac{e^{-\beta \sum_{\vec p} (\omega^h_{\vec p} - (\mu_h - \mu/2)) \, 
 b^\dag_{h, \vec p} b_{h,\vec p}} } {Z_{{\rm R},h}} \, .
  \end{eqnarray}
Here, $Z_{{\rm R},e}$ and $ Z_{{\rm R},h}$ are the partition functions for the electronic baths with 
\begin{equation}
Z_{{\rm R},e/h} = 
{\rm Tr}_{e/h}\big( {e^{-\beta \sum_{\vec p} (\omega^{e/h}_{\vec p} - (\mu_{e/h} - \mu/2)) \, 
b^\dag_{e/h,\vec p} b_{e/h, \vec p}}} \big )
\end{equation}
such that ${\rm Tr}_{e/h} \rho_{{\rm R},e/h} =1$.

Note that both electronic bath energies $\omega^\alpha_{\vec p} (\alpha =e,h)$ from Eq.~\eqref{19b} include energy shifts $-(\mu/2)$ 
which however 
cancel in Eqs.~\eqref{81a} and \eqref{81b}. Therefore  $\rho_{{\rm R}, e}$ and $\rho_{{\rm R}, h}$ 
describe thermal equilibrium situations for the electronic baths with temperature $1/\beta$
and chemical potentials $\mu_e$ and $\mu_h$.
As aforementioned both electronic baths were assumed to be huge, 
i.e., they always stay in thermal equilibrium, even in the steady state.  
In contrast, the quantity $\mu$, introduced by Eqs.~\eqref{15}--\eqref{16c},  
will generally not act as a chemical potential since photons  
may ``escape'' from the e-h-p subsystem due to the leakage into the free-photon space. For vanishing coupling $\Gamma^\psi_{\vec q \vec p}$ 
of cavity photons to free-space photons the e-h-p subsystem together with the electronic baths 
will reach a new thermal equilibrium with  
 $\mu$ acting as the usual chemical potential again.


\subsection{Electronic expectation values in the long-time limit}
\label{IV.B}
Let us consider  the long-time
behavior of a general expectation value, 
\begin{equation}
\label{82}
\langle \mathcal A(t) \rangle= {\rm Tr}( \mathcal A(t) \, \rho_0)  \, ,
\end{equation}
 where--within the Heisenberg picture--the time dependence is governed by Hamiltonian $\mathcal H$ from
 Eq.~\eqref{20}. Since the total $\mathcal H$ does not commute with
 the initial density matrix $\rho_0$,  
 $[\rho_0, \mathcal H] \neq 0$, the expectation value 
 $\langle  \mathcal A(t) \rangle$ is  intrinsically  time-dependent. 
The steady-state properties are found from the  time-independent solutions of $\langle \mathcal A(t)\rangle$ 
for $t \rightarrow \infty$, which must obey 
 \begin{eqnarray}
\label{83}
&&\lim_{t \rightarrow \infty} \frac{\rm d}{\rm dt} \langle \mathcal A(t) \rangle 
= 0 \, .
\end{eqnarray}
Remember, an
explicit time-dependent factor $e^{i \mu t}$ was already extracted from Eqs.~\eqref{15}, \eqref{16a} and~\eqref{16b}.
To evaluate $\langle \mathcal A(t) \rangle$, we use the invariance property of operator expressions against unitary transformations under a trace: 
 \begin{eqnarray}
\label{84}
\langle {\mathcal A}(t) \rangle &=& {\rm Tr}(\mathcal A_{\lambda}(t) \,  \rho_{0,\lambda} ) 
=  {\rm Tr}(\tilde{\mathcal A}(t) \,  \tilde \rho_{0} ) 
  \, .
\end{eqnarray}
Here, $\mathcal A_\lambda$ and $\rho_{0,\lambda} $ are transformed operators 
at cutoff $\lambda$ 
\begin{equation}
\label{85}
 \mathcal A_\lambda= e^{X_\lambda} {\mathcal A}e^{-X_\lambda} 
 \quad {\rm and} \quad 
  \rho_{0,\lambda}  = e^{X_\lambda} {\rho_0}e^{-X_\lambda} \,.
 \end{equation} 
The exponential function $e^{X_\lambda}$  stands for a compact notation of the 
unitary transformation operator between cutoffs $\Lambda$ and $\lambda$. 
In the last equation of \eqref{84} the operators $\tilde{\mathcal A}$ and $\tilde \rho_{0}$ denote 
the fully renormalized operators at cutoff $\lambda=0$, and the time dependence is now
governed by $\tilde{\mathcal H}$. 
By contrast, the time dependence of $\mathcal A_\lambda(t)$ is given 
by the transformed  Hamiltonian ${\mathcal H}_\lambda = {\mathcal H}_{\rm S, \lambda} + \mathcal H_{\rm R} + \mathcal H_{\rm SR}$:
\begin{equation}
\label{86}
\mathcal A_{\lambda}(t)= e^{i{\mathcal H}_\lambda t} \mathcal A_\lambda e^{-i{\mathcal H}_\lambda t}\,.
\end{equation}

We now derive the steady-state results for
the electronic quantities   
\begin{eqnarray}
\label{88}
 d^*_{\vec k} &=& d^*_{\vec k}(t \rightarrow \infty)\,, \\
 \label{88.1}
 n^e_{\vec k} &=& n^e_{\vec k}(t \rightarrow \infty) \,, \\
 \label{88.2}
 n^h_{-\vec k} &=&  n^h_{-\vec k}(t\rightarrow \infty)\,.
\end{eqnarray}
Starting point are the time dependent expectation values, 
\begin{eqnarray}
\label{89a} 
 d^*_{\vec k}(t)&=& \langle(e_{\vec k}^\dag h^\dag_{-\vec k})(t) \rangle 
= \langle(\tilde e_{\vec k}^\dag \tilde h^\dag_{-\vec k})(t) \rangle_{\tilde \rho_0} 
\,, \\\label{89b} 
 n^e_{\vec k}(t) &=&\langle (e^\dag_{\vec k} e_{\vec k})(t) \rangle 
 =  \langle (\tilde e^\dag_{\vec k} \tilde e_{\vec k})(t) \rangle_{\tilde \rho_0}  \,,\\\label{89c} 
 n^h_{-\vec k}(t) &=&  \langle (h^\dag_{-\vec k} h_{-\vec k})(t) \rangle  
 = \langle (\tilde h^\dag_{-\vec k} \tilde h_{-\vec k})(t) \rangle_{\tilde \rho_0} \, ,
\end{eqnarray}
where  relation \eqref{84} was used on the right hand sides. 
The expectation values $\langle \cdots \rangle_{\tilde \rho_0}$ 
are formed with $\tilde \rho_0$, $\tilde e^\dag_{\vec k}$ 
and $\tilde h^\dag_{-\vec k}$  are the fully transformed one-particle operators, and the time dependence 
of the last expressions is formed with $\tilde{\mathcal H}$.  
According to Appendix \ref{C} an appropriate {\it ansatz}
for $\tilde e^\dag_{\vec k}$ and $\tilde h^\dag_{-\vec k}$ is   
\begin{eqnarray}
\label{90}
\tilde e_{\vec k}^\dag &=& \tilde{x}_{\vec k} e_{\vec k}^\dag
+\frac{1}{\sqrt N}   \sum_{\vec q}  \tilde t_{\vec k-\vec q,\vec q}   h_{\vec q -\vec k} \, 
:{\psi}_{\vec q}^{\dagger}: \\
&&+ \frac{1}{N}\sum_{\vec k_1\vec k_2}
\tilde{\alpha}_{\vec k_1 \vec k \vec k_2} e_{\vec k_1}^\dag
:h_{\vec k_2}^\dag h_{\vec k_1+ \vec k_2-\vec k}: \, ,\nonumber\\ 
\label{91}
\tilde h_{- \vec k}^\dag  &=& \tilde y_{\vec k} h_{-\vec k}^\dag
+\frac{1}{\sqrt{N}} \sum_{\vec q} \tilde u_{\vec k,\vec q} e_{\vec{q}+\vec k} \,
 :\psi_{\vec q}^\dag: \\
 &&+ \frac{1}{N}\sum_{\vec k_1 \vec k_2}
 \tilde \beta_{\vec k_1 \vec k_2,\vec k_2-\vec k_1-\vec k}
 :e_{\vec k_1}^\dag e_{\vec k_2}:  h_{\vec k_2-\vec k_1- \vec k}^\dag \, .\nonumber
\end{eqnarray}
Here, the operator structure is caused by the electron-photon and the
electron-electron interaction. Again, the parameters with tilde symbols are the  
fully renormalized quantities which 
result from the solution of the corresponding renormalization 
equations given in  Appendix \ref{C}. 
Inserting Eqs.~\eqref{90} and \eqref{91} into Eqs.~\eqref{89a}-\eqref{89c} one finds 
\begin{eqnarray}
\label{92}
d^*_{\vec k}(t)&=&  
 \tilde x_{\vec k} \tilde y_{\vec k}\, \hat d^*_{\vec k}(t) 
\nonumber \\
&& + \frac{1}{N} \sum_{\vec k_1} 
\big[\tilde x_{\vec k}  \tilde \beta_{\vec k_1, \vec k, - \vec k_1}
\hat n^e_{\vec k}(t)   \nonumber \\
&& +  \tilde y_{\vec k} \tilde{\alpha}_{\vec k_1,  \vec k, -\vec k_1} \times (1- \hat n^h_{-\vec k}(t)) \big]  
   \, \hat d^*_{\vec k_1}(t)
   \nonumber \\
  && 
  - \frac{1}{N^2} \sum_{\vec k_1 \vec k_2} \tilde \alpha_{\vec k_1 \vec k, -\vec k_2}
  \tilde \beta_{\vec k_2 \vec k_1, \vec k_1 -\vec k_2 -\vec k}   \, \hat n^e_{\vec k_1}(t) 
  \nonumber \\
&& \times \, (1- \hat n^h_{\vec k_1-\vec k_2-\vec k}(t) ) \, \hat d^*_{\vec k_2}(t) \,,
\end{eqnarray}
\begin{align}
\label{93}
 &n^e_{\vec k}(t) = |\tilde x_{\vec k}|^2 \hat n^e_{\vec k}(t)  \\
&\qquad + \frac{1}{N} \sum_{\vec q} |\tilde t_{\vec k -\vec q, \vec q}|^2
\,  (1- \hat n^h_{\vec k-\vec q}(t)) \,  \hat n^\psi_{\vec q}(t) \,
   \nonumber \\
&\qquad +\frac{1}{N^2} \sum_{ \vec k_1 \vec k_2} |\tilde \alpha_{\vec k_1 \vec k \vec k_2}|^2
\hat n^e_{\vec k_1}(t) \hat{n}^h_{\vec k_2}(t) (1- \hat n^h_{\vec k_1+ \vec k_2-\vec k}(t))
\nonumber \,,
\end{align}
and
\begin{eqnarray}
\label{94}
n^h_{-\vec k}(t) &=& |\tilde y_{\vec k}|^2 \hat n^h_{-\vec k}(t)  \\
&& + \frac{1}{N} \sum_{\vec q} |\tilde u_{\vec k, \vec q}|^2 \, 
\hat n^\psi_{\vec q}(t) (1- \hat n^e_{\vec k +\vec q}(t)) 
  \nonumber \\
&& + \frac{1}{N^2} \sum_{ \vec k_1 \vec k_2} |\tilde \beta_{\vec k_1 \vec k_2,  \vec k_2 -\vec k_1 -\vec k}|^2
\hat n^e_{\vec k_1}(t) (1- \hat n^e_{\vec k_2}(t) ) \nonumber \\
&&  \times  \hat{n}^h_{\vec k_2- \vec k_1 - \vec k}(t) \, ,\nonumber
\end{eqnarray}
where an additional  factorization approximation was used.
Here  $ \hat n^e_{\vec k}(t)$, $\hat n^h_{-\vec k}(t)$, and $\hat n^\psi_{\vec q}(t)$ are time-dependent  occupation numbers for electrons, holes, and photons, which  are formed with $\tilde \rho_0$:  
 \begin{eqnarray}
\label{95}
&& \hat n^e_{\vec k}(t) = \langle (e^\dag_{\vec k} e_{\vec k})(t)\rangle_{\tilde{\rho}_0} \,,
\\
\label{96}
&& \hat n^h_{-\vec k}(t) = \langle (h^\dag_{-\vec k} h_{-\vec k})(t)\rangle_{\tilde{\rho}_0}\,,  
\\
\label{97}
&& \hat n^\psi_{\vec q}(t) = \langle (:\psi^\dag_{\vec q}: \, :\psi_{\vec q}:)(t) \rangle_{\tilde \rho_0} \, .
\end{eqnarray}
The quantity $\hat n_{\vec q}^\psi$  will be evaluated in Appendix \ref{C}.
Moreover, $\hat d^*_{\vec k}(t)$ accounts for the order parameter of exciton  formation  
\begin{equation}
\label{98}
\hat d^*_{\vec k}(t) =  \langle (e^\dag_{\vec k} h^\dag_{-\vec k})(t)\rangle_{\tilde{\rho}_0}  \, .
\end{equation}
The time dependence in Eqs.~\eqref{95}--\eqref{98} is determined by $\tilde{\mathcal H}$. 
Let us clarify the factorization approximations used in Eqs.~\eqref{92}--\eqref{94} in more detail. 
As an example, we consider expression \eqref{93} for $n^e_{\vec k}(t)$.  
Starting point is Eq.~\eqref{89b}. Inserting expression \eqref{90} for $\tilde e^\dag_{\vec k}$ 
we find 
\begin{eqnarray}
\label{98a} 
&&n^e_{\vec k}(t) = |\tilde x_{\vec k}|^2  \langle (e^\dag_{\vec k} e_{\vec k})(t)\rangle_{{\tilde \rho}_0}  \nonumber \\
&&+  \frac{1}{N} \sum_{\vec q \vec q'} \tilde t_{\vec k - \vec q, \vec q} \tilde t^*_{\vec k -\vec q', \vec q'}
\langle  \big( h_{\vec q -\vec k} :\psi^\dag_{\vec q}: \, : \psi_{\vec q'}: h^\dag_{\vec q'-\vec k}\big)(t) \rangle_{{\tilde \rho}_0}  \nonumber \\
&&+ \frac{1}{N^2} \sum_{\vec k_1 \vec k_2 \vec k_1' \vec k_2'} \tilde{\alpha}^*_{\vec k_1 \vec k \vec k_2} \tilde{\alpha}_{\vec k_1' \vec k \vec k_2'} 
\nonumber \\
&& \quad \times \langle \big(  e_{\vec k_1}^\dag
:h_{\vec k_2}^\dag h_{\vec k_1+ \vec k_2-\vec k}: \, :h_{\vec k_1'+ \vec k_2'-\vec k}^\dag h_{\vec k_2'}: e_{\vec k_1'} \big)(t) \rangle_{{\tilde \rho}_0}\,. \nonumber \\
&&
\end{eqnarray}
Obviously, result \eqref{93} is obtained by factorizing corresponding operators in the expectation values  of \eqref{98a}. 
For instance, in the second term the operator $h_{\vec q -\vec k}$ is factorized   
with $ h^\dag_{\vec q'-\vec k}$,  and $:\psi_{\vec q}^\dag:$ with
$:\psi_{\vec q'}:$. This leads to the second term in expression \eqref{93}. Note 
however that in Eq.~\eqref{93} the following small contribution to second order in the  order parameter $d_{\vec k}$ was neglected 
\begin{eqnarray}
\label{98b} 
&&(-1)  \frac{1}{N^2} \sum_{\vec k_1 \vec k_2} \tilde{\alpha}^*_{\vec k_1 \vec k \vec k_2} \tilde{\alpha}_{\vec k-(\vec k_1 +\vec k_2),  \vec k \vec k_2} \nonumber \\ 
&& \quad  \times \hat n^h_{\vec k_2}(t) \hat d_{\vec k_1}(t) \hat d_{\vec k-(\vec k_1 +\vec k_2)}(t) \, .   
\end{eqnarray}
 It results from an additional factorization of the last term in Eq.~\eqref{98a}, where $e_{\vec k_1}^\dag$ was factorized 
 with $ h_{\vec k_1'+ \vec k_2'-\vec k}^\dag$ and $h_{\vec k_1+ \vec k_2-\vec k}$ with $e_{\vec k_1'}$.  

 In principle  the  factorization \eqref{92}-\eqref{94}  implies two approximations: (i) According to Sec.~\ref{II.B}
  the initial density $\rho_0$ is a product  of the density  $\rho_{\rm S}$ for the e-h-p subsystem and of $\rho_{\rm R}$ for the reservoirs. Thereby
 the capacity of the reservoirs was assumed to be infinitely large so that only the density   $\rho_{\rm S}$ of the e-h-p system is changed under the influence of the unitary transformations. Therefore,  the renormalized  density  $\tilde \rho_{\rm S}$  should differ from  the initial $\rho_{\rm S}$. However, these errors are of higher order in the interaction parameters $g$ and $U$ and should in principle be negligible. Moreover, which is more important, it turns out  that  the final results \eqref{124a}--\eqref{124c} 
  in the steady state for $\hat d_{\vec k}^*$, $\hat n^e_{\vec k}$, and $\hat n^h_{\vec k}$ are  independent of the initial density $\rho_0$, as expected. Therefore the renormalization of $\tilde \rho_{\rm S}$ seems not to be important.
    (ii) The time dependence of $d_{\vec k}^*(t)$, $n_{\vec k}^e(t)$, 
 and $n^h_{-\vec k}(t)$ is governed not only by the e-h-p Hamiltonian $\tilde{\mathcal H}_{\rm S}$  but also by the coupling $\mathcal H_{\rm SR}$ to the 
 electronic and photonic reservoirs. Therefore the correct time dependence might be influenced by the factorization in Eqs.~\eqref{92}--\eqref{94}.


In the next step, following the steady-state condition \eqref{83}, equations of motion for
$\hat d^*_{\vec k}(t)$, $\hat n^e_{\vec k}(t)$, and $\hat n^h_{-\vec k}(t)$  have to be derived. 
This is best done by expressing  the operators in Eqs.~\eqref{95}--\eqref{98} 
by the Bogolyubov quasiparticles $C^{(\dag)}_{1, \vec k}$
and $C^{(\dag)}_{2, \vec k}$ from Eqs.~\eqref{70} and \eqref{71}.
According to Appendix \ref{C}  one first finds
\begin{eqnarray}  
\label{99}
 \hat d^*_{\vec k}(t) &=&  \xi^*_{\vec k} \eta^*_{\vec k} ( A^{11}_{\vec k}(t)-  A^{22}_{\vec k}(t)) 
\nonumber \\
&&+ {\xi^*_{\vec k}}^2   A^{12}_{\vec k}(t) -  {\eta^*_{\vec k}}^2 A^{21}_{\vec k}(t) 
 \,,\\
&&   \nonumber \\
\label{100}
\hat n^e_{\vec k}(t)&=& |\xi_{\vec k}|^2 A^{11}_{\vec k}(t)  
+ |\eta_{\vec k}|^2  A^{22}_{\vec k}(t) \nonumber \\
&&- ( \xi_{\vec k}^* \eta_{\vec k} A^{12}_{\vec k}(t) 
+ \xi_{\vec k} \eta_{\vec k}^* A^{21}_{\vec k}(t) )  
 \,,\\
&& \nonumber \\ 
\label{101}
 \hat n^h_{-\vec k}(t) &=& |\eta_{\vec k}|^2(1- A^{11}_{\vec k}(t))  
 + |\xi_{\vec k}|^2(1-  A^{22}_{\vec k}(t)) 
\nonumber \\
&-& (\xi_{\vec k}^* \eta_{\vec k} A^{12}_{\vec k}(t) + 
\xi_{\vec k} \eta^*_{\vec k} A^{21}_{\vec k}(t) )  
\end{eqnarray}
with ($n,m = 1,2$)
\begin{eqnarray}
\label{102}
&& A^{nm}_{\vec k}(t) = \langle (C^\dag_{n\vec k} C_{m \vec k})(t)\rangle_{\tilde \rho_0}  \, .
\end{eqnarray}
The equations of motion for $A^{nm}_{\vec k}(t)$ are found by applying the Mori-Zwanzig projection operator formalism~\cite{Mo65b,Zw60}. 
According to Appendix~\ref{C} they read 
\begin{eqnarray} 
\label{103}
 \frac{\rm d }{\rm dt} A^{12}_{\vec k}(t) &=& - \big[2 \gamma  - i ( \tilde E_{1 \vec k} - \tilde E_{2 \vec k})  
 \big] \,  A^{12}_{\vec k}(t)  \nonumber \\
  &&- \gamma \, \xi_{\vec k}  \eta_{\vec k}^*  
 \big( f_e(\tilde E_{1 \vec k}) +  f_e(\tilde E_{2 \vec k})\big)     
 \nonumber \\
 &&-  \gamma \, \xi_{\vec k}   \eta_{\vec k}^* 
 \big(f_h(-\tilde E_{1 \vec k}) + f_h(-\tilde E_{2 \vec k}) -2
\big)
\nonumber \\
&=& \Big(\frac{\rm d }{\rm dt}  A^{21}_{\vec k}(t)\Big)^\dag \,,
\end{eqnarray}
 \begin{eqnarray}
\label{104}
 \frac{\rm d }{\rm dt} A^{11}_{\vec k}(t) &=&  - 2 \gamma \,  A^{11}_{\vec k}(t)   
 + 2 \gamma \, |\xi_{\vec k}|^2  f_e(\tilde E_{1 \vec k})
 \nonumber \\
 &&+ 2 \gamma \, |\eta_{\vec k}|^2 \big(1- f_h(-\tilde E_{1 \vec k}) \big)\,,
\end{eqnarray}
\begin{eqnarray} 
\label{105}
 \frac{\rm d }{\rm dt}  A^{22}_{\vec k}(t) &=&  - 2 \gamma  \,  A^{22}_{\vec k}(t) 
  + 2 \gamma \,   |\eta_{\vec k}|^2 f_e(\tilde E_{2 \vec k})
 \nonumber \\
 &&+ 2 \gamma \, |\xi_{\vec k}|^2 \big( 1- f_h(-\tilde E_{2 \vec k}) \big)\, .
\end{eqnarray}
The  damping rate $\gamma$, appearing in Eqs.~\eqref{103}--\eqref{105}, results from the coupling to the electronic reservoirs    
and is assumed to be the same for electrons and holes [see  App. C.1, \eqref{C14}]. 
The functions $f_e(\omega)$ and $f_h(\omega)$ give the  
occupation numbers of bath electrons and bath holes in thermal equilibrium:
\begin{align}
\label{106a}
& f_e(\omega_{\vec p}^e) = \langle b^\dag_{e \vec p}  b_{e \vec p}\rangle_{\rho_{\rm R}} = 
\frac{1}{1+ e^{\beta [ \omega^e_{\vec p} -(\mu_e - \mu/2)]}} \,, \\
\label{10ba}& f_h(\omega_{-\vec p}^h) = \langle b^\dag_{h, -\vec p}  b_{h, -\vec p}\rangle_{\rho_{\rm R}} = 
\frac{1}{1+ e^{\beta [ \omega^h_{-\vec p} - (\mu_h - \mu/2)]}} \,,
\end{align}
[compare Eqs.~\eqref{81a} and \eqref{81b}]. The first contribution in each of the Eqs.~\eqref{103}--\eqref{105} is a  relaxation term
for e-h-p quasiparticle pairs, while the last two terms stand for the relaxation of 
quasiparticle pairs into the electronic baths.    
 The damping rate  $\gamma$ for all contributions is caused by that part of the interaction
  $\mathcal H_{\rm SR}$ which couples electrons and holes  of the e-h-p system with the 
  respective electronic baths. As above mentioned, we have adapted   
the usual assumption that the rates for electrons and holes are equal [compare Eq.~\eqref{C12}]. 
 Furthermore, the second term in Eq.~\eqref{103}, being proportional to $ i (\tilde E_{1\vec k} - \tilde E_{2\vec k}) $,  is a frequency term
 and enters from the dynamics of  $\tilde{\mathcal H}_{\rm S}$. 

We are now in the position to study the steady-state expressions for $A_{\vec k}^{nm}(t)$. 
Defining the steady-state values in analogy to Eqs.~\eqref{88}--\eqref{88.2},  
\begin{equation}
\label{110}
A^{nm}_{\vec k} = A^{nm}_{\vec k}(t \rightarrow \infty)\,,
\end{equation}
we arrive, for $t \rightarrow \infty$, with condition \eqref{83} at  
\begin{align}
\label{107}
 &\big[ i (\tilde E_{1 \vec k} - \tilde E_{2 \vec k}) -2 \gamma  \big] A^{12}_{\vec k} =
  \gamma \, \xi_{\vec k} \eta_{\vec k}^*  \\
 & \times 
 \big[  f_e(\tilde E_{1 \vec k}) + f_e(\tilde E_{2\vec k})  
  + f_h(-\tilde E_{1 \vec k})   + f_h(-\tilde E_{2 \vec k}) -2 \big]\,, \nonumber
\end{align}
$A^{21}_{\vec k} = (A^{12}_{\vec k})^*$, and  (for $\gamma \neq 0$)
\begin{eqnarray}
\label{108}
 A^{11}_{\vec k}  &=&  |\xi_{\vec k}|^2 f_e(\tilde E_{1 \vec k}) +
  |\eta_{\vec k}|^2  \big[1 - f_h(-\tilde E_{1 \vec k}) \big]\,,   \\
\label{109}
 A^{22}_{\vec k}  &=&  |\eta_{\vec k}|^2 \, f_e(\tilde E_{2 \vec k}) +
 |\xi_{\vec k}|^2 \big[1 - f_h(-\tilde E_{2 \vec k}) \big]  \, ,
\end{eqnarray}
where the common prefactor $\gamma$ on both sides of  Eqs.~\eqref{108} and \eqref{109}  has dropped.
Therefore,  both equations are only valid for finite $\gamma$. 
If $\gamma=0$, no term would drive the system into a steady state.

To sum up, the steady-state quantities $d_{\vec k}^*, n^e_{\vec k}$,
and $n^h_{-\vec k}$ can be first expressed by means of Eqs.~\eqref{92}--\eqref{94}:
\begin{eqnarray}
\label{111}
 d^*_{\vec k}&=&  
 \tilde x_{\vec k} \tilde y_{\vec k}\, \hat d^*_{\vec k} + \frac{1}{N} \sum_{\vec k_1} 
\Big[\tilde x_{\vec k}  \tilde \beta_{\vec k_1, \vec k, - \vec k_1}
\hat n^e_{\vec k}   \nonumber \\
&& +  \tilde y_{\vec k} \tilde{\alpha}_{\vec k_1,  \vec k, -\vec k_1} (1- \hat n^h_{-\vec k}) \Big]  
   \, \hat d^*_{\vec k_1}
   \nonumber \\
  && 
  - \frac{1}{N^2} \sum_{\vec k_1 \vec k_2} \tilde \alpha_{\vec k_1 \vec k, -\vec k_2}
  \tilde \beta_{\vec k_2 \vec k_1, \vec k_1 -\vec k_2 -\vec k}   \, \hat n^e_{\vec k_1} 
  \nonumber \\
&& \times \, (1- \hat n^h_{\vec k_1-\vec k_2-\vec k}(t) ) \, \hat d^*_{\vec k_2} \, ,
  \end{eqnarray}
\begin{eqnarray}
\label{112}
 n^e_{\vec k} &=& |\tilde x_{\vec k}|^2 \hat n^e_{\vec k}  \nonumber \\
&& + \frac{1}{N} \sum_{\vec q} |\tilde t_{\vec k -\vec q, \vec q}|^2
\,  (1- \hat n^h_{\vec k-\vec q}) \,  \hat n^\psi_{\vec q} \,
   \nonumber \\
&& +\frac{1}{N^2} \sum_{ \vec k_1 \vec k_2} |\tilde \alpha_{\vec k_1 \vec k \vec k_2}|^2
\hat n^e_{\vec k_1} \hat{n}^h_{\vec k_2} (1- \hat n^h_{\vec k_1+ \vec k_2-\vec k})\,,
\nonumber \\
\end{eqnarray}
\begin{eqnarray}
\label{113}
n^h_{-\vec k} &=& |\tilde y_{\vec k}|^2 \hat n^h_{-\vec k}   \nonumber \\
&& + \frac{1}{N} \sum_{\vec q} |\tilde u_{\vec k, \vec q}|^2 \, 
\hat n^\psi_{\vec q} (1- \hat n^e_{\vec k +\vec q}) 
  \nonumber \\
&& + \frac{1}{N^2} \sum_{ \vec k_1 \vec k_2} |\tilde \beta_{\vec k_1 \vec k_2,  \vec k_2 -\vec k_1 -\vec k}|^2
\hat n^e_{\vec k_1} (1- \hat n^e_{\vec k_2} ) \nonumber \\
&&  \times  \hat{n}^h_{\vec k_2- \vec k_1 - \vec k} \, .
\end{eqnarray}
Thereby, the quantities  with hat symbols  $\hat d_{\vec k}^*, \hat n^e_{\vec k}$,
and $\hat n^h_{-\vec k}$, are written in terms of $A^{nm}_{\vec k}$:
\begin{eqnarray}  
\label{114}
 \hat d^*_{\vec k} &=&  \xi^*_{\vec k} \eta^*_{\vec k} ( A^{11}_{\vec k} -  A^{22}_{\vec k}) 
+ {\xi^*_{\vec k}}^2   A^{12}_{\vec k} -  {\eta^*_{\vec k}}^2 A^{21}_{\vec k}\, ,
 \\\nonumber\\
\label{115}
\hat n^e_{\vec k}&=& |\xi_{\vec k}|^2 A^{11}_{\vec k} 
+ |\eta_{\vec k}|^2  A^{22}_{\vec k} 
- ( \xi_{\vec k}^* \eta_{\vec k} A^{12}_{\vec k} 
+ \xi_{\vec k} \eta_{\vec k}^* A^{21}_{\vec k} )  \,,\nonumber \\&&\\
\label{116}
 \hat n^h_{-\vec k} &=& |\eta_{\vec k}|^2(1- A^{11}_{\vec k})  
 + |\xi_{\vec k}|^2(1-  A^{22}_{\vec k}) 
\nonumber \\
&&- (\xi_{\vec k}^* \eta_{\vec k} A^{12}_{\vec k} + 
\xi_{\vec k} \eta^*_{\vec k} A^{21}_{\vec k} ) \, , 
\end{eqnarray}
where the steady-state results for $A^{nm}_{\vec k}$ are given by Eqs.~\eqref{107}--\eqref{109}.

\subsection{Reformulation of the system dynamics}
\label{IV.C}

It makes sense to express the equations of motion \eqref{103}--\eqref{105} in terms of the 
variables with hat symbols $\hat d^*_{\vec k}(t)$, $\hat n^e_{\vec k}(t)$, and $\hat n^h_{-\vec k}(t)$.
Let us start from  $\hat d^*_{\vec k}(t)$. Using Eqs.~\eqref{99}--\eqref{101}  
we find
\begin{eqnarray} 
\label{117}
 \frac{\rm d}{\rm dt} \hat d^*_{\vec k} &=& 
i(\tilde E_{1 \vec k} - \tilde E_{2 \vec k}) ( \xi^{*2}_{\vec k} A^{12}_{\vec k} + \eta^{*2}_{\vec k} A^{21}_{\vec k} )  \nonumber \\
&&- 2\gamma \, \big[  \xi_{\vec k}^* \eta_{\vec k}^* (A^{11}_{\vec k} -  A^{22}_{\vec k}) +
\xi^{*2}_{\vec k}A^{12}_{\vec k} -\eta^{*2}_{\vec k} A^{21}_{\vec k} \big] \nonumber \\
&&+ 2\gamma \,  \hat d^{0*}_{\vec  k} \, ,
\end{eqnarray}
where on the right hand side we have defined 
\begin{eqnarray} 
\label{118}
 \hat d^{0*}_{\vec  k}  
  &=&   \frac{1}{2} \xi_{\vec k}^* \eta_{\vec k}^* \Big\{  
   f_e(\tilde E_{1 \vec k}) - f_h(-\tilde E_{1 \vec k})  \nonumber \\
 && -     [  f_e(\tilde E_{2 \vec k}) -   f_h(-\tilde E_{2 \vec k})  ]  \Big\}\,.
\end{eqnarray}
Moreover, using the Bogolyubov transformation \eqref{70}--\eqref{74}, as well as  Eqs.~\eqref{95}, \eqref{96}, and \eqref{98}, 
we obtain
\begin{eqnarray} 
\label{119}
 \frac{\rm d}{\rm dt} \hat d^*_{\vec k}(t) &=&  i (\tilde \varepsilon^e_{\vec k} + \tilde \varepsilon^h_{\vec k}) \,
 \hat d^*_{\vec k}(t)
+ i \tilde\Delta_{\vec k}^* (1 - \hat n^e_{\vec k}(t) - \hat n^h_{\vec k}(t)) \nonumber \\
&&- 2\gamma \, \big( \hat d^*_{\vec k}(t) - \hat d^{0*}_{\vec  k} \big) \, .
\end{eqnarray}
Similarly we derive the equations of motions for $\hat n^e_{\vec k}(t)$ and  $\hat n^h_{-\vec k}(t)$:
\begin{eqnarray}
\label{120}
 && \frac{\rm d}{\rm dt} \hat n^e_{\vec k}(t) =  2 \Im[\tilde \Delta_{\vec k} \hat d^*_{\vec k}(t)]   \\
&& \qquad\;\;   -2 \gamma \big[ \hat n^e_{\vec k}(t) -  |\xi_{\vec k}|^2 f_e(\tilde E_{1 \vec k}) -  
 |\eta_{\vec k}|^2 f_e(\tilde E_{2 \vec k})  \big] \,, \nonumber \\
 \label{121}
&&  \frac{\rm d}{\rm dt} \hat n^h_{-\vec k}(t) =   2 \Im[\tilde \Delta_{\vec k} \hat d^*_{\vec k}(t)]   \\
&& \qquad\;\; -2 \gamma \big[ \hat n^h_{-\vec k}(t) -  |\eta_{\vec k}|^2 f_h(- \tilde E_{1 \vec k}) -  
 |\xi_{\vec k}|^2 f_h(-\tilde E_{2 \vec k})  \big]  \, ,    \nonumber 
\end{eqnarray}
where $2 \Im[\tilde \Delta_{\vec k} \hat d^*_{\vec k}]  =  -i (\tilde \Delta_{\vec k} \hat d^*_{\vec k}  -
\tilde \Delta^*_{\vec k} \hat d_{\vec k})$ was used.  The steady-state expectation values of 
$\hat d^*_{\vec k}$, $\hat n_{\vec k}^e$, and  $\hat n_{-\vec k}^h$  are obtained from 
Eqs.~\eqref{119}--\eqref{121} by setting the left hand sides equal to zero 
 \begin{eqnarray}
 \label{124a}
  && \hat d_{\vec k}^* = - \frac{1}{(\tilde \varepsilon_{\vec k}^e + \tilde \varepsilon_{\vec k}^h) + 2i \gamma}
 \big[ \tilde \Delta_{\vec k}^* (1 - \hat n_{\vec k}^e - \hat n_{\vec k}^k) -2i \gamma \, \hat d_{\vec k}^{0*} 
 \big] \,,\nonumber \\
 && 
 \end{eqnarray}
and 
\begin{eqnarray}
 \label{124b}
 && \hat n_{\vec k}^e =  |\xi_{\vec k}|^2 f_e(\tilde E_{1\vec k}) + |\eta_{\vec k}|^2 f_e(\tilde E_{2\vec k}) +
 \frac{1}{\gamma} \Im[\tilde \Delta_{\vec k} \hat d_{\vec k}^*] \,,
   \\
  \label{124c}
 && \hat n_{\vec k}^h =  |\eta_{\vec k}|^2 f_h(-\tilde E_{1\vec k}) + |\xi_{\vec k}|^2 f_h(-\tilde E_{2\vec k}) +
 \frac{1}{\gamma} \Im[\tilde \Delta_{\vec k} \hat d_{\vec k}^*] \,.\nonumber \\
 && 
 \end{eqnarray}
Of course, this result is equivalent to
the former equations \eqref{114}--\eqref{116}. 
The steady-state expressions \eqref{124a}--\eqref{124c} can further be
simplified by using definition \eqref{118} and Eqs.~\eqref{72}--\eqref{75}. According to Appendix C.2  
one finds:
\begin{eqnarray}
 \label{124d}
 \hat d^*_{\vec k} &=& \frac{\tilde \Delta^*_{\vec k}}{(\tilde \varepsilon^e_{\vec k} +\tilde \varepsilon^h_{\vec k}) +2i  \, \gamma}
\Big[ (\hat n_{\vec k}^e + \hat n_{\vec k}^h -1)   \nonumber \\
&+& 
 i \gamma \, {\rm sgn }(\tilde \varepsilon^e_{\vec k} +\tilde \varepsilon^h_{\vec k})  \frac{F^+_{1 \vec k}}{W_{\vec k}}
\Big]
 \end{eqnarray}
and 
\begin{eqnarray}
 \label{124e}
  \hat n^e_{\vec k} + \hat n^h_{\vec k} -1  &=& 
\frac{|\tilde \varepsilon^e_{\vec k} +\tilde \varepsilon^h_{\vec k}|}{2W_{\vec k}} F_{1\vec k}^+ + \nonumber  \\
&+&  \frac{1}{2} \frac{ F_{2 \vec k}^+ -2}{1+ \displaystyle 
  \frac{4|\tilde \Delta_{\vec k}|^2}{(\tilde \varepsilon^e_{\vec k} +\tilde \varepsilon^h_{\vec k})^2 + (2 \gamma)^2}   }  \,,
\end{eqnarray} 
\begin{eqnarray} 
 \label{124f}
 &&  \hat n_{\vec k}^e - \hat n_{\vec k}^h  = \frac{1}{2} F_{1\vec k}^- + \frac{|\tilde \varepsilon_{\vec k}^e + 
 \tilde \varepsilon_{\vec k}^h|}{2W_{\vec k}} F_{2\vec k}^-\,.
 \end{eqnarray}  
The quantities $F_{1\vec k}^\pm$ and $F_{2\vec k}^\pm$  are defined by 
 \begin{align}
 \label{124g}
& F_{1\vec k}^\pm =  f_e(\tilde E_{1\vec k}) - f_h(-\tilde E_{1\vec k}) \mp
 [f_e(\tilde E_{2\vec k})- f_h(-\tilde E_{2\vec k}) ]\,,  \nonumber \\
 & \\
 \label{124h}
 & F_{2\vec k}^\pm = f_e(\tilde E_{1\vec k}) + f_h(-\tilde E_{1\vec k}) \pm
 [f_e(\tilde E_{2\vec k})+ f_h(-\tilde E_{2\vec k})]\, . 
   \end{align}
   
 Let us look again at the symmetric case $\tilde \varepsilon^e_{\vec k} = \tilde \varepsilon^h_{\vec k}$ with charge neutrality $\mu_e = \mu_h$  
 (compare Sec.~\ref{IV.B}). Here, the quasiparticle energies $\tilde E_{1,2 \vec k}$ reduce to 
 \begin{equation}
\label{77a}
\tilde E_{1\vec{k}}= - \tilde E_{2 \vec k}= {\rm sgn}( \tilde{\varepsilon}^e_{\vec{k}})\frac{W_{\vec k}}{2}    \,,
\end{equation}
 and $F_{1 \vec k}^{\pm}$ and $F_{2 \vec  k}^{\pm}$ to
  \begin{eqnarray}
\label{124i}
&& F_{(1,2) \vec k}^+ = 2 \big[ f(\tilde E_{1 \vec k}) \mp f(\tilde E_{2 \vec k}) \big] \\
\label{124j}
&&  F_{(1,2) \vec k}^- =0 \, .
\end{eqnarray}
with
\begin{equation}
\label{124k}
f(E) = \frac{1}{\displaystyle 1 + e^{\beta [E -(\mu_B -\mu)/2]}} = f_e(E) = f_h(E) 
\end{equation}
and $\mu_B= \mu_e + \mu_h = 2 \mu_e  = 2 \mu_h$.  From the relations ~\eqref{124j} and~\eqref{124f} immediately follows  $\hat n^e_{\vec k} = \hat n^h_{\vec k}$,  which is  
a natural property  of the symmetric case with charge neutrality.

\subsection{Photon condensation}
\label{IV.D}
Next, let us  study the steady-state expression 
$\langle \psi^\dag_{\vec q} \rangle= \langle \psi^\dag_{\vec q}(t\rightarrow \infty) \rangle$
for the photonic expectation value $ \langle \psi^\dag_{\vec q}(t) \rangle$.  
Starting from Eq.~\eqref{84}, we first rewrite 
\begin{eqnarray}
 \label{125}
 \langle \psi^\dag_{\vec q}(t) \rangle =  \langle \tilde\psi^\dag_{\vec q}(t) \rangle_{\tilde \rho_0} \, ,
 \end{eqnarray}
 where $\tilde\psi^\dag_{\vec q}$  
 is the renormalized photon operator [cf. Eq.~\eqref{B23}]. The dynamics on the right hand side
 is governed by $\tilde{\mathcal H}$ and the expectation value is formed with $\tilde \rho_0$.  
 According to Appendix C.3  an appropriate representation 
for  $\tilde\psi^\dag_{\vec q}$ is 
 \begin{eqnarray}
\label{126}
\tilde \psi^\dag_{\vec q} = \tilde z_{\vec q}\psi^\dag_{\vec q}
+ \frac{1}{\sqrt N} \sum_{\vec k} \tilde v_{\vec k \vec q} \, :e^\dag_{\vec k + \vec q} h^\dag_{-\vec k}: \, ,
\end{eqnarray}
leading, with Eq.~\eqref{125}, to 
\begin{eqnarray}
\label{127}
\langle\psi^\dag_{\vec q}(t) \rangle &=&  \tilde z_{\vec q}  \langle\psi^\dag_{\vec q}(t)\rangle_{\tilde \rho_0} 
 \nonumber \\
&+& \frac{1}{\sqrt N} \sum_{\vec k} \tilde v_{\vec k \vec q} \,
 \langle :e^\dag_{\vec k + \vec q} h^\dag_{-\vec k}: (t) \rangle_{\tilde \rho_0} \, ,
\end{eqnarray}
where  $ \tilde z_{\vec q}$  and $\tilde v_{\vec k \vec q}$ are  the renormalized coefficients.
 An equation  of motion for the expectation value $\langle \psi^\dag_{\vec q}(t)\rangle_{\tilde \rho_0}$ 
can be derived from the generalized Langevin equations~\eqref{C5}
\begin{equation} 
\label{128}
\frac{\rm d}{\rm dt} \psi^\dag_{\vec q}(t )  = i \tilde \omega_{\vec q}   \psi^\dag_{\vec q}(t)  
+ i \sqrt N \tilde \Gamma^* \delta_{\vec q, 0} 
- \kappa  \psi^\dag_{\vec q}(t)  + \mathcal F_{\vec q}^\psi \, ,
\end{equation}
 from which one finds for ${\bf q} =0$: 
\begin{eqnarray}
\label{129}
&& \frac{\rm d}{\rm dt} \langle \psi^\dag_{0}(t ) \rangle_{\tilde \rho_0} =
 i  \omega_{0} \Big( \langle \psi^\dag_{0}(t) \rangle_{\tilde \rho_0} 
 + \frac{\sqrt N \tilde \Gamma^*}{\omega_0}  \Big)
- \kappa  \langle \psi^\dag_0(t) \rangle_{\tilde \rho_0}\,.  \nonumber \\
&&
\end{eqnarray}
Thereby $\kappa \sim  \pi \sum_{\vec p}|\Gamma^\psi_{\vec q \vec p}|^2 \delta(\omega^\varphi_{\vec p}) $
is the damping rate for cavity photons into the free space due to a nonvanishing leakage. Moreover, 
$\tilde \Gamma^*$ is the renormalized field parameter which accounts for a possible photon condensation 
[cf. Eq.~\eqref{23}].  Using condition~\eqref{83}, Eq.~\eqref{129} leads to  the steady-state result,
$ \langle \psi^\dag_{0}\rangle_{\tilde \rho_0} =\langle \psi^\dag_{0}(t \rightarrow \infty) \rangle_{\tilde \rho_0}$, 

\begin{eqnarray}
\label{130}
\langle  \psi^\dag_{\vec q} \rangle_{\tilde \rho_0}  = 
-    \frac{\sqrt N \tilde \Gamma^*}{\tilde \omega_0 + i\kappa} \delta_{\vec q 0} \, .
\end{eqnarray}
Finally, neglecting the fluctuation term being proportional to $\langle :e^\dag_{\vec q+ \vec k} h^\dag_{-\vec k}:\rangle_{\tilde \rho_0}$ on the right hand side~of Eq.~\eqref{127},  the steady-state result for 
$\langle  \psi^\dag_{\vec q} \rangle$ becomes
\begin{eqnarray}
\label{131}
\langle  \psi^\dag_{\vec q} \rangle  = 
-   \tilde z_0  \frac{\sqrt N \tilde \Gamma^*}{\tilde \omega_0 + i\kappa} \delta_{\vec q 0}     \,  .
\end{eqnarray}
A corresponding expression for $n_{\vec q}^\psi =\langle  :\psi^\dag_{\vec q}: : \psi_{\vec q}: \rangle $  is found in Appendix ~C.3.

\subsection{Comparison with previous results} 
\label{IV.E}
It may be worthwhile to compare our results~\eqref{119}--\eqref{121}
with those obtained by the Yamamoto group~\cite{YKOY12,YNKOY15}.  
Using a generating functional approach, the following equations were derived by these authors: 
\begin{eqnarray}
 \label{122}
 \frac{\rm d}{\rm dt} d^*_{\vec k}(t) &=& i (\varepsilon^{{\rm HF}, e}_{ \vec k} 
+  \varepsilon^{{\rm HF}, h}_{\vec k}) d^*_{\vec k}(t) 
-2\gamma (d^*_{\vec k}(t) -d^{0*}_{\vec k}) 
\nonumber \\
 && - i  \Delta_{\vec k}^{\rm HF *} (1 - n^e_{\vec k}(t) - n^h_{\vec k}(t))  \,,  \\
\label{123}
 \frac{\rm d}{\rm dt} n^{e}_{\vec k}(t) &=& 2 \Im{[ \Delta^{\rm HF}_{\vec k} d^*_{\vec k}(t)]} 
- 2\gamma \big(n^{e}_{\vec k}(t) - n^0_{h, \vec k} \big) \,, \\
\label{124}
 \frac{\rm d}{\rm dt} n^{h}_{\vec k}(t) &=& 2 \Im{[ \Delta^{\rm HF}_{\vec k} d^*_{\vec k}(t)]}
- 2\gamma \big(n^{h}_{\vec k}(t) - n^0_{h, \vec k} \big)\,.
\end{eqnarray}
 In Eqs.~\eqref{122}--\eqref{124}, the time-independent quantities $d^{0*}_{\vec k}$ 
and $n^0_{e,\vec k}, n^0_{h,\vec k}$ on the right hand side
are given in an integral formulation. In principle, the time-dependent quantities 
$d^*_{\vec k}(t), n^e_{\vec k}(t)$ and $n^h_{\vec k}(t)$ in Eqs.~\eqref{122}--\eqref{124} should agree with our
previous quantities \eqref{89a}--\eqref{89c},  
however, there are differences. Calculating the PRM quantities $d^*_{\vec k}(t), n^e_{\vec k}(t)$ and $n^h_{\vec k}(t)$
via Eqs.~\eqref{92}--\eqref{94}, fluctuation processes from $\mathcal H_{g}$ and $\mathcal H_{U}$
will be included to infinite order, while in Eqs.~\eqref{122}--\eqref{124} the interactions $\mathcal H_{g}$ 
and $\mathcal H_{U}$  enter only  in mean-field approximation. 
Hence the latter result can not directly be compared with the true PRM dynamics of 
$d^*_{\vec k}(t), n^e_{\vec k}(t)$ and $n^h_{\vec k}(t)$. However, one might compare 
equations \eqref{122}--\eqref{124} with equations  
\eqref{119}--\eqref{121} for the PRM quantities $\hat d^*_{\vec k}(t),  
\hat n^e_{\vec k}(t)$ and $\hat n^h_{\vec k}(t)$ (with hat symbols):
\begin{eqnarray} 
\label{124l}
 \frac{\rm d}{\rm dt} \hat d^*_{\vec k}(t) &=&  i (\tilde \varepsilon^e_{\vec k} + \tilde \varepsilon^h_{\vec k}) \,
 \hat d^*_{\vec k}(t)
+ i \tilde\Delta_{\vec k}^* (1 - \hat n^e_{\vec k}(t) - \hat n^h_{\vec k}(t)) \nonumber \\
&& - 2\gamma \, \big( \hat d^*_{\vec k}(t) - \hat d^{0*}_{\vec  k} \big) \,, \\
\label{124m}
\frac{\rm d}{\rm dt} \hat n^e_{\vec k}(t) &=&  2 \Im[\tilde \Delta_{\vec k} \hat d^*_{\vec k}(t)]  
   -2 \gamma \big[ \hat n^e_{\vec k}(t) -  \hat n_{e, \vec k}^0 \big] \,,  \\
 \label{124n}
  \frac{\rm d}{\rm dt} \hat n^h_{-\vec k}(t)& =&   2 \Im[\tilde \Delta_{\vec k} \hat d^*_{\vec k}(t)]  
-2 \gamma \big[ \hat n^h_{-\vec k}(t) -  \hat n_{h,\vec k}^0 \big]  \,,     \quad
\end{eqnarray}
where
\begin{eqnarray}
\label{124o}
 \hat d^{0*}_{\vec  k} &=& \frac{1}{2} \xi_{\vec k}^* \eta_{\vec k}^* F^+_{1\vec k} \,,\\
\label{124p}
 \hat n_{e, \vec k}^0 &= & |\xi_{\vec k}|^2 f_e(\tilde E_{1 \vec k}) +  
 |\eta_{\vec k}|^2 f_e(\tilde E_{2 \vec k}) \,, \\
 \label{124pp}
 \hat n_{h,\vec k}^0 &=&  |\eta_{\vec k}|^2 f_h(- \tilde E_{1 \vec k}) +  
 |\xi_{\vec k}|^2 f_h(-\tilde E_{2 \vec k}) =  \hat n_{e, \vec k}^0 \,.\;\; 
\end{eqnarray}
Here, the additional  fluctuation terms following from  
Eqs.~\eqref{111}--\eqref{113} are absent. However there are differences between 
Eqs.~\eqref{122}--\eqref{124} and \eqref{124l}--\eqref{124n} which still remain:
The energies $\tilde \varepsilon^e_{\vec k}$,
 $\tilde \varepsilon^h_{\vec k}$, and $\tilde \Delta^*_{\vec k}$ in \eqref{124l}--\eqref{124pp}    
 are renormalized quantities, whereas the energies $ \varepsilon^{{\rm HF},e}_{\vec k}$, 
 $ \varepsilon^{{\rm HF},h}_{\vec k}$ 
 and $ \Delta^{\rm HF}_{\vec k}$ from Eqs.~\eqref{122}--\eqref{124} are not. 
It remains for us to compare the time-independent quantities 
$d_{\vec k}^{0*}$, $n^0_{e,\vec k}$, and $n^0_{h,\vec k}$
in Eqs.~\eqref{122}--\eqref{124} with the corresponding quantities 
$\hat d^{0*}_{\vec k}$, $\hat n_{e, \vec k}^0$, and $\hat n_{h, \vec k}^0$
in Eqs.~\eqref{124o}--\eqref{124pp}.

 For this reason let us consider two limiting cases 
from  Refs.~\cite{YKOY12,YKNOY13,YNKOY15}. 
Thereby,  we use slightly modified conditions and restrict ourselves
again to the symmetric case and charge neutrality.  
%
\subsubsection{${\rm min} |2 \tilde E_{(1,2) \vec k}| \geq \mu_B -\mu$}
%
Here ${\rm min} |2 \tilde E_{(1,2) \vec k}|$ is  the minimal excitation energy of electron-hole  
pairs and the difference  $\mu_B- \mu$ can be considered as being responsible for the particle supply from the 
pumping baths to the e-h-p system. According to Eq.~\eqref{124k}  the Fermi function $f(E)$ can then be approximated by 
\begin{eqnarray}
\label{124q}
f(E) \simeq \frac{1}{1 + e^{-\beta E}}\,.
\end{eqnarray}
Using $\tilde E_{2 \vec k}= -\tilde E_{1 \vec k}$, $f(\tilde E_{2 \vec k}) = f(-\tilde E_{1 \vec k})
 =1- f(\tilde E_{1\vec k})$ and  Eq.~\eqref{124i}
one has:
\begin{eqnarray}
\label{124r}
&& F^+_{1\vec k} \simeq 2\big[ 2f(\tilde E_{1k}) -1\big]= - 2 \tanh{\frac{\beta \tilde E_{1 \vec k}}{2}} \,,\\
&&F^+_{2\vec k} \simeq  2\,.
\end{eqnarray}
Hence,  with $\tilde E_{(1,2)\vec k}= \pm {\rm sgn}(\tilde \varepsilon^e_{\vec k}+\tilde \varepsilon^h_{\vec k})
W_{\vec k}/2$ 
 and relation \eqref{74}, one obtains for Eq.~\eqref{124o}
\begin{eqnarray}
\label{124s}
\hat d^{0*}_{\vec k} = -\frac{\tilde \Delta_{\vec k}^*}{2 (W_{\vec k}/2)}
 \tanh{\frac{\beta (W_{\vec k}/2)}{2}} \,,
 \end{eqnarray}
and similarly
\begin{eqnarray}
\label{124t}
\hat n^0_{e, \vec k} = \hat n^0_{h, \vec k} = \frac{1}{2} 
- \frac{\tilde \varepsilon^e_{\vec k}+\tilde \varepsilon^h_{\vec k} }{2W_{\vec k}}  
\tanh{\frac{\beta (W_{ \vec k}/2)}{2}}\,.
 \end{eqnarray}
These results 
 agree with the corresponding expressions from the Japanese group~\cite{YKOY12,YKNOY13,YNKOY15}. 
 
 The same results are also obtained with Eqs.~\eqref{124d}--\eqref{124f}
 for the steady-state  expressions of $\hat d^*_{\vec k}$ and $\hat n^{e,h}_{\vec k}$:
 \begin{eqnarray}
 \label{124tt}
&& \hat d^*_{\vec k} =  -\frac{\tilde \Delta_{\vec k}^*}{2 (W_{\vec k}/2)}
 \tanh{\frac{\beta (W_{\vec k}/2)}{2}}\,, \\
 && \hat n^e_{ \vec k} = \hat n^h_{\vec k} = \frac{1}{2} 
- \frac{ \tilde \varepsilon^e_{\vec k}+\tilde \varepsilon^h_{\vec k} }{2W_{\vec k}}  
\tanh{\frac{\beta  (W_{\vec k}/2)}{2}}\,,
 \end{eqnarray}
 and moreover (see App.~C.3)
 \begin{equation}
 \label{124ttt}
 \langle \psi^\dag_{\vec q} \psi_{\vec q} \rangle = |\tilde z_{\vec q}|^2 \frac{N |\tilde \Gamma|^2} 
 {\tilde \omega^2_0 + \kappa^2} \delta_{\vec q,0}
 + \frac{1}{N} \sum_{\vec k} |\tilde v_{\vec k \vec q}|^2 \hat n^e_{\vec k + \vec q} \hat n^h_{-\vec k}.
 \end{equation}
 Note that the damping rate $\gamma$ does not enter the 
equations for $\hat d^*_{\vec k}$  and $\hat n^{e,h}_{\vec k}$. 
Using a mean-field approximation,
in Refs.~\cite{YKOY12,YKNOY13,YNKOY15}
also a gap equation for the order parameter $\Delta_{\vec k}$ was derived,   
which was formally equivalent to a BCS gap equation. 
 Therefore, $\beta = 1/ k_B T$ and $\mu$ 
 can be regarded as the inverse temperature and the 
 chemical potential of the e-h-p system, even  though $\beta$ and $\mu$ were originally introduced as the inverse temperature of the pumping baths and the oscillation frequency of the photon  and polarization fields. 
  In other words, in case of  vanishing damping $\kappa$ [see Eq.~\eqref{124ttt}] the system can be considered 
  as being in a quasi-equilibrium, because thermodynamic variables 
 are defined. Thus, for $\kappa=0$ the region with ${\rm min} |2\tilde E_{(1,2) \vec k}| \geq \mu_B -\mu$ is equivalent
 to the thermodynamic  equilibrium theory of Ref.~\cite{PBF16} for the isolated e-h-p system, apart from the explicit 
 factor $e^{i\mu t}$ in Eqs.~\eqref{15} and \eqref{16a}. For non-vanishing damping $\kappa$ the number of 
 cavity photons is only slightly reduced as long as $\kappa$ is small compared to the cavity photon 
 frequency $\tilde \omega_{\vec q=0}$.

%
 \subsubsection{$\mu_B -\mu \geq {\rm min} |2\tilde E_{(1,2)\vec k}|$}
 %
 In this case  the second term in the exponential of Eq.~\eqref{124k} dominates, i.e.: 
 \begin{eqnarray}
 \label{124u1}
 f(E) \simeq \frac{1}{1+ e^{- (\beta/2)(\mu_B -\mu)}}  =: f_0 \, ,
 \end{eqnarray}
 and
  \begin{eqnarray}
 \label{124u2}
F^+_{1\vec k} \simeq 0 \, ,\qquad F^+_{2 \vec k} \simeq 4 f_0  \, .
  \end{eqnarray}
 With Eqs.~\eqref{124o}--\eqref{124pp} one finds for the time-independent quantities 
 in Eqs.~\eqref{124l}--\eqref{124n}, 
 \begin{eqnarray}
 \label{124w}
 && \hat d^{0*}_{\vec k} =0 \,, \quad 
 \hat n^0_{e, \vec k} = \hat n^0_{h, \vec k} = f_0\simeq  1 \, ,\nonumber \\
 &&
 \end{eqnarray}
where additionally in the last relation  $(\beta/2)(\mu_B -\mu) \gg 1$ was used (low-temperature approximation).\\

 The steady-state results for $\hat d^*_{\vec k}$ and $\hat n_{\vec k}^e=\hat n_{\vec k}^h$ are found from 
 Eqs.~\eqref{124u1}--\eqref{124u2} and \eqref{124d}--\eqref{124f}:
\begin{align}
 \label{124x}
\hat d^*_{\vec k} &= \frac{ \tilde \Delta^*_{\vec k}} 
{(\tilde \varepsilon_{\vec k}^e + \tilde \varepsilon_{\vec k}^h) + 2i \gamma} \, 
(\hat n_{\vec k}^e + \hat n_{\vec k}^h -1) \,,\\
 \label{124y}
 \hat n_{\vec k}^e +& \hat n_{\vec k}^h -1 =    2 f_0 -1 + \frac{2}{\gamma} \Im{[\tilde \Delta_{\vec k} \hat d^*_{\vec k}]}
  \,,
 \end{align}
 from which also follows
 \begin{eqnarray}
\label{124yy}
\hat n^e_{\vec k} +\hat n^h_{\vec k} -1  = \frac{2f_0 -1}{1 + 
\displaystyle \frac{4 |\tilde \Delta_{\vec k}|^2}
{(\tilde \varepsilon^e_{\vec k}+\tilde \varepsilon^h_{\vec k} )^2 + 4\gamma^2}} \, . 
\end{eqnarray}
In the considered regime  $\mu_B -\mu \geq {\rm min} |2\tilde E_{(1,2)\vec k}|$ the e-h-p system can no longer be perceived as being in a quasi-equilibrium, solely formed by  the isolated e-h-p system. This can be concluded from relation  \eqref{124yy},
assuming a small influence of the numerator (\small $|\tilde \Delta|^2$). Then for low temperatures the right hand side of 
\eqref{124yy} indicates  that electrons and holes are strongly excited and are in the high-density regime. Thus, increasing further the concentration of the 
total particle number of the e-h-p system, 
\begin{eqnarray}  
\label{124zz}
n_{exc} = \frac{1}{N}\Big[ \frac{1}{2} \sum_{\vec k}( n_{\vec k}^e +  n^h_{\vec k} )
+ \sum_{\vec q} \langle \psi^\dag_{\vec q} \psi_{\vec q} \rangle \Big]\,,
 \end{eqnarray}
only the number of cavity photons will mainly increase since possible electron-hole excitations 
tend already to be used up. 

Cavity photons are also affected by a non-vanishing  leakage ($\kappa \neq 0$) to the 
external  photonic free-space. Then, the e-h-p system is no longer in an equilibrium situation 
with the electronic pumping baths.
Therefore, one of the main differences between the two regimes 
$\mu_B -\mu \leq {\rm min} |2\tilde E_{(1,2)\vec k}|$ and $\mu_B -\mu \geq {\rm min}|2\tilde E_{(1,2)\vec k}|$ is the    
 relative importance of electron-hole and photonic excitations.  Whereas in the first regime particle-hole excitations are 
dominant this is not the case for the second regime. 
 
As said before, photon excitations are less 
 pronounced in regime $\mu_B -\mu \leq  {\rm min}|2\tilde E_{(1,2)\vec k}|$. This means that the system 
 is less affected by the photon leakage. In contrast, for $\mu_B -\mu \geq  {\rm min}|2\tilde E_{(1,2)\vec k}|$ the system is 
in a high-density regime and is  strongly affected by the photon leakage, which suggests that a large degree of non-equilibrium is achieved.

\subsection{Self-consistency of the steady-state solution }
\label{IV.F}
Above we have derived the renormalization equations   
 for the order parameters $\Delta_{\vec k,\lambda}$ and $\Gamma_\lambda$ and  found 
 a compact  re\-presentation for the exciton-condensation parameter  $d^*_{\vec k}$. The equations can be numerically 
  solved, provided  $\langle \psi_0\rangle$ and $\mu$ are known.
However, these quantities are not yet determined since $\tilde \Gamma$ in Eq.~\eqref{131} 
depends implicitly on $\langle \psi_{0}\rangle$ and $\mu$ as well as on the sets of quantities 
 $d_{\vec k}$ and $\Delta_{\vec k,\lambda}$.
In particular, 
$\mu$ is not a chemical potential, since the total number of particles of the e-h-p system 
together with the particle number of the  electronic baths is not fixed due to the leakage of cavity photons into the free space. To determine $\mu$ and $\langle \psi_0\rangle$ 
a ``way out'' has been  discussed in the literature~\cite{SKL06}.  The starting point  is Eq.~\eqref{131}, 
 \begin{equation}
 \label{132}
 (\tilde \omega_0 - i \kappa) \, \langle\psi_{0}\rangle= - \tilde z_0 \sqrt N \tilde \Gamma \, ,
 \end{equation}
 which is a complex equation due to the presence of the damping rate $\kappa$.  
 The final solution for  $\tilde \Gamma$  results
 from the renormalization equation \eqref{A20}:
  \begin{eqnarray}
 \label{133}
 && \Gamma_{\lambda -\Delta \lambda} = \Gamma_\lambda 
 + \delta \Gamma^{(0)}_\lambda + \delta \Gamma^{(c)}_\lambda
 + \delta \Gamma^{(g)}_\lambda \, ,
 \end{eqnarray}
 where the $\delta \Gamma^{(0)}_\lambda$, $\delta \Gamma^{(c)}_\lambda$, and  $\delta \Gamma^{(g)}_\lambda$ 
defined in Eqs.~\eqref{A3}, \eqref{A6a} and \eqref{A9d} become: 
 \begin{align}
\label{134}
 \delta {\Gamma}_{\lambda}^{(0)}  &=  \frac{g}{N} \sum_{\vec k} A_{\vec k 0}(\lambda, \Delta \lambda) 
(\varepsilon^e_{\vec k,\lambda} + \varepsilon^h_{-\vec k, \lambda}) \,  d_{\vec k}\,,  \\
\label{135}
 \delta\Gamma_\lambda^{(c)} &= \frac{g}{N} \sum_{\vec k} 
  A_{\vec k 0}(\lambda, \Delta \lambda) 
( 1- n^e_{\vec k} - n^h_{-\vec k}) {\Delta}_{\vec k, \lambda}  \,, \\ 
\label{136}
\delta \Gamma^{(g)}_\lambda &= \frac{2 g^2}{N \sqrt N} \sum_{\vec k} A_{\vec k 0}(\lambda, \Delta \lambda)
(1- n^e_{\vec k} -n^h_{-\vec k})  \langle\psi_{0}\rangle \, .
\end{align}
The  initial value  of $\Gamma_\lambda$ is
  \begin{eqnarray}
\label{137}
&& \Gamma_\Lambda = \hat \Gamma=   \Gamma - (g/N) \sum_{\vec k} d_{\vec k} 
\end{eqnarray}
($\Gamma=0^+$).  
Note that  the contribution $\delta \Gamma^{(g)}_\lambda $ is proportional to 
$\langle\psi_{0}\rangle$ as expected, whereas 
$\delta\Gamma_\lambda^{(0)}$ and $\delta\Gamma_\lambda^{(c)}$
depend on the order parameters $d_{\vec k}$ and $\Delta_{\vec k,\lambda}$. 
Similarly, from Eqs.~\eqref{111} and \eqref{114} one concludes
that  $d_{\vec k}$ is fixed if the  order parameters $\tilde \Delta_{\vec k}$ are known. What remains to be shown 
is that $\Delta_{\vec k,\lambda}$ is fixed for given $\langle \psi_0 \rangle$ and $d_{\vec k}$, which follows from  renormalization equation \eqref{65}. Thus, putting everything together, 
 $\tilde \Gamma$ can be considered as an implicit function of $\langle \psi_0 \rangle$ and $\mu$, 
 i.e.,~$\tilde\Gamma= \tilde\Gamma[\langle\psi_0\rangle, \mu]$:
\begin{equation}
\label{138}
\frac{\langle \psi_0\rangle}{\sqrt N} = 
- \tilde z_0 \frac{\tilde\Gamma[\langle\psi_0\rangle, \mu]}{\tilde \omega_0 -i \kappa} \,.
 \end{equation}
However,  
the number of coupled equations by \eqref{138} is one less than the number of unknown variables,  
since $\langle \psi_0\rangle$ and  $\tilde{\Gamma}$
are  in general complex quantities. This can be seen from equation \eqref{107} for
 $A^{12}_{\vec k}$ and $A^{21}_{\vec k}$.  Since the denominator in Eq.~\eqref{107} is complex 
also $A^{12}_{\vec k}$ and $A^{21}_{\vec k}$ will be complex.
 Assuming $\langle \psi_0 \rangle$ is complex, Eq.~\eqref{138} would contain three unknown quantities, 
the real and the imaginary parts of $\langle \psi_0 \rangle$ as well as the energy parameter $\mu$, whereas  the complex equation only fixes two of them.  
However the number of unknown variables  can be reduced by fixing the phase of $\langle \psi_0 \rangle$. 
Taking a phase for which the imaginary part of $\langle \psi_0\rangle$ vanishes, 
the number of coupled equations becomes equal 
to the number of unknown variables and the complex equation \eqref{138} only represents 
two independent equations  for $\langle  \psi_0\rangle$ and $\mu$:
\begin{eqnarray}
\label{139a}
 \frac{\langle \psi_0\rangle}{\sqrt N} &=& - \frac{\tilde z_0}{\tilde\omega_0^2 + \kappa^2} 
 \big(\tilde \omega_0  \, \Re\tilde \Gamma  - \kappa \, \Im \tilde \Gamma \big) \,, \\
 \label{139b}
 0 &=&    \frac{\tilde z_0}{\tilde\omega_0^2 + \kappa^2} 
 \big(  \kappa \, \Re\tilde \Gamma  +  \tilde \omega_0 \, \Im \tilde \Gamma  \big)
  \, , 
\end{eqnarray}
where $\tilde \Gamma = \Re\tilde \Gamma + i \Im \tilde \Gamma$ is complex. 
 From equation \eqref{139b} one obtains 
\begin{equation}
\label{140}
\Im \tilde \Gamma= - (\kappa/ \tilde \omega_0) \Re \tilde \Gamma, 
\end{equation}
which leads for the first equation to 
\begin{eqnarray}
\label{141}
 \frac{\langle \psi_0\rangle}{\sqrt N} &=& - \frac{\tilde z_0}{\tilde\omega_0} 
 \Re\tilde \Gamma \, .
\end{eqnarray}
Note that the last relation agrees with what is known for a closed system in thermal equilibrium, 
though it is now valid also for the general case of an open system.  
Eqs.~\eqref{140} and \eqref{141} have to be solved self-consistently for $\mu$ and $\langle \psi_0\rangle$.

\subsection{Limit of vanishing damping rate $\kappa$}
\label{IV.G}
%
In this subsection, we study the limit of a vanishing damping rate $\kappa$ between the cavity photons and 
the free space photons. 
As stressed before, a finite leakage to external photons implies that the quantity $\mu$ does not act
 as a common chemical potential of the total system. The reason is, that  photons  
can  escape from the e-h-p system into the free-photon space.
On the other hand, for vanishing $\kappa$ thermal equilibrium should develop. Then $\mu$ should become 
the usual chemical potential for the remaining system, which is composed of the e-h-p subsystem 
and the two electronic baths.  In this context, we are mostly interested  in the case of strong 
damping rate $\gamma$ for the coupling rate to the electronic baths.

Analyzing the limit $\kappa\to 0$, we start from Eq.~\eqref{140} which states that the imaginary part of $\tilde \Gamma$ must vanish:
\begin{equation}
\label{141a}
\Im \tilde \Gamma =0 \, .
\end{equation}
Thereby $\tilde \Gamma$ results from the solution of the renormalization equation \eqref{133}
 for $\Gamma_\lambda$,
  \begin{equation}
 \label{141b}
 \Gamma_{\lambda -\Delta \lambda} - \Gamma_\lambda =
  \delta \Gamma^{(0)}_\lambda + \delta \Gamma^{(c)}_\lambda
 + \delta \Gamma^{(g)}_\lambda\,,
 \end{equation}
with renormalization contributions $\delta {\Gamma}_{\lambda}^{(0)}$, $\delta \Gamma^{(g)}_\lambda$, 
and $\delta \Gamma^{(g)}_\lambda$, given by Eqs.~\eqref{134}--\eqref{136}. One finds for the imaginary parts:
\begin{align}
\label{141c}
 \Im \Gamma_{\lambda -\Delta \lambda} -& \Im\Gamma_\lambda = \frac{g}{N} \sum_{\vec k} A_{\vec k 0}(\lambda, \Delta \lambda) \\
  &\times \big[
 (\varepsilon_{\vec k,\lambda}^e + \varepsilon_{\vec k,\lambda}^h) \Im d_{\vec k}
 - (n_{\vec k}^e + n_{\vec k}^h -1) \Im \Delta_{\vec k, \lambda} \big]   \nonumber 
\end{align}
with  initial value 
$\Gamma_{\lambda = \Lambda} = \hat \Gamma=   \Gamma - \frac{g}{N} \sum_{\vec k} d_{\vec k}$,  
($\Gamma=0^+$). Here  we have already exploited that $\delta \Gamma_\lambda^{(g)}$ is real. 
 In the following  we again neglect all renormalization contributions to $d_{\vec k}$ and $n_{\vec k}^{e,h}$ from
 Eqs.~\eqref{111}--\eqref{113}, thereby replacing $d_{\vec k}$ and $n_{\vec k}^{e,h}$ by $\hat d_{\vec k}$ and $\hat n_{\vec k}^{e,h}$. Then, according to
Eqs.~\eqref{124d} and \eqref{124e} we obtain
\begin{align}
 \label{141d}
 \hat d_{\vec k} =& \frac{\tilde \Delta_{\vec k} {\rm sgn}(\tilde \varepsilon_{\vec k}^e + \tilde \varepsilon_{\vec k}^h)}{2 W_{\vec k}} F_{1 \vec k}^+ 
\nonumber \\
& + \frac{1}{2} \frac{\tilde \Delta_{\vec k}}{(\tilde \varepsilon_{\vec k}^e + \tilde \varepsilon_{\vec k}^h) - 2i \gamma} 
\,  \frac{ F_{2 \vec k}^+ -2}{1+ \displaystyle 
  \frac{4|\tilde \Delta_{\vec k}|^2}{(\tilde \varepsilon^e_{\vec k} +\tilde \varepsilon^h_{\vec k})^2 + (2 \gamma)^2}   } 
 \end{align}
and 
\begin{equation}
 \label{141e}
  \hat n^e_{\vec k} + \hat n^h_{\vec k} -1  = 
\frac{|\tilde \varepsilon^e_{\vec k} +\tilde \varepsilon^h_{\vec k}|}{2W_{\vec k}} F_{1\vec k}^+ 
+  
\frac{1}{2} \frac{ F_{2 \vec k}^+ -2}{1+ \displaystyle 
  \frac{4|\tilde \Delta_{\vec k}|^2}{(\tilde \varepsilon^e_{\vec k} +\tilde \varepsilon^h_{\vec k})^2 + (2 \gamma)^2}   } .
\end{equation} 
Inserting Eqs.~\eqref{141d} and \eqref{141e} into Eq.~\eqref{141c} we find
\begin{align}
\label{141f}
 &\Im \Gamma_{\lambda -\Delta \lambda} - \Im\Gamma_\lambda =   \frac{g}{N} \sum_{\vec k} A_{\vec k 0}(\lambda, \Delta \lambda)\nonumber \\
 & \times \Big[
(\varepsilon_{\vec k,\lambda}^e + \varepsilon_{-\vec k,\lambda}^h) \Im \tilde \Delta_{\vec k} 
-  (\tilde \varepsilon_{\vec k}^e + \tilde \varepsilon_{-\vec k}^h) \Im \Delta_{\vec k, \lambda}
\Big]   \frac{{\rm sgn}(\tilde \varepsilon_{\vec k}^e + 
 \tilde \varepsilon_{-\vec k}^h) }{2W_{\vec k}} F^+_{1\vec k}  \nonumber \\
 & + \frac{g}{2N} \sum_{\vec k} A_{\vec k 0}(\lambda, \Delta \lambda) 
  \Big[(\varepsilon_{\vec k,\lambda}^e + \varepsilon_{-\vec k,\lambda}^h)
  \Im \Big( \frac{\tilde \Delta_{\vec k}}{ (\tilde \varepsilon_{\vec k}^e + \tilde \varepsilon_{-\vec k}^h) -2i \gamma} 
  \Big) \nonumber \\
  & \qquad\qquad\qquad -  \Im \Delta_{\vec k, \lambda} \Big]
 \frac{F^+_{2\vec k} -2}{1 +  \displaystyle 
 \frac{4|\tilde \Delta_{\vec k}|^2}{(\varepsilon_{\vec k,\lambda}^e + \varepsilon_{-\vec k,\lambda}^h)^2 + (2\gamma)^2} }\,.
\end{align}
This result can further be simplified. First of all,  we neglect the first term in Eq.~\eqref{141f}, which is  small.   
It consists of the difference of two contributions  which are of quite similar 
character. In particular,  for small $\lambda$ (almost full renormalization) the cancellation of the two terms is exact, and for $\lambda= \Lambda$ (initial point) contributions from the  renormalization are small.  Thus 
\begin{align}
\label{141fa}
 &\Im \Gamma_{\lambda -\Delta \lambda} - \Im\Gamma_\lambda \approx  \frac{g}{2N} \sum_{\vec k} A_{\vec k 0}(\lambda, \Delta \lambda) \nonumber \\
 & \times \Big[(\varepsilon_{\vec k,\lambda}^e + \varepsilon_{-\vec k,\lambda}^h)
  \Im \Big( \frac{\tilde \Delta_{\vec k}}{ (\tilde \varepsilon_{\vec k}^e + \tilde \varepsilon_{-\vec k}^h) -2i \gamma} 
  \Big) -  \Im \Delta_{\vec k, \lambda} \Big] \nonumber \\
  & \times
 \frac{F^+_{2\vec k} -2}{1 +  \displaystyle 
 \frac{4|\tilde \Delta_{\vec k}|^2}{(\varepsilon_{\vec k,\lambda}^e + \varepsilon_{-\vec k,\lambda}^h)^2 + (2\gamma)^2} }\,.
\end{align}
Next, let us consider the limit of large damping $\gamma$, thereby assuming that the following conditions are fulfilled: 
\begin{eqnarray}
\label{141fb}
&&2\gamma \gg [\varepsilon_{\vec k,\lambda}^e + \varepsilon_{-\vec k,\lambda}^h| \quad \mbox{and}
\quad  2\gamma \gg 2|\tilde \Delta_{\vec k}|
\end{eqnarray} 
for most values of $\vec k$. The first condition is met easier for a semimetal than for a semiconductor. 
As a consequence of the conditions \eqref{141fb}, an expansion of Eq.~\eqref{141fa} for large $\gamma$  gives to 
leading order  $\gamma^{-1}$:
\begin{eqnarray}
\label{141g}
 &&\Im \Gamma_{\lambda -\Delta \lambda} - \Im\Gamma_\lambda  \approx 
- \frac{g}{2N} \sum_{\vec k} A_{\vec k 0}(\lambda, \Delta \lambda)  \nonumber \\
 &&\qquad \times \big(  \Im \Delta_{\vec k, \lambda} -  \frac{\varepsilon^e_{\vec k, \lambda} + 
\varepsilon^h_{-\vec k, \lambda} }{2 \gamma} \Re{\tilde \Delta_{\vec k}} \big)
(F^+_{2\vec k} -2) \,.
\end{eqnarray}
Here the term $\sim (1/\gamma)\Re \tilde \Delta_{\vec k}$ followed from the first contribution in the squared brackets
of Eq.~\eqref{141fa} and the denominator of the common factor behind the brackets was replaced by one. 
Expanding Eq.~\eqref{141d} to the same order as Eq.~\eqref{141g}, the imaginary part of  the initial condition,
$\Gamma_\Lambda = \Gamma - \frac{g}{N} \sum_{\vec k} d_{\vec k}\ (\Gamma=0^+)$,  
  becomes for large $\gamma$
\begin{equation}
\label{141h}  
 \Im \Gamma_\Lambda = \Im \Gamma 
 - \frac{g}{N} \sum_{\vec k} \frac{(\Im \tilde \Delta_{\vec k}) 
 {\rm sgn}(\tilde \varepsilon_{\vec k}^e + \tilde \varepsilon_{\vec k}^h)}{2 W_{\vec k}} F_{1 \vec k}^+ 
+ \frac{\Re \tilde \Delta_{\vec k}}{2\gamma} \frac{F^+_{2\vec k}-2}{2} \,.
\end{equation}
Our aim is to study under which conditions the renormalized quantity $\tilde \Gamma$ is real so that  
$\Im \tilde \Gamma =0$ is valid.  For this, according to Eq.~\eqref{141g}, one also has to  study the renormalization of 
$\Im \Delta_{\vec k, \lambda}$.  As is easily seen, $\Im \Delta_{\vec k, \lambda}$ renormalizes to zero 
in dominant order, whereas $\Re \tilde \Delta_{\vec k}$ stays finite. 
Therefore, in order to arrive at the desired result $\Im \tilde \Gamma =0$, including the less dominant contribution 
on the right hand side  of Eq.~\eqref{141g}, the common factor $(F^+_{2\vec k} -2)$  must vanish. 
This condition can only be met by fixing the value of $\mu$ to
the chemical potential $\mu_B$ of the electronic baths,  
$\mu= \mu_B$, so that $F_{2\vec k}^+ -2 =0$.

The remaining equations for $\Re \Gamma_\lambda$ and $\Re \Delta_{\vec k, \lambda}$
 are easily found from Eqs.~\eqref{134} and \eqref{65}, and 
 completely agree with those of the equilibrium case.  Also the quantities $\hat d_{\vec k}$ and 
 $\hat n_{\vec k}^e + \hat n_{\vec k}^h -1$ for large $\gamma$  agree with the corresponding equilibrium expressions:  
 \begin{align}
\label{141ij}
& \hat d_{\vec k} = \frac{\tilde \Delta_{\vec k} {\rm sgn}(\tilde \varepsilon_{\vec k}^e + \tilde \varepsilon_{\vec k}^h)}{2 W_{\vec k}} F_{1 \vec k}^+\,, \\
 &  \label{141jjj}
  \hat n^e_{\vec k} + \hat n^h_{\vec k} -1  =
\frac{|\tilde \varepsilon^e_{\vec k} +\tilde \varepsilon^h_{\vec k}|}{2W_{\vec k}} F_{1\vec k}^+ \, .
\end{align}
 
To sum up, we have shown that the present extension of the PRM 
leads back to the usual thermodynamic equilibrium approach of Ref.~\cite{PBF16}.  
The  equilibrium is mainly of electronic nature with $\mu= \mu_B$ for the case 
that the following two conditions are fulfilled:
(i) the damping $\kappa$ of cavity photons to free space photons is zero and (ii) the coupling
$\gamma$ of the e-h-p subsystem to the electronic baths is sufficiently large in accordance with 
Eq.~\eqref{141fb}.   In particular, the second condition is only fulfilled, 
when $\tilde \Delta_{\vec k}$ is sufficiently small. 
However, as shown in Fig.~\ref{fig4},  its photonic part $\Delta_{ph}$  may tremendously increase 
 at larger values of $n_{exc}$ in the case of large detuning $d=3.5$, whereas at small detuning $d=-0.5$ 
 the quantity  $\Delta_{ph}$ already starts to increase at comparatively small values of $n_{exc}$. As 
 discussed in more detail below, this behavior of $\Delta_{ph}$ can be understood as a phase space filling and Pauli blocking effect. 
At small $n_{exc}$ additional excitations are either of excitonic or polaritonic nature until the electronic bands 
are completely filled. Then, for even larger $n_{exc}$ photonic excitations dominate.
Note, however, that  even for large $n_{exc}$, when $\Delta_{ph}$ and $\tilde \Delta_{\vec k}$ become large, 
the total e-h-p subsystem, together with the electronic reservoirs, has to realize a thermodynamic equilibrium state for the 
case that $\kappa$ is zero or sufficiently small.

\section{Single-particle spectral function }
\label{V}

The one-particle spectral function $A(\vec k, \omega)$ for the steady state is 
defined by the 
Laplace transform of the time-dependent electron anti-commutator correlation function in the limit $t \rightarrow \infty$:
\begin{eqnarray}
\label{165a}
 A_e(\vec k, \omega) &=& \frac{1}{\pi} \lim_{t \rightarrow \infty} \Re \int_0^\infty   d\tau e^{i\omega \tau}
\langle [ e^\dag_{\vec k}(t), e_{\vec k}(t + \tau)  ]_+\rangle \nonumber \\
&& \\
\label{166a}
&=& \frac{1}{\pi} \lim_{t \rightarrow \infty} \Re \int_0^\infty  d\tau e^{i\omega \tau}
\langle [ \tilde e^\dag_{\vec k}(t), \tilde e_{\vec k}(t + \tau)  ]_+\rangle_{\tilde \rho_0} . \nonumber 
\end{eqnarray}
In the first equation the time dependence is governed by the original 
Hamiltonian $\mathcal H$, whereas in the second line the dynamics is again given by
$\tilde{\mathcal H}$, and also the expectation value is formed with the transformed 
density operator $\tilde \rho_0$.  Note that in Eq.~\eqref{165a} the lower integration limit $\tau=0$ is a time much larger than $\tau_0$.  
For that time a steady state has already been reached, with properties that do not depend on the details of the initial state. 
Likewise the two-time correlation function
$\langle  \tilde e_{\vec q}^\dag(t) \tilde e_{\vec q}(t + \tau) \rangle_{\tilde \rho_0}$ depends only on the 
relative time difference $\tau$ and not on $t$.
Thus, the stationary spectrum can be calculated at any fixed time  
$t$. At the end, $t$ is shifted to infinity. Furthermore,  $\tilde e^\dag_{\vec k}$ is the transformed one-particle
 operator \eqref{90}:
\begin{eqnarray}
\label{167}
\tilde e_{\vec k}^\dag &=& \tilde{x}_{\vec k} e_{\vec k}^\dag
+\frac{1}{\sqrt N}   \sum_{\vec q}  \tilde t_{\vec k-\vec q,\vec q}   h_{\vec q -\vec k} \, 
:{\psi}_{\vec q}^{\dagger}: \\
&&+ \frac{1}{N}\sum_{\vec k_1\vec k_2}
\tilde{\alpha}_{\vec k_1 \vec k \vec k_2} e_{\vec k_1}^\dag
:h_{\vec k_2}^\dag h_{\vec k_1+ \vec k_2-\vec k}: \, .\nonumber
\end{eqnarray}

Let us consider the coherent part of the spectrum, which  results from the first term on the right hand side of Eq.~\eqref{167}:
\begin{align}
\label{168}
 &A_{e, \rm coh}(\vec k, \omega) =\frac{|\tilde x_{\vec k}|^2}{\pi}\\ &\qquad\times \lim_{t \rightarrow \infty} \Re \int_0^\infty  d\tau e^{i\omega \tau}
\langle [ e^\dag_{\vec k}(t), e_{\vec k}(t + \tau)  ]_+\rangle_{ \rho_0} \,.\nonumber
\end{align}
Decomposing $e^\dag_{\vec k}$ and $e_{\vec k}$ into eigenmodes $C^{(\dag)}_{1\vec k}$ and 
 $C^{(\dag)}_{2\vec k}$ of $\tilde{\mathcal H}_S$, according to Eqs.~\eqref{70} and \eqref{71},
$A_{\rm coh}(\vec k, \omega)$  transforms to
\begin{align}
\label{169}
& A_{e, \rm coh}(\vec k, \omega) =\frac{|\tilde x_{\vec k}|^2}{\pi} \lim_{t \rightarrow \infty} \Re \int_0^\infty  d\tau e^{i\omega \tau} \nonumber \\
 & \qquad \quad \times \Big\{ |\xi_{\vec k}|^2 \langle [ C^\dag_{1\vec k}(t), C_{1 \vec k}(t + \tau)  ]_+\rangle_{ \rho_0} 
 \nonumber \\
 & \qquad \quad+ |\eta_{\vec k}|^2 \langle [ C^\dag_{2\vec k}(t), C_{2 \vec k}(t + \tau)  ]_+\rangle_{ \rho_0} \nonumber \\
& \qquad \quad- \xi^*_{\vec k} \eta_{\vec k} \langle [ C^\dag_{1\vec k}(t), C_{2 \vec k}(t + \tau)  ]_+\rangle_{ \rho_0}
\nonumber \\
 & \qquad \quad - \xi_{\vec k} \eta^*_{\vec k} \langle [ C^\dag_{2\vec k}(t), C_{1 \vec k}(t + \tau)  ]_+\rangle_{ \rho_0}
  \Big\}\,.
\end{align}
  The $\tau$-dependence will be treated by employing the Mori-Zwanzig projection formalism described in Appendix \ref{C}.  
  Using the fermionic anti-commutator relations we find
  \begin{eqnarray}
\label{170}
\langle [ C^\dag_{1\vec k}(t), C_{1 \vec k}(t + \tau)  ]_+\rangle_{ \rho_0} &=& e^{(- i \tilde E_{1\vec k} -\gamma) \tau}\,,\\
\label{171}
 \langle [ C^\dag_{1\vec k}(t), C_{2 \vec k}(t + \tau)  ]_+\rangle_{ \rho_0} &=& 0\,.
  \end{eqnarray}
This leads for $A_{\rm coh}(\vec k, \omega)$ to 
\begin{eqnarray}
\label{172}
&& A_{e, \rm coh}(\vec k, \omega) =\frac{|\tilde x_{\vec k}|^2}{\pi} \lim_{t \rightarrow \infty} \Re \int_0^\infty  d\tau e^{i\omega \tau}  \\
&& \qquad\qquad \times \Big\{ |\xi_{\vec k}|^2 e^{(- i \tilde E_{1\vec k} -\gamma) \tau} +
       |\eta_{\vec k}|^2 e^{(- i \tilde E_{2\vec k} -\gamma) \tau}   \Big\}\nonumber\,.
 \end{eqnarray}
Finally, by  integrating over $\tau$ and taking into account only the dissipative part of the integral, one finds
 \begin{eqnarray}
\label{173}
&& A_{e, \rm coh}(\vec k, \omega) =\frac{|\tilde x_{\vec k}|^2}{\pi}  \\
&& \quad \times \Big\{ |\xi_{\vec k}|^2  \frac{\gamma}{(\tilde E_{1\vec k} -\omega)^2 + \gamma^2}
 +  |\eta_{\vec k}|^2  \frac{\gamma}{(\tilde E_{2\vec k} -\omega)^2 + \gamma^2}   \Big\} \nonumber \,,.
 \end{eqnarray} 
  Thus, the spectrum $A_{e, \rm coh}(\vec k, \omega)$ consists of resonances at the quasiparticle 
  energies $\tilde E_{1\vec k}$ and $\tilde E_{2\vec k}$ with damping $\gamma$ and weights 
  which are determined by 
  $|\xi_{\vec k}|^2$ and $|\eta_{\vec k}|^2$, respectively. The spectral function 
  $A_{h, \rm coh}(\vec k, \omega)$ for holes can be written in the form~\eqref{173} as well, however, with
the  weights $|\xi_{\vec k}|^2$ and $|\eta_{\vec k}|^2$ interchanged. The incoherent part of $A_{e}(\vec k, \omega)$
  can be obtained by help of the second and third term in Eq.~\eqref{167}, and  is expected to lead 
  to a background spectrum for the coherent part.

\section{Steady-state Luminescence}
\label{VI}

The steady-state emission spectrum 
is obtained from the Laplace transform of the photon
correlation function~\cite{SZ97}: 
\begin{equation}
\label{142}
S(\vec q, \omega) = \frac{1}{\pi} \lim_{t \rightarrow \infty} \Re \int_{0}^\infty d\tau \, e^{ i\omega \tau}
\langle  \psi_{\vec q}^\dag(t) \psi_{\vec q}(t + \tau) \rangle 
\end{equation}
or---with the help of relation \eqref{84}---by
\begin{eqnarray}
\label{143}
&& S(\vec q, \omega) = \frac{1}{\pi} \lim_{t \rightarrow \infty} \Re \int_{0}^\infty d\tau \, e^{i\omega \tau}
\langle  \tilde \psi_{\vec q}^\dag(t) \tilde \psi_{\vec q}(t + \tau) \rangle_{\tilde \rho_0} \,. \nonumber \\
&&
\end{eqnarray}
Again, in Eq.~\eqref{142}, the time dependence  is governed by the original 
Hamiltonian $\mathcal H$, whereas in Eq.~\eqref{143} the dynamics is given by the transformed 
Hamiltonian $\tilde{\mathcal H}$. Moreover,  $\tilde \psi_{\vec q}^\dag$ is the transformed photon 
operator \eqref{126}, and the expectation value in Eq.~\eqref{143} is formed with the transformed initial
density operator $\tilde \rho_0$. We note that  a quite similar photon correlation function $B(\vec q, \omega)$ 
was studied  in Ref.~\cite{PBF16} for thermal equilibrium. However,  in contrast to $S(\vec q, \omega)$ the function 
$B(\vec q, \omega)$ was a response function, that is
a photon commutator correlation function. We would like to point out here that the  
renormalization equations (B41) and (B36) in Ref.~\cite{PBF16} are not  completely correct. 
The correct equations are given by the present Eqs.~\eqref{B20} and \eqref{B21}.

\subsection{Coherent part}
\label{V.A}
 Also the luminescence spectrum consists of two parts. 
 The coherent part results from the first term on the right hand side  of Eq.~\eqref{126}:
\begin{equation}
\label{145}
S_{\rm coh}(\vec q, \omega) =\frac{|\tilde z_{\vec q}|^2}{\pi} \lim_{t \rightarrow \infty} \Re \int_{0}^\infty d\tau \, e^{i\omega \tau}
\langle   \psi_{\vec q}^\dag(t)  \psi_{\vec q}(t + \tau) \rangle_{\tilde \rho_0} .
\end{equation}
The time dependence of $\psi_{\vec q}^\dag(t)$  on the right hand side  of Eq.~\eqref{145}
is found from the solution of the equation of 
motion \eqref{128}:
\begin{equation}
\label{146}
\psi^\dag_{\vec q}(t) = - \frac{i \sqrt N \tilde \Gamma^* }{ i \tilde \omega_0 - \kappa} \delta_{\vec q,0}
+  \Big(  \psi^\dag_{\vec q} +  \frac{ i \sqrt N \tilde \Gamma^* }{ i \tilde \omega_0 - \kappa} \delta_{\vec q,0}
\Big) e^{( i \tilde \omega_{\vec q} - \kappa) t }  \, .
\end{equation}
Substituting \eqref{146} into Eq.~\eqref{145} leads to
\begin{equation}
\label{147}
S_{\rm coh}(\vec q, \omega) = \frac{N |\tilde z_0|^2|\tilde \Gamma|^2}{\hat \omega_0^2 + \kappa^2} \delta_{\vec q,0}
\, \delta(\omega) \,,
\end{equation}
which shows that  in the condensed phase a delta-function peak appears at $\omega=0$. 

\subsection{Incoherent part}
\label{V.B}

The incoherent part of $S(\vec q, \omega)$ is given by 
\begin{equation}
\label{148}
 S_{\rm inc}(\vec q, \omega) = \frac{1}{\pi} \lim_{t \rightarrow \infty} \Re \int_{0}^\infty d\tau \, e^{i\omega \tau}
\langle   :b_{\vec q}^\dag(t): \,  :b_{\vec q}(t + \tau): \rangle_{\tilde \rho_0} \,, 
\end{equation}
where $b^\dag_{\vec q}$ creates an exciton with wave vector $\vec q$ which is modified by 
the coefficients $\tilde v_{\vec k \vec q} $:
\begin{equation}
\label{149}
 b^\dag_{\vec q}(t) = 
\frac{1}{\sqrt N} \sum_{\vec k} \tilde v_{\vec k \vec q} \, 
(e^\dag_{\vec k  + \vec q} h^\dag_{-\vec k})(t) \, .
\end{equation}
Thus 
\begin{align}
\label{150}
&S_{\rm inc}(\vec q, \omega) = \frac{1}{N\pi} \lim_{t \rightarrow \infty} \sum_{\vec k \vec k'}  \tilde v_{\vec k \vec q}
 \tilde v_{\vec k' \vec q}^*
\Re \int_{0}^\infty d\tau \, e^{i\omega \tau} \nonumber \\
&\quad\times \langle   :(e^\dag_{\vec k  + \vec q} h^\dag_{-\vec k})(t) : \,  
:(h_{-\vec k'} e_{\vec k' + \vec q})(t + \tau):   \rangle_{\tilde \rho_0}  \, .
\end{align}
In a factorization approximation this simplifies to 
\begin{align}
\label{151}
& S_{\rm inc}(\vec q, \omega) = \frac{1}{N\pi} \lim_{t \rightarrow \infty} \sum_{\vec k }  |\tilde v_{\vec k \vec q}|^2
\Re \int_{0}^\infty d\tau \, e^{i\omega \tau} \nonumber \\
&\;\times \langle e^\dag_{\vec k  + \vec q}(t) e_{\vec k + \vec q}(t + \tau) \rangle_{\tilde \rho_0} 
\langle h^\dag_{-\vec k}(t) h_{-\vec k}(t + \tau)   \rangle_{\tilde \rho_0} \, .
\end{align}
Note that expectation values $\langle  :(e^\dag_{\vec k  + \vec q} h^\dag_{-\vec k})(t): \rangle_{\tilde \rho_0}$ and $\langle:(h^\dag_{-\vec k'} e_{\vec k' + \vec q})(t + \tau) :\rangle_{\tilde \rho_0}$ drop out in Eq.~\eqref{150} so that only the pairwise factorization of Eq.~\eqref{151} survives.   

What remains to be done is the $\tau$ integration in Eq.~\eqref{151}.  Again, this can best be  
achieved by using Bogolyubov quasiparticles in accordance with
Eqs.~\eqref{70} and \eqref{71}. With
\begin{eqnarray}
\label{152}
 h_{-\vec k}^\dag &=& \eta_{\vec k}^*C_{1\vec k} +
\xi_{\vec k}^* C_{2 \vec k} \,, \\
e^\dag_{\vec k} &=&  \xi_{\vec k}^* C_{1 \vec k }^\dag - \eta_{\vec k }^* C_{2 \vec k}^\dag \,,
\end{eqnarray}
one finds
\begin{align}
\label{153}
 &\langle e^\dag_{\vec k}(t) e_{\vec k}(t+\tau)  \rangle_{\tilde \rho_0} = 
|\xi_{\vec k}|^2  \langle C^\dag_{1\vec k}(t) C_{1 \vec k}(t+\tau)  
\rangle_{\tilde \rho_0} \nonumber \\
&\hspace*{3cm}+ |\eta_{\vec k}|^2  \langle C^\dag_{2\vec k}(t) C_{2 \vec k}(t+\tau)  \rangle_{\tilde \rho_0}\nonumber\\
& \hspace*{3cm}- \eta_{\vec k} \xi_{\vec k}^*  \langle C^\dag_{1\vec k}(t) C_{2 \vec k}(t+\tau)  \rangle_{\tilde \rho_0} 
\nonumber \\
&\hspace*{3cm}- \eta_{\vec k}^* \xi_{\vec k}  \langle C^\dag_{2\vec k}(t) C_{1 \vec k}(t+\tau)  \rangle_{\tilde \rho_0}
\end{align}
and 
\begin{align}
\label{154}
& \langle h^\dag_{-\vec k}(t) h_{-\vec k}(t+\tau)  \rangle_{\tilde \rho_0} =|\eta_{\vec k}|^2  \langle C_{1\vec k}(t) C_{1 \vec k}^\dag(t+\tau)  \rangle_{\tilde \rho_0} \nonumber \\
& \hspace*{3cm}+ |\xi_{\vec k}|^2  \langle C_{2\vec k}(t) C_{2 \vec k}^\dag(t+\tau)  \rangle_{\tilde \rho_0} \nonumber \\
& \hspace*{3cm}+ \eta_{\vec k}^* \xi_{\vec k}  \langle C_{1\vec k}(t) C_{2 \vec k}^\dag(t+\tau)  \rangle_{\tilde \rho_0} 
\nonumber \\
& \hspace*{3cm}+ \eta_{\vec k} \xi_{\vec k}^*  \langle C_{2\vec k}(t) C_{1 \vec k}^\dag(t+\tau)  \rangle_{\tilde \rho_0}\,.
\end{align}
As before, the $\tau$-dependence  in Eqs.~\eqref{153} and \eqref{154} is treated 
by employing the Mori-Zwanzig projection formalism.
From the corresponding equations of motion one finds
\begin{align}
\label{155}
&\langle e^\dag_{\vec k}(t) e_{\vec k}(t+\tau)  \rangle_{\tilde \rho_0} =|\xi_{\vec k}|^2  e^{(- i\tilde E_{1\vec k} - \gamma)\tau}
\langle C^\dag_{1\vec k}(t) C_{1 \vec k}(t)  
\rangle_{\tilde \rho_0} \nonumber \\
&\hspace*{2cm} + |\eta_{\vec k}|^2  e^{(- i\tilde E_{2\vec k} - \gamma)\tau}
\langle C^\dag_{2\vec k}(t) C_{2 \vec k}(t)  \rangle_{\tilde \rho_0} \nonumber \\
&\hspace*{2cm}  - \eta_{\vec k} \xi_{\vec k}^*  e^{(- i\tilde E_{2\vec k} - \gamma)\tau}
\langle C^\dag_{1\vec k}(t) C_{2 \vec k}(t)  \rangle_{\tilde \rho_0} 
\nonumber \\
&\hspace*{2cm} - \eta_{\vec k}^* \xi_{\vec k}  e^{(-i\tilde E_{1\vec k} - \gamma)\tau}
\langle C^\dag_{2\vec k}(t) C_{1 \vec k}(t)  \rangle_{\tilde \rho_0}
\end{align}
and 
\begin{align}
\label{156}
& \langle h^\dag_{-\vec k}(t) h_{-\vec k}(t+\tau)  \rangle_{\tilde \rho_0} 
=|\eta_{\vec k}|^2 e^{(i\tilde E_{1\vec k} - \gamma)\tau} \langle C_{1\vec k}(t) C_{1 \vec k}^\dag(t)  \rangle_{\tilde \rho_0} \nonumber \\
& \hspace*{2cm}+ |\xi_{\vec k}|^2  e^{(i\tilde E_{2\vec k} - \gamma)\tau} 
\langle C_{2\vec k}(t) C_{2 \vec k}^\dag(t)  \rangle_{\tilde \rho_0} \nonumber \\
& \hspace*{2cm}+ \eta_{\vec k}^* \xi_{\vec k} e^{(i\tilde E_{2\vec k} - \gamma)\tau} 
\langle C_{1\vec k}(t) C_{2 \vec k}^\dag(t)  \rangle_{\tilde \rho_0} 
\nonumber \\
& \hspace*{2cm}+ \eta_{\vec k} \xi_{\vec k}^* e^{(i\tilde E_{1\vec k} - \gamma)\tau} 
\langle C_{2\vec k}(t) C_{1 \vec k}^\dag(t)  \rangle_{\tilde \rho_0} 
\end{align}
with $\gamma$ being the damping rate of the electrons and holes of the e-h-p system due to the coupling 
 to the fermionic baths.  Combining all parts of the correlation functions 
 with the same $\tau$-dependence  one obtains:
 \begin{equation}
\label{157}
 \langle e^\dag_{\vec k}(t) e_{\vec k}(t+\tau)  \rangle_{\tilde \rho_0} =   a^e_{1\vec k}(t) \, 
 e^{(- i\tilde E_{1\vec k} - \gamma)\tau} 
 +  a^e_{2 \vec k}(t) \, 
  e^{(-i\tilde E_{2\vec k} - \gamma)\tau}
\end{equation}
 and
   \begin{equation}
\label{158}
\langle h^\dag_{-\vec k}(t) h_{-\vec k}(t+\tau)  \rangle_{\tilde \rho_0} =  
a^h_{1 \vec k} (t)  e^{(i\tilde E_{1\vec k} - \gamma)\tau} 
 + a^h_{2 \vec k} (t)  e^{(i\tilde E_{2\vec k} - \gamma)\tau} \, .
  \end{equation}
Here, we have introduced coefficients
\begin{eqnarray}
\label{159}
 a^e_{1 \vec k}(t) &=& |\xi_{\vec k}|^2 A^{11}_{\vec k}(t) -  \eta^*_{\vec k} \xi_{\vec k} A^{21}_{\vec k}(t)\,,
\\[0.1cm]
  a^e_{2 \vec k}(t) &=&    |\eta_{\vec k}|^2 A^{22}_{\vec k}(t) -  \eta_{\vec k} \xi^*_{\vec k} A^{12}_{\vec k}(t)\,,\\[0.1cm]
\label{160}
 a^h_{1 \vec k}(t) &=& |\eta_{\vec k}|^2 (1-A^{11}_{\vec k}(t)) -  \eta_{\vec k} \xi_{\vec k}^* A^{12}_{\vec k}(t)\,,
 \\[0.1cm]
  a^h_{2 \vec k}(t) &=&  |\xi_{\vec k}|^2 (1-A^{22}_{\vec k}(t)) -  \eta^*_{\vec k} \xi_{\vec k} A^{21}_{\vec k}(t) \,,
\end{eqnarray}
 and $ A^{nm}_{\vec k}(t) = \langle (C^\dag_{n\vec k} C_{m\vec k})(t) \rangle_{\tilde \rho_0}$ 
 [compare Eq.~\eqref{101}]. 
Finally, inserting  the relations \eqref{157} and \eqref{158} into 
Eq.~\eqref{151} and performing the integration over $\tau$ one finds
\begin{eqnarray}
\label{161}
&& S_{\rm inc}(\vec q, \omega) =  \frac{1}{N\pi}   \Re \sum_{\vec k} |\hat v_{\vec k \vec q}|^2   \\
&& \quad \times
\Big[
 \frac{2 \gamma \, }
{(\tilde E_{1 \vec k + \vec q} - \tilde E_{1 \vec k} - \omega)^2 + (2 \gamma)^2} \, 
a^e_{1 \vec k + \vec q}(t)\, a^h_{1 \vec k}(t)
\nonumber \\
&& \quad + 
 \frac{2 \gamma }
{(\tilde E_{2 \vec k + \vec q} - \tilde E_{2 \vec k} - \omega)^2 + (2 \gamma)^2} 
\, a^e_{2 \vec k + \vec q}(t)\, a^h_{2 \vec k}(t)
\nonumber \\
&& \quad +
 \frac{2 \gamma }
{(\tilde E_{1 \vec k + \vec q} - \tilde E_{2 \vec k} - \omega)^2 + (2 \gamma)^2} 
\, a^e_{1 \vec k + \vec q}(t)\, a^h_{2 \vec k}(t)
\nonumber \\
&&  \quad +
 \frac{2 \gamma }
{(\tilde E_{2 \vec k + \vec q} - \tilde E_{1 \vec k} - \omega)^2 + (2 \gamma)^2} 
\, a^e_{2 \vec k + \vec q}(t)\, a^h_{1 \vec k}(t)
\Big] \nonumber \, , 
\end{eqnarray}
where again only the dissipative part of the integral was considered. 
In Eq.~\eqref{161} the coefficients $a^{e,h}_{1\vec k}(t)$ and $a^{e,h}_{2\vec k}(t)$ still depend on time $t$.
The result for the steady state is obtained in the limit  $t \rightarrow \infty$. Thus,
\begin{eqnarray}
\label{165}
&& S_{\rm inc}(\vec q, \omega) =  \frac{1}{N\pi} \sum_{\vec k} |\tilde v_{\vec k \vec q}|^2    \\
&& \quad \times
\Big[
 \frac{2 \gamma \, }
{(\tilde E_{1 \vec k + \vec q} - \tilde E_{1 \vec k} - \omega)^2 + (2 \gamma)^2} \, 
a^e_{1 \vec k + \vec q}\, a^h_{1 \vec k}
\nonumber \\
&& \quad + 
 \frac{2 \gamma }
{(\tilde E_{2 \vec k + \vec q} - \tilde E_{2 \vec k} - \omega)^2 + (2 \gamma)^2} 
\, a^e_{2 \vec k + \vec q}\, a^h_{2 \vec k}
\nonumber \\
&& \quad+
 \frac{2 \gamma }
{(\tilde E_{1 \vec k + \vec q} - \tilde E_{2 \vec k} - \omega)^2 + (2 \gamma)^2} 
\, a^e_{1 \vec k + \vec q}\, a^h_{2 \vec k}
\nonumber \\
&&  \quad+
 \frac{2 \gamma }
{(\tilde E_{2 \vec k + \vec q} - \tilde E_{1 \vec k} - \omega)^2 + (2 \gamma)^2} 
\, a^e_{2 \vec k + \vec q}\, a^h_{1 \vec k}
\Big] \, . \nonumber 
\end{eqnarray}
Choosing the coefficients $a^e_{(1,2) \vec k}$ and    $a^h_{(1,2) \vec k}$ to be real is 
compatible with Eqs.~\eqref{157} and \eqref{158}. We obtain: 
\begin{eqnarray}
\label{162}
 a^e_{1\vec k + \vec q} &=& 
|\xi_{\vec k+ \vec q}|^2 A^{11}_{\vec k+ \vec q} -  
\Re\big(\eta^*_{\vec k+ \vec q} \xi_{\vec k+ \vec q} A^{21}_{\vec k+ \vec q}\big)
\,, \\
  a^e_{2 \vec k+ \vec q} &=&    |\eta_{\vec k+ \vec q}|^2 A^{22}_{\vec k+ \vec q} 
 - \Re \big(\eta_{\vec k+ \vec q} \xi^*_{\vec k+ \vec q} A^{12}_{\vec k+ \vec q}\big) \,,\\
 a^h_{1 \vec k} &=& |\eta_{\vec k}|^2 (1-A^{11}_{\vec k}) 
-  \Re \big(\eta_{\vec k} \xi_{\vec k}^* A^{12}_{\vec k} \big)
 \,,\\
  a^h_{2 \vec k} &=&  |\xi_{\vec k}|^2 (1-A^{22}_{\vec k}) 
- \Re \big( \eta^*_{\vec k} \xi_{\vec k} A^{21}_{\vec k} \big)
\label{162c}
\end{eqnarray}
with $A^{nm}_{\vec k} = A^{nm}_{\vec k}(t \rightarrow \infty)$
\begin{eqnarray}
\label{163}
A^{11}_{\vec k}  &=&  |\xi_{\vec k}|^2 f_e(\tilde E_{1 \vec k}) +
  |\eta_{\vec k}|^2  \big(1 - f_h(-\tilde E_{1 \vec k}) \big) \,, \\
 A^{22}_{\vec k}  &=&  |\eta_{\vec k}|^2 \, f_e(\tilde E_{2 \vec k}) +
 |\xi_{\vec k}|^2 \big(1 - f_h(-\tilde E_{2 \vec k}) \big)  \, ,  
\end{eqnarray}
and 
\begin{align}
\label{164}
\Re &\big(\eta_{\vec  k} \xi_{\vec k}^*A^{12}_{\vec k}\big) =
\frac{-2\gamma^2 \, |\xi_{\vec k}|^2 |\eta_{\vec k}|^2}{ (\tilde E_{1 \vec k} - \tilde E_{2 \vec k})^2 +(2 \gamma)^2  }
 \\
 &\quad \times 
 \big(  f_e(\tilde E_{1 \vec k}) + f_e(\tilde E_{2\vec k})  
  + f_h(-\tilde E_{1 \vec k})   + f_h(-\tilde E_{2 \vec k}) -2 \big) \nonumber
\end{align}
with $A^{21}_{\vec k} = (A^{12}_{\vec k})^*$. 
%
Obviously, the denominators of Eq.~\eqref{165} describe 
the frequency dependence of  $S_{\rm inc}(\vec q, \omega)$.  It is caused by 
transitions between energy levels of the quasiparticle Hamiltonian $\hat {\mathcal H}$. 
Whereas the first two excitations in \eqref{165} are due to transitions  within the same 
quasiparticle bands,
$\tilde E_{1 \vec k + \vec q} \rightarrow \tilde E_{1 \vec k}$ and 
$\tilde E_{2 \vec k + \vec q} \rightarrow \tilde E_{2 \vec k}$, the last two excitations result 
from transitions between the two
bands.   The factors $a^{e}_{(1,2)\vec k + \vec q}$ and  $a^{h}_{(1,2)\vec k}$  in~\eqref{165} 
determine the weight of the transitions. Note that all transitions  are broadened by $2 \gamma$, i.e., twice the damping 
rate $\gamma$ of single electrons or holes into their respective baths.  
In particular, for the case $\vec q=0$ one finds two quasi-elastic excitations around $\omega=0$ with
a broadening of  $2\gamma$ as well.

\section{Numerical results}
\label{VII}
Evaluating the theory developed so far, we assume, for simplicity, $\varepsilon^e_{\vec k}= \varepsilon^h_{\vec k}$ 
and charge neutrality $ \mu_e = \mu_h$. We then self-consistently solve the set of renormalization equations~\eqref{A16}--\eqref{A20},~\eqref{B5}--\eqref{B10}, and~\eqref{B20}--\eqref{B21}, together with Eqs.~\eqref{111}--\eqref{113}, and~\eqref{131} for the expectation values, in momentum space (on a grid with $N=160$ lattice sites), for a one-dimensional system.  Convergence is assumed to be achieved when the relative error of all quantities is less than $10^{-10}$.  

In the numerical work, we fix the interaction parameters $g=0.2$, $U=2.0$, the zero-point cavity photon frequency  $\omega_c=0.5$, and consider a finite but very low temperature $T=0.001$. All energies will be measured in units of the particle transfer amplitude $t$  and the wave vectors in units of the lattice constant $a$, where we take as typical values $t \simeq 2 eV$ and $a \simeq  5 {\rm \AA} $, yielding $c \simeq  0.4\,  c_0$ for the speed of light of the microcavity 
        ($c_0$ is  the speed of light  in vacuum).  We found that the physical properties only slightly depend on the actual value of $c$~\cite{PBF16}. 

Since the coupling between electrons, holes and photons is most effective in case the excitation energy of an electron-hole pair (exciton) matches a photonic excitation, we introduce, for the following discussion,  the so-called detuning 
\begin{equation}
\label{167c}
d = \omega_c-E_g\,,
\end{equation}
where $E_g$ denotes the minimum distance (gap) between the bare electron and hole bands~\cite{PBF16}. A positive (negative) $E_g$ indicates a semiconducting (semimetallic) setting.
 
 \subsection{Expectation values}

\begin{figure}[t]
\includegraphics[width=0.48\textwidth]{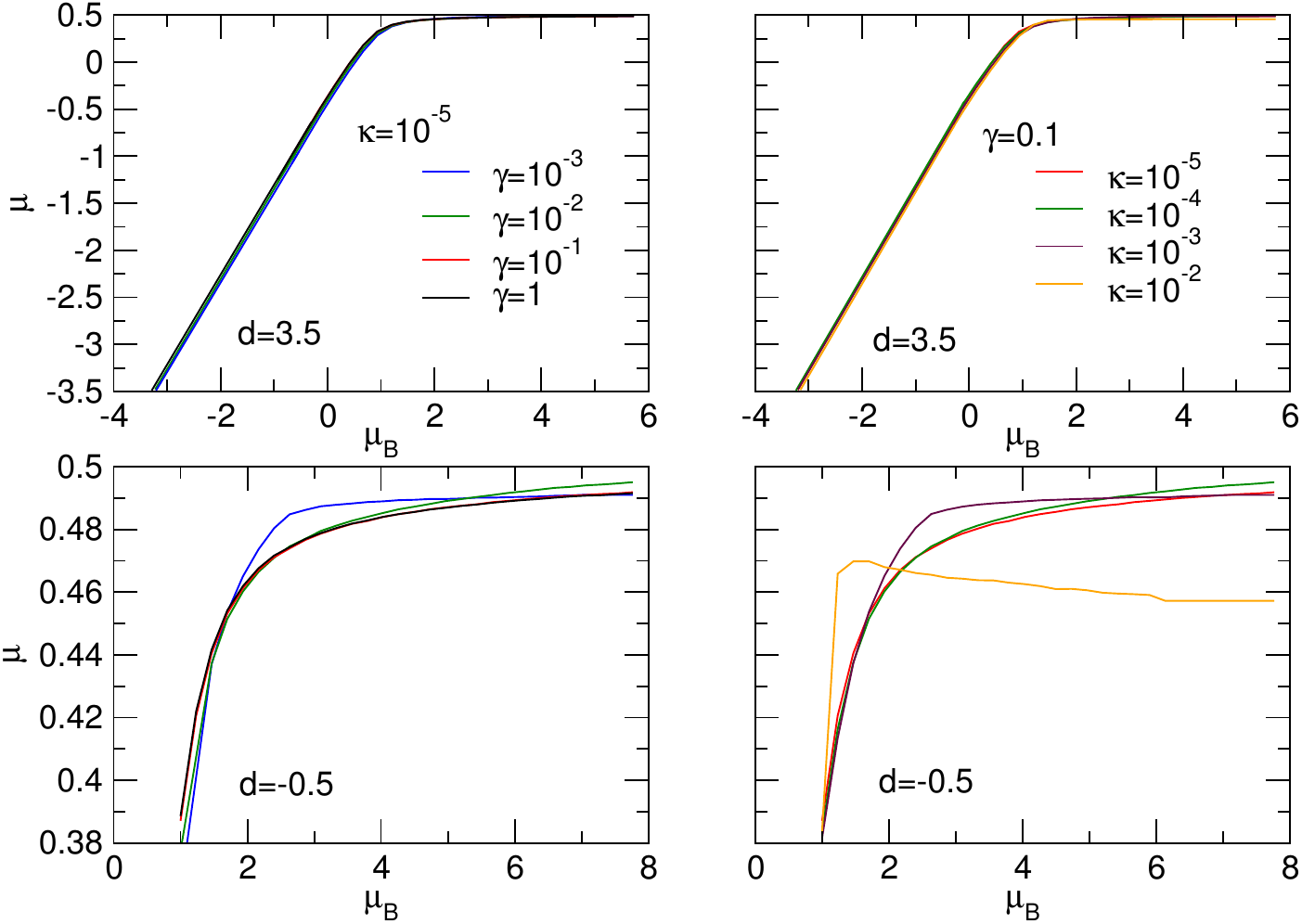}
\caption{(Color online) Parameter $\mu$ characterizing the steady state description~\eqref{15}--\eqref{16a}, where $\mu$ becomes the chemical potential in equlibrium.    Pictured here is $\mu$ as a function of $\mu_B$ for different values of $\gamma$ at  fixed $\kappa=10^{-5}$ (left panels) and, likewise, for different values of $\kappa$ at fixed $\gamma=0.1$  (right panels). The two upper panels refer to
detuning $d=3.5$, the lower panels refer to $d=-0.5$; note the different scales of the ordinates.} 
\label{fig1}
\end{figure}

We will start by examining the relation between $\mu$ and $\mu_B$. Remember that $\mu_e=\mu_h$ is the common 
chemical potential of both electronic baths, a parameter that is fixed from  outside. The quantity $\mu$, on the other side, gets a physical meaning in (quasi-) equilibrium only, where it becomes the chemical potential of the system. Therefore a difference 
between $\mu$ and $\mu_B$ can be taken as a measure for an increased importance of cavity photons 
(compare Sec.~\ref{IV.E}). Figure~\ref{fig1} gives $\mu$ as a function of $\mu_B$ at fixed damping rate $\kappa$ (left panels) and $\gamma$ (right panels), describing the coupling of the system to the photonic and electronic baths, respectively.  The upper and lower panels of  Fig.~\ref{fig1} reflect  large  and small detuning, where $d=3.5$, $E_g=-3$ and $d=-0.5$, $E_g=1$, respectively. 
In the former case, we observe  a linear dependence of $\mu$ on $\mu_B$ over 
almost the whole energy range of the electron band (bare bandwidth $4t$); 
the saturation when $\mu$ approaches the upper band edge originates from electron phase space filling. 
In the latter case, $\mu_B$ has to overcome the band gap $E_g$ first, thereafter  $\mu$ grows monotonously. 
If the self-consistently calculated $\mu$ reaches $\omega_c$, any further excitation 
is photonic in nature in both cases. This is the range where $\mu$ notably deviates from $\mu_B$ and non-equilibrium effects become important. These are more prominent for small detuning and less photon leakage.

\begin{figure}[t]
\includegraphics[width=0.48\textwidth]{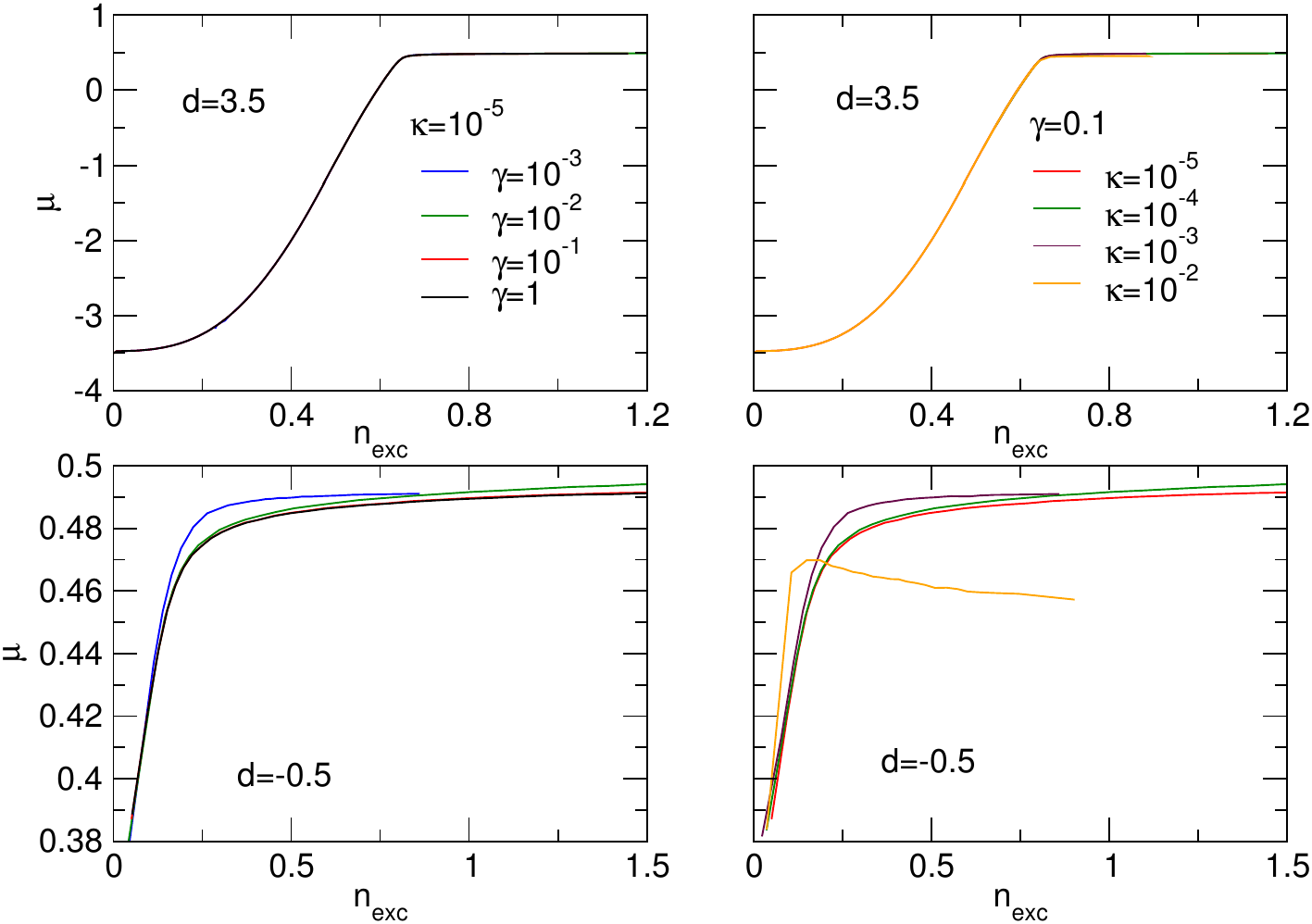}
\caption{(Color online) Steady-state parameter $\mu$ as a function of the total density of excitations  in the e-h-p microcavity system, $n_{exc}$ from~\eqref{124zz}. Shown are results for $d=3.5$ (top) and $d=-0.5$ (bottom) at a fixed value of $\kappa$ (left) and $\gamma$ (right).}
\label{fig2}
\end{figure}

Figure~\ref{fig2} directly relates $\mu$ to the total number $n_{exc}$  of excitations in the electron-hole-photon system, which is given by Eq.~\eqref{124zz}. At small-to-moderate  excitation densities and large (small) detuning, the excitations are excitons (polaritons) for the most part~\cite{PBF16}. Here, the system is close to (quasi-) equlibrium and $\mu$ takes over the role of a true chemical potential.    When $n_{exc}$ increases, the photons play a major role, and the system moves away from the former equilibrium configuration, which was described by $\mu= \mu_B$. This is why the curves $\mu(n_{exc})$ flatten for large
$n_{exc}$. Of course, above $n_{exc}=1$ any further excitation has to be photonic.  Again, for large detuning, the overall behavior of $\mu(n_{exc})$ only weakly depends on the damping/coupling parameters $\kappa$ and~$\gamma$.

\begin{figure}[t]
\includegraphics[width=0.48\textwidth]{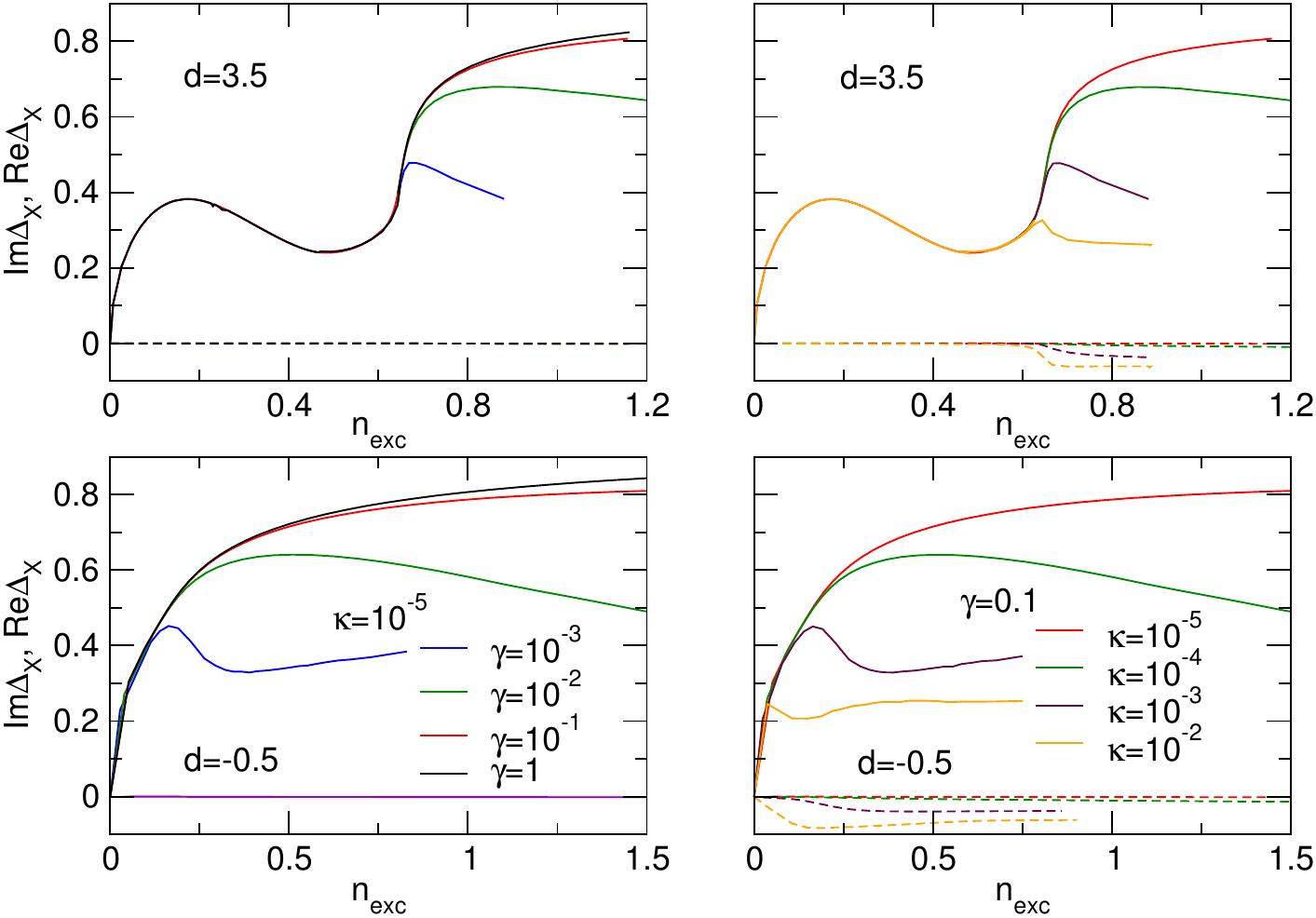}
\caption{(Color online) Excitonic order parameter $\Delta_X$ [see~Eq.~\eqref{167d}], reflecting the exciton contribution to the condensate in the steady state.  Depicted are the real parts (solid lines) and imaginary parts (dashed lines) of $\Delta_X$ as a function of $n_{exc}$  for large (top) and small (bottom) detuning at fixed  $\kappa$ (left) and $\gamma$ (right).
}
\label{fig3}
\end{figure}

\begin{figure}[htp]
\includegraphics[width=0.48\textwidth]{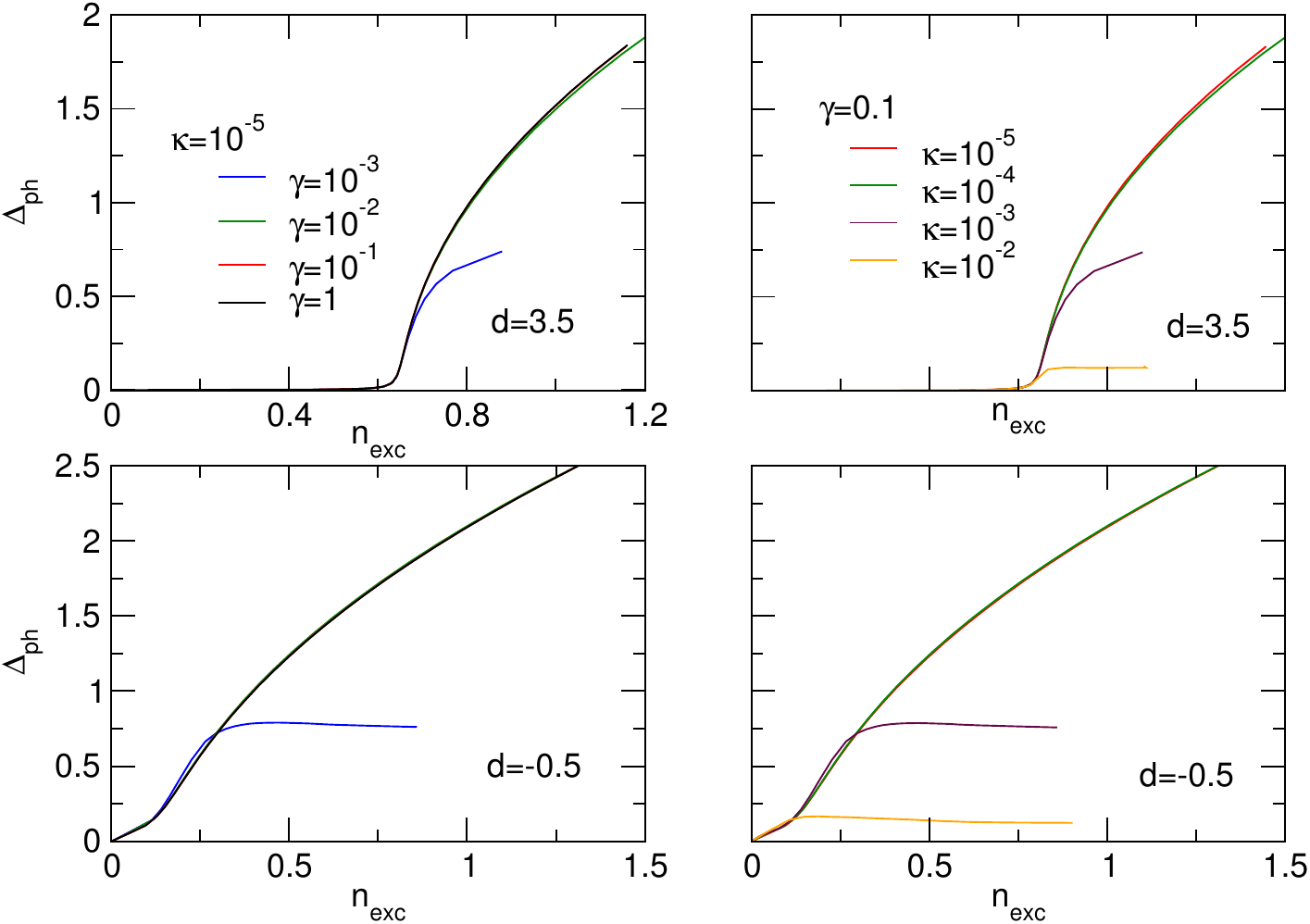}\\
\caption{(Color online) Photonic order parameter $\Delta_{ph}$ [see~Eq.~\eqref{167d}], reflecting the photon contribution to the condensate in the steady state.  Here, 
$\Delta_{ph}$ is given  as a function of $n_{exc}$ for $d=3.5$ (top) and $d=-0.5$ (bottom) at fixed
$\kappa=10^{-5}$ (left) and $\gamma=0.1$ (right). }
\label{fig4}
\end{figure}

We now aim at a characterization of the possible condensed phases of our e-h-p system.  In Fig.~\ref{fig3}  we show the excitonic order parameter, 
\begin{equation}
\label{167d}
\Delta_X= - (U/N) \sum_{\vec k} d_{\vec k}\,, 
\end{equation}
in dependence on  the density of excitations. Figure~\ref{fig4} gives the corresponding photonic order parameter 
\begin{equation}
\label{167f}
\Delta_{ph}=- (g/\sqrt N) \langle \psi_0\rangle 
\end{equation}
[compare Eq.~\eqref{34a}].   
At large detuning ($d=3.5$; upper panels),  valence and conduction bands will penetrate each other and---for the considered values of the Coulomb interaction between electrons and holes ($U=2$) and exciton-photon coupling ($g=0.2$)---a gapful renormalized band structure develops, just as  for a BCS-type excitonic insulator state~\cite{PBF10}. Here, the condensate formed at low and intermediate excitation densities is mainly triggered by the Coulomb attraction between electrons and holes and therefore is predominantly an excitonic one; cf. the vanishing value of $\Delta_{ph}$ in Fig.~\ref{fig4}.  If we would have strengthened the Coulomb interaction at fixed $n_{exc}$,  we would be able to observe a BCS-BEC crossover in the excitonic condensate~\cite{BF06,ZIBF12}.  Increasing the density of excitation $n_{exc}$ the location of the correlation-induced gap is shifted to larger $k$ values, and phase-space and Fermi-surface effects become increasingly important. This is indicated by the downturn of $\Re \Delta_X$.  At still  larger values of $n_{exc}$,  photonic excitations come into play  more and more. As a consequence, the condensate turns from excitonic to polaritonic, and finally to a purely photonic one (lasing regime~\cite{YNKOY15}).  For small detuning ($d=-0.5$; lower panels), where the system is in the semiconducting regime from the very beginning, both excitonic and photonic order parameters are finite, even at small excitation densities, which can be taken as a clear signature of a strong coupling between the light and matter degrees of freedom. As a result, a BEC of polaritons forms. Again the photons are dominant at large $n_{exc}$ (especially in the lasing regime). Obviously the influence of the bath degrees of freedom on the results is more pronounced for smaller (larger) values of  $\gamma$ ($\kappa$).  This is in accord with the analytical results of Sec.~IV~G, indicating that an equilibrium description is appropriate in the limit of large (vanishing) $\gamma$ ($\kappa$).  When $\gamma$ gets smaller, we found self-consistent solutions of the renormalization equations in a smaller range of $n_{exc}$ only. Note that the excitonic order parameter receives a finite imaginary part only for sufficiently large values of $\kappa$, almost irrespective of $\gamma$.

\begin{figure}[t]
\includegraphics[width=0.2\textwidth]{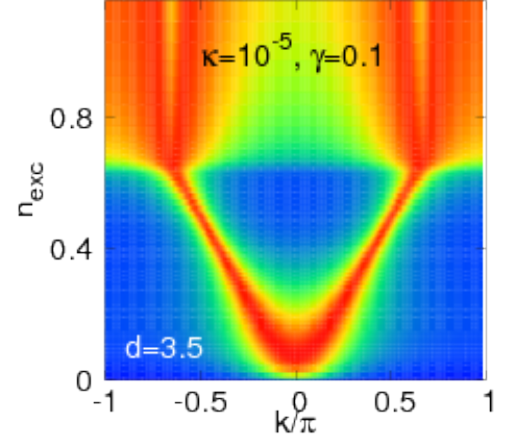}
\includegraphics[width=0.23\textwidth]{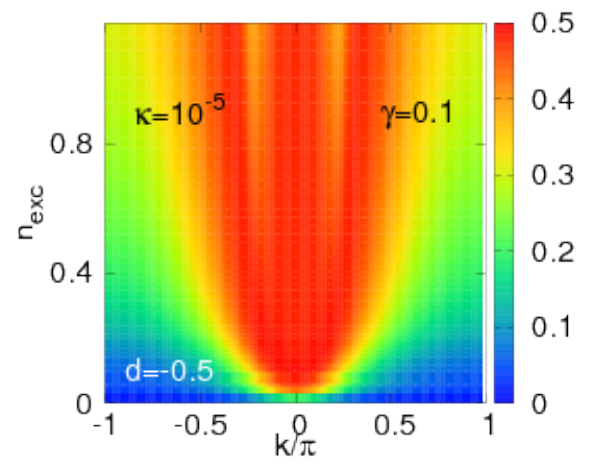}\\
\includegraphics[width=0.2\textwidth]{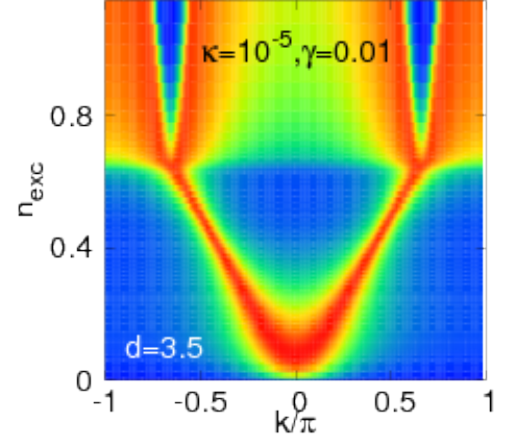}
\includegraphics[width=0.23\textwidth]{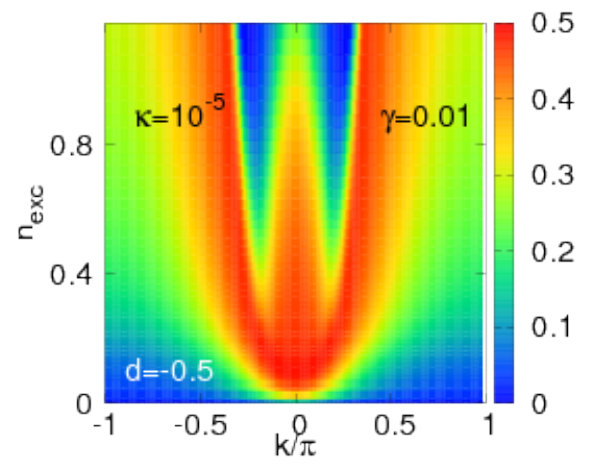}\\
\includegraphics[width=0.2\textwidth]{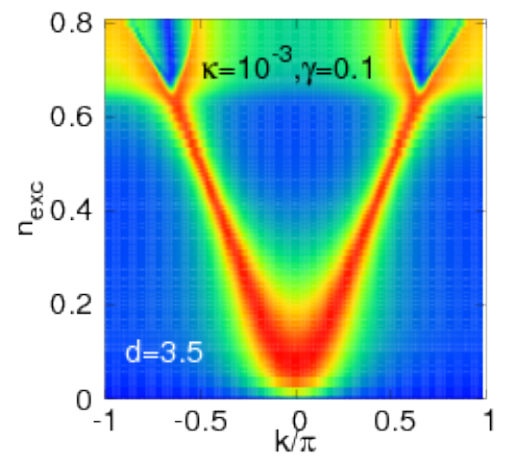}
\includegraphics[width=0.23\textwidth]{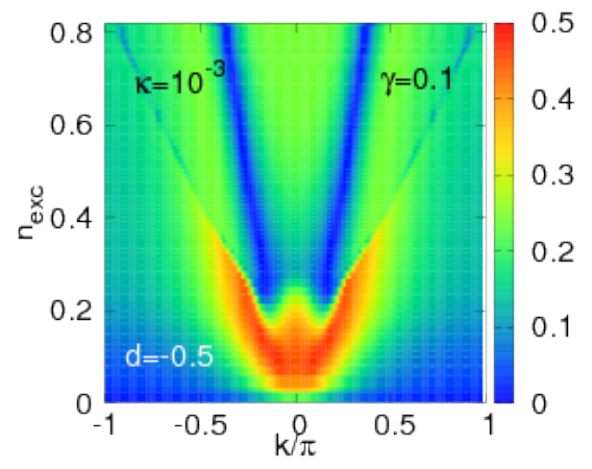}
\caption{(Color online) Electron-hole pairing amplitude $d_{\bf k}$   [see Eq.\eqref{30}], indicating exciton condensation.  Shown is an intensity plot of its real part  in the momentum-density plane for semimetallic ($d=3.5$; left panels) and semiconducting ($d=-0.5$; right panels) situations.}
\label{fig5}
\end{figure}

\begin{figure}[htp]
\includegraphics[width=0.19\textwidth]{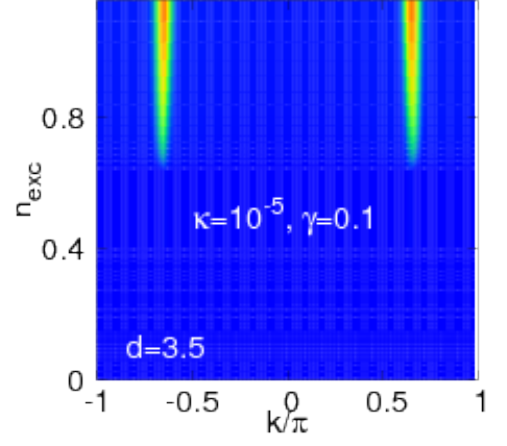}
\includegraphics[width=0.23\textwidth]{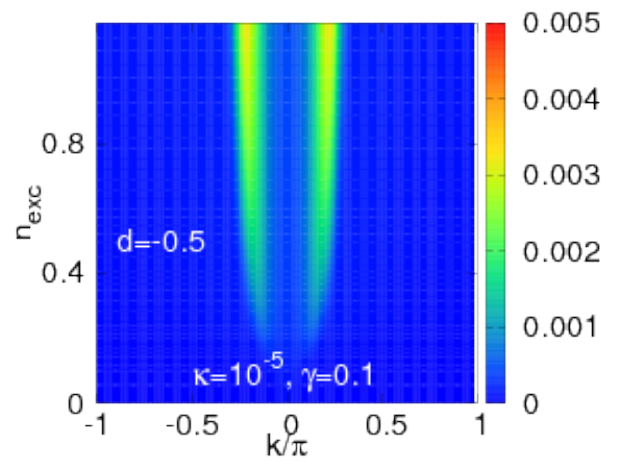}\\
\includegraphics[width=0.19\textwidth]{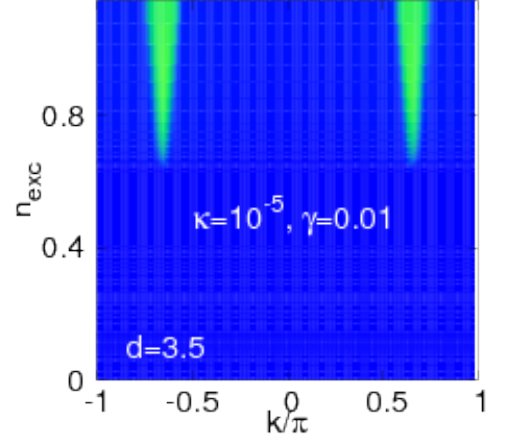}
\includegraphics[width=0.23\textwidth]{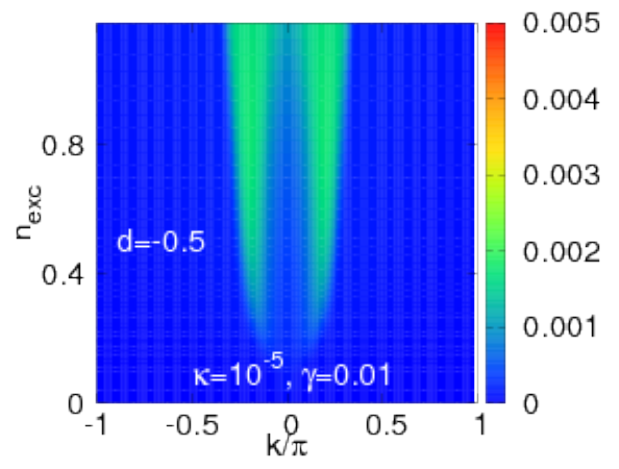}\\
\includegraphics[width=0.19\textwidth]{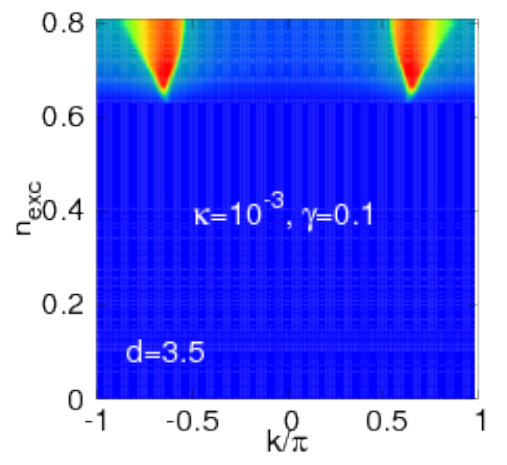}
\includegraphics[width=0.23\textwidth]{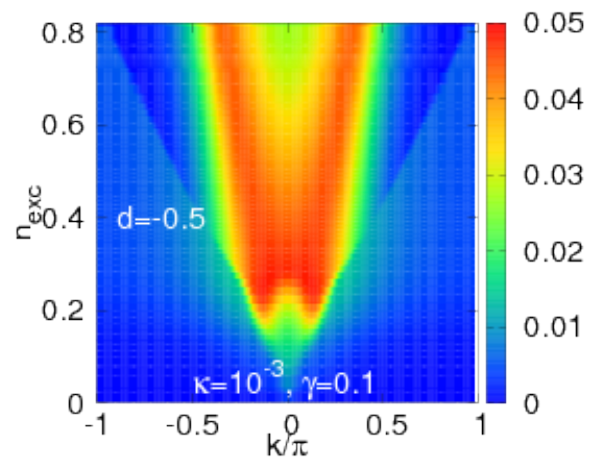}
\caption{(Color online) Exciton pairing amplitude $d_{\bf k}$. Shown is the intensity plot of its imaginary part in the momentum-density plane for $d=3.5$ (left) and $d=-0.5$ (right).}
\label{fig6}
\end{figure}

\begin{figure}[htp]
\includegraphics[width=0.21\textwidth]{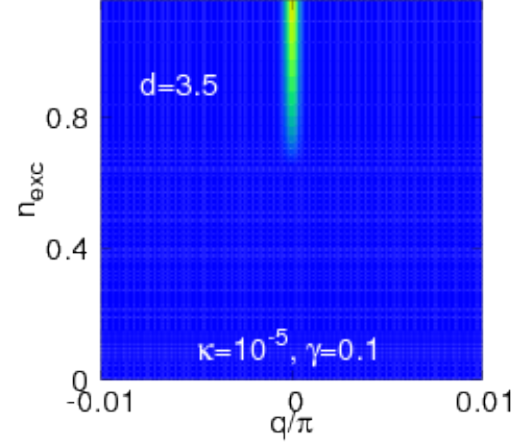}
\includegraphics[width=0.23\textwidth]{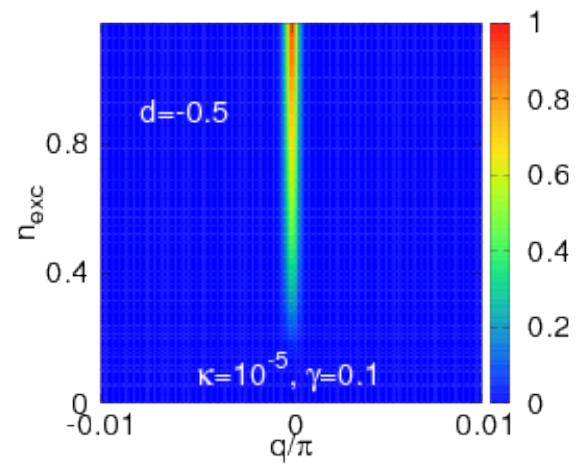}\\
\includegraphics[width=0.21\textwidth]{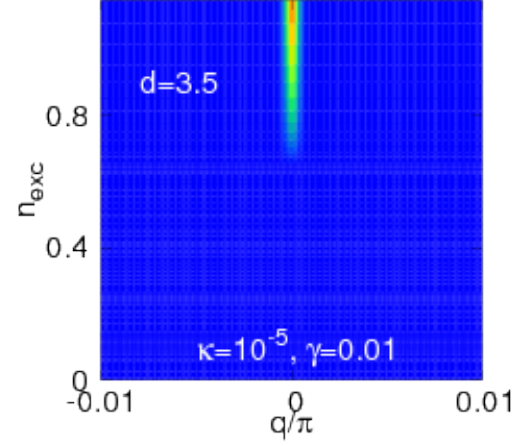}
\includegraphics[width=0.23\textwidth]{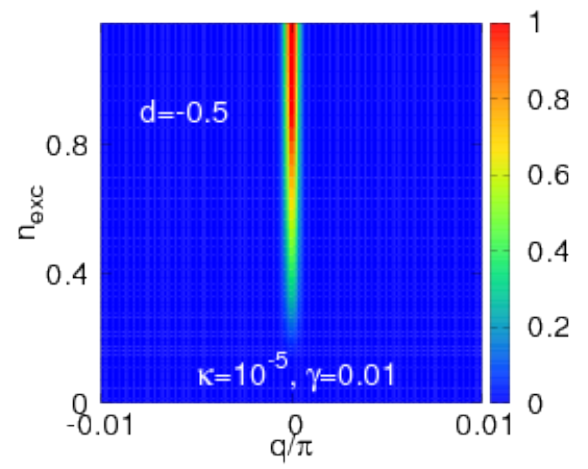}\\
\includegraphics[width=0.21\textwidth]{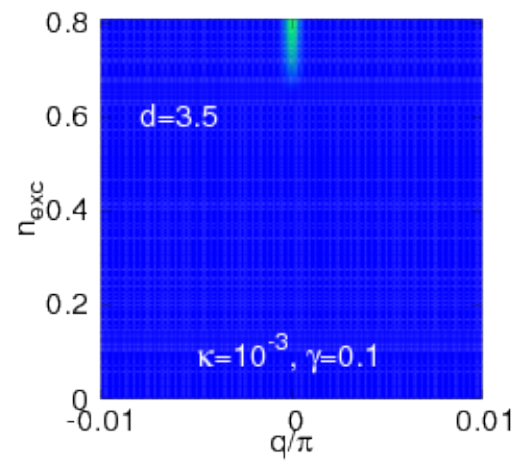}
\includegraphics[width=0.23\textwidth]{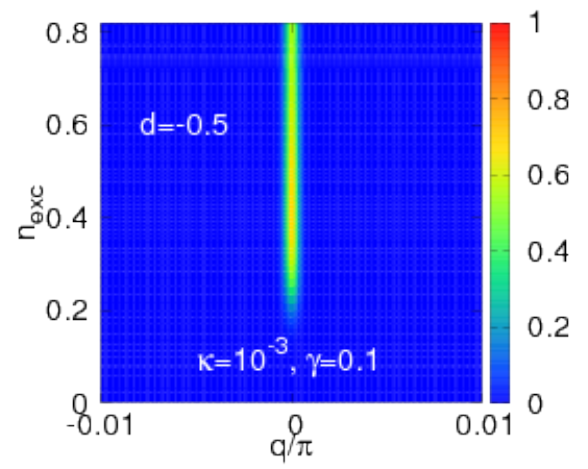}\\
\caption{ (Color online) Intensity of the photon field $\langle \psi^\dagger_{\bf q}\psi_{\bf q}\rangle$ in the momentum-density plane for detunings $d=3.5$ (left) and $d=-0.5$ (right).} 
\label{fig7}
\end{figure}

Figures \ref{fig5},~\ref{fig6}, and~\ref{fig7} show the wave-vector resolved, excitation-density dependent  intensity of the real and imaginary parts of the electron-hole pairing  amplitude $d_{\bf k}$ and the photon density expectation value $\langle \psi_{\bf q}^\dagger  \psi_{\bf q}^{}\rangle$. Not surprisingly, the results for small photon leakage $\kappa$ and relatively large coupling to the electronic baths (upper panels) are more or less the same as in equilibrium~\cite{PBF16}. In both the semimetallic (left panels) and semiconducting (right panels) regimes the amplitude for electron-hole pairing is largest at $k=0$ if $n_{exc}\to 0$. Increasing the excitation density at large detuning, the maximum is shifted to larger $k$-values in the course of exciton formation, respecting the band structure, phase space filling and Pauli blocking, until, when $\mu$ approaches $\omega_c=0.5$ near $n_{exc}\simeq 2/3$, the  photon field  severely interferes. From this moment, the real (imaginary) part of the pairing amplitude is substantially reduced (enhanced) and the photon density becomes finite. Clearly, in view of the above, this effect gets stronger the smaller $\gamma$ (see middle panels) or the larger $\kappa$  will be (see lower panels). For $n_{exc}\gtrsim 2/3$  the pairing amplitude $d_{\bf k}$ is enhanced for almost all $k$- values, with the exception of the momenta (energies) where the photons interfere. Here, the system is more or less characterized by its large photon loss in the environment, whereby the leakage strengthens at larger $\kappa$ (see lower left panels).  At even larger $n_{exc}$ one expects to enter the lasing regime~\cite{YNKOY15}.  For small detuning, exciton formation is intimately related to electron-hole excitation across the bare band gap, i.e., the coupling to the photons affects the properties of the system from the very beginning and, as a consequence, a broad maximum in $d_{\bf k}$ develops when $n_{exc}$ increases.  The strong signatures emerging in the imaginary part of $d_{\bf k}$ can be attributed to polariton formation. As a matter of course the  maximum intensity of the photon field is always at $q=0$, but the abrupt increase of the photon density changes to larger excitation densities for larger detuning. 

\subsection{Spectral properties}

\begin{figure}[t]
\includegraphics[width=0.19\textwidth]{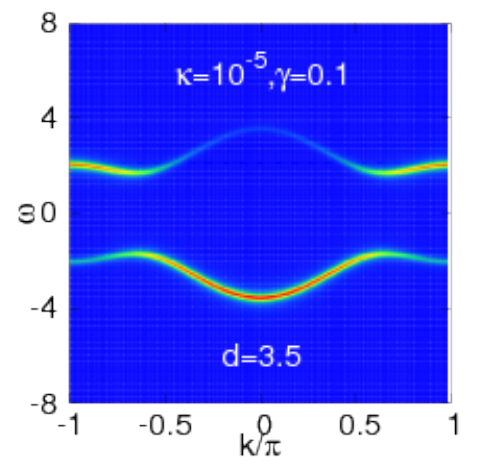}
\includegraphics[width=0.21\textwidth]{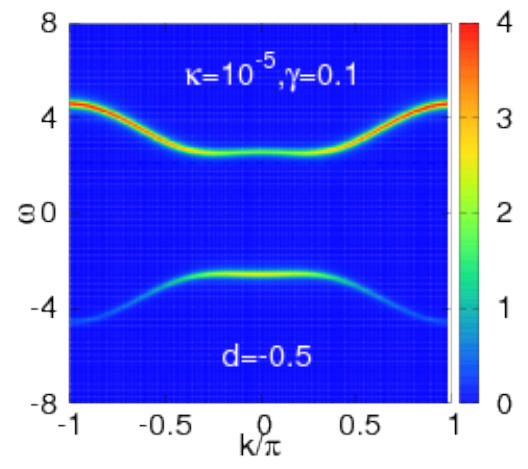}\\
\includegraphics[width=0.185\textwidth]{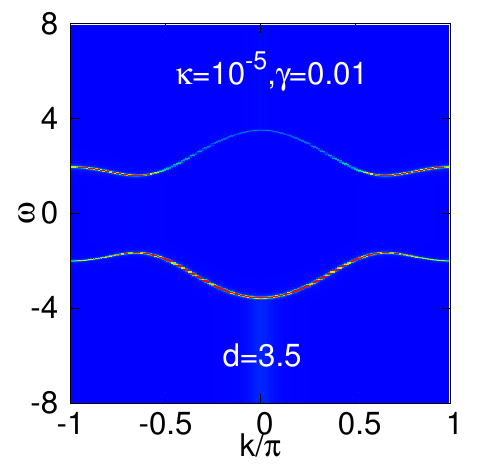}
\includegraphics[width=0.21\textwidth]{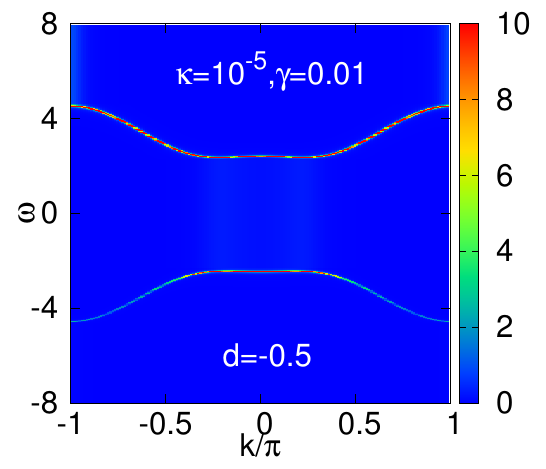}\\
\includegraphics[width=0.19\textwidth]{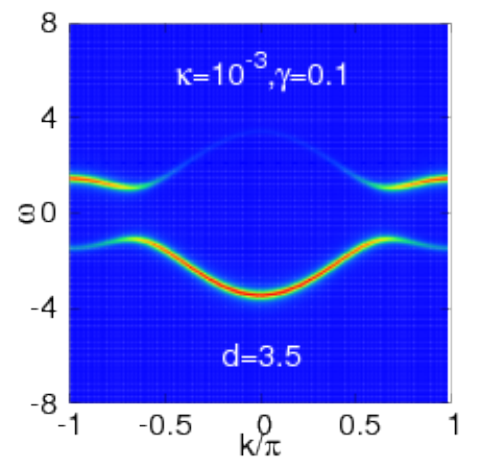}
\includegraphics[width=0.21\textwidth]{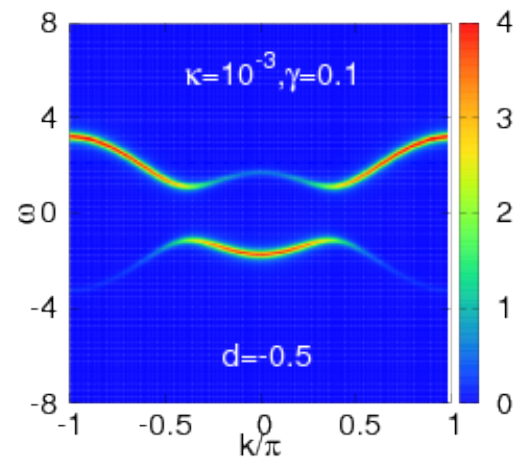}
\caption{(Color online) Electron single-particle spectrum in the steady state of the considered e-h-p microcavity system.  The quasiparticle band dispersion clearly appears in the 
intensity plot of the coherent part of the fully renormalized spectral function, $A_{e, {\rm coh}} ({\bf k},\omega)$ given by Eq.~\eqref{173}.  Results are given for typical  semimetallic  ($d=3.5$, left) and semiconducting ($d=-0.5$, right) situations. Here, the excitation density $n_{exc}=0.8$. Note that  the frequency is measured from~$\mu$.}
\label{fig8}
\end{figure}

\begin{figure}[h]
\includegraphics[width=0.20\textwidth]{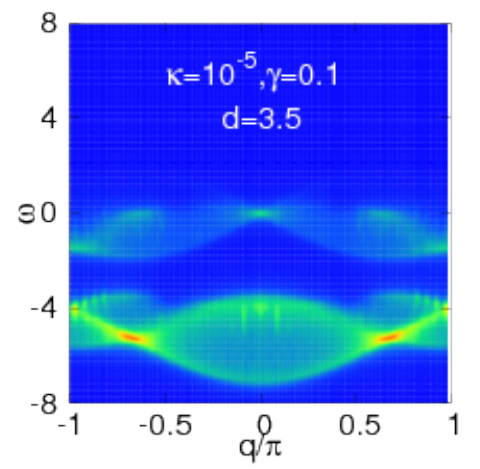}
\includegraphics[width=0.25\textwidth]{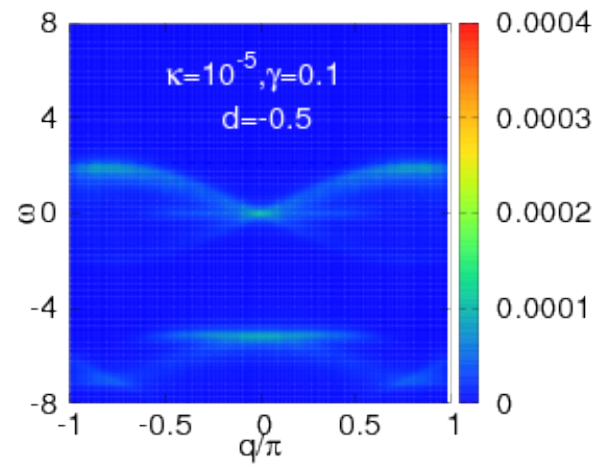}\\
\includegraphics[width=0.20\textwidth]{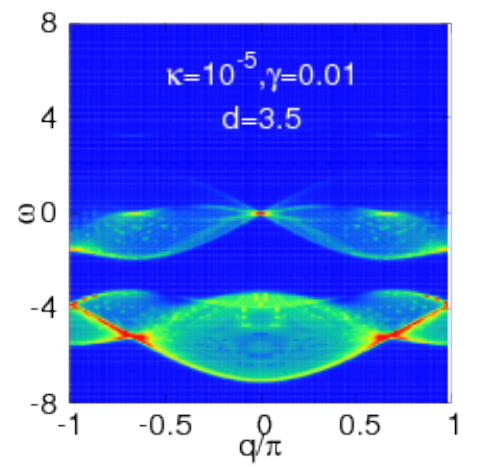}
\includegraphics[width=0.25\textwidth]{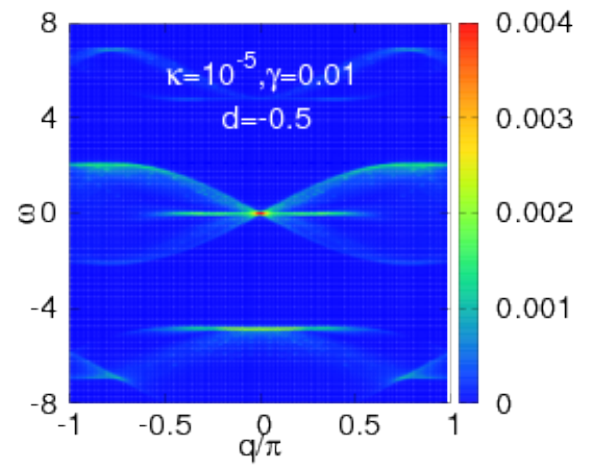}\\
\includegraphics[width=0.20\textwidth]{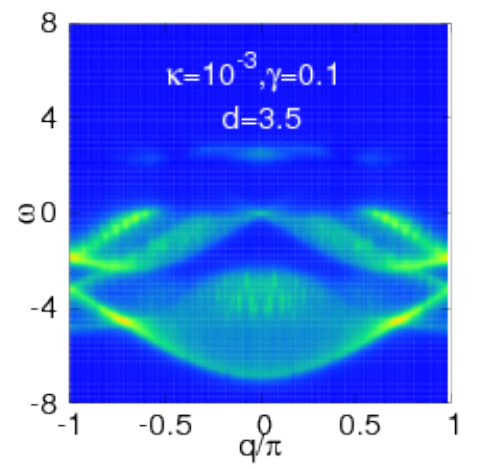}
\includegraphics[width=0.25\textwidth]{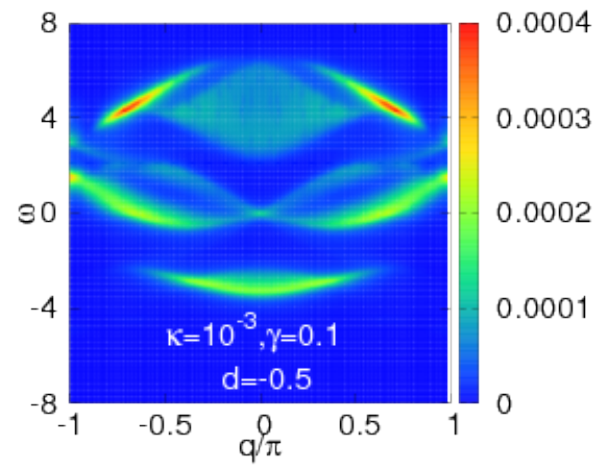}\\
\caption{(Color online) Steady-state luminescence of the e-h-p microcavity system under consideration. Shown is an intensity plot of its incoherent part, $S_{\rm inc}({\bf q},\omega)$ given by Eq.~\eqref{165}, at $n_{exc}=0.8$, for $d=3.5$ (left) and $d=-0.5$ (right).}
\label{fig9}
\end{figure}

We now consider selected spectral quantities characterizing the physical properties of the e-h-p system if it is coupled to 
electronic and photonic baths. Thereby, we first examine how the correlations and fluctuations resulting from the Coulomb and light-matter interactions will renormalize the band structure. Of course, this band structure has to be calculated in a self-consistent way for a given excitation density since the electron and hole contributions to the spectral function are interrelated in the PRM scheme.  Hereafter we consider $n_{exc}=0.8$. The quasiparticle band dispersion shows up in the coherent part of the single-particle spectrum, $A_{e,{\rm coh}} ({\bf k},\omega)$ in Fig.~\ref{fig8},  which probes both the occupied and unoccupied states as it is defined via the anticommutator (Green) function in Eq.~\eqref{169}.  As briefly mentioned already above, at large detuning the bare bands interpenetrate and the electron-hole Coulomb attraction favors the formation of a macroscopic quantum-coherent excitonic insulator state, in formal analogy to the occurrence of the BCS-type superconducting phase.  This becomes evident by looking at  the quasiparticle bands shown in Fig.~\ref{fig8} left panels): Here the (correlation induced) band gaps open at finite momenta (around $k_F$), where an almost complete backfolding of the bands is observed. At the considered large $n_{exc}$ the gap appears at the momenta where the real (imaginary) parts of the electron-hole pairing amplitude is substantially suppressed (enhanced), cf. Figs.~\ref{fig5} and~\ref{fig6}, which indicates the importance of photons in both the polartion BEC and lasing phases. A so-called "lasing gap",  where (light-induced) electron-hole pairs will be formed around the laser frequency (momentum of the kinetic hole burning), recently has been predicted theoretically~\cite{YNKOY15}, but has not been observed experimentally so far.  For small detuning the renormalized  band structure is different in nature (right panels). In principle, the  quasiparticle bands retain their (bare) semiconductor-like arrangement, but the particle-photon coupling causes a noticeable flattening (plateau structure) of the conduction band bottom and valence band top, thereby enlarging the single-particle spectral gap. The plateau structure can be attributed to a  polariton (photonic) BEC. For both detunings, a smaller value of $\gamma$ will reduce the spreading of the coherent signal while it enhances its intensity (see middle panels). A larger value of $\kappa$, keeping  $\gamma$ at fixed, will reduce the magnitude of the gap in the renormalized band structure.  In this case the leakage to the external photon vacuum is enlarged, leading to a weakening of the excitonic  order parameter $\tilde \Delta_{\vec k}$ and thus to a weakening of the quasiparticle band gap. 

Finally, we will look at the steady-sate luminescence of the e-h-p system, considering the same parameters as for the single-particle spectra. Clearly the coherent part of the luminescence spectrum is the dominant one, however, $S_{\rm coh}(\vec q, \omega)$  has neither a nontrivial ${\bf q}$- nor a nontrivial  $\omega$-dependence [see Eq.~\eqref{147}].  Therefore, in Fig.~\ref{fig9}, we only display the behavior of the incoherent part  of the luminescence, $S_{\rm inc}(\vec q, \omega)$, which is characterized by particle-hole excitations according to Eq.~\eqref{149}.  The results shown in Fig.~\ref{fig9} include all possible annihilation and creation processes of electron-hole pairs inside and in-between the fully renormalized quasiparticle bands $\tilde E_{(1,2) \vec k}$ without any additional photons involved. From Eq.~\eqref{162}--\eqref{162c} it is evident that the interband contributions between the two bands between the two bands $\tilde E_{1 \vec k}$ and $\tilde E_{2 \vec k}$, caused  by terms proportional to $|\xi_{\vec k+ \vec q}|^2 |\xi_{\vec k}|^2$  in the prefactor of the next to last term, are the dominant ones.  Special attention deserves the significant flattening of the excitonic response at small momentum transfer for small detuning and $\kappa=10^{-5}$, which is due to a strong light-matter interaction and indicates the formation of an exciton-polariton condensate. 

\section{Summary and Conclusions}
\label{VIII}
The projector-based renormalization method (PRM) is a reliable and powerful analytical technique that has already been successfully applied to a wide range of {\it equilibrium} solid-state physics problems in the past; examples are magnetism, superconductivity, charge density wave formation, phonon-softening, or valence and metal-insulator transitions. The main purpose of this work was to provide a consistent extension of the PRM  for dealing with more general {\it non-equilibrium} situations in open systems, as they appear when quantum systems are coupled to external reservoirs. A prime example for this is the light-matter coupling in semiconductor microcavities, where electrons and holes---for example, after being excited with light---can form excitonic bound states due to their Coulomb interaction, or can recombine into photons, when cavity photons can escape into the vacuum (e.g., because of mirrors with imperfect reflectivity). Furthermore, in such systems coherent quantum condensates may arise, realized as BCS or BEC equilibrium states, but  also manifest nonequilibrium (lasing-like)  phases. The PRM framework, we developed can treat, if combined with the Mori-Zwanzig projection technique, these equilibrium and nonequilibrium situations in a rather unified way. The steady-state properties of the system are thereby obtained from the long-term behavior of appropriate expectation values and, equally important, the many-body correlations and fluctuations processes are taken into account beyond mean-field in the whole range of excitation densities. Besides expectation values also spectral properties can be evaluated in the steady state. From a theoretical point of view, this ensures diverse future application possibilities of the proposed approach.

Other examples, where the newly developed PRM approach might be applied, coming from the very topical and promising field of ultracold atomic physics~\cite{RDBE13}. In these systems  particles (atoms, molecules) or even BECs are loaded into optical lattices created by dynamic cavity fields and are studied  in connection with different ordering phenomena, quantum phase transitions, superradiance phase transitions, driving and dissipation~\cite{BAF13,MPR17,LHDLMDE18,LDKHDE18,KGVKL18,VGKBKKL18}. Here the particles in a quantum many-body correlated phase of matter strongly influence the properties of light and vice versa, whereby the tunable interplay between rather short-ranged direct particle-particle interaction and long-range interaction mediated by the coupling to the optical cavity mode is of particular importance. Then in particular the quantum properties of light scattered from the emergent structured cold-atom phases will require a non-equilibrium or at least steady-state description~\cite{BM15}. The Hamiltonians being normally discussed in this context are extended Bose-Hubbard-type models supplemented by an atom-field interaction part or Dicke-type models, and take into account dissipation due to photon leakage (coupling to reservoirs).  The  proposed PRM, adjusted correspondingly, is definitely suitable for treating such models. 

In this paper we considered a rather generic open model system consisting of interacting electrons, holes and cavity photons and their corresponding reservoirs. The  focus was on exciton and polariton formation, and their possible condensation in the course of spontaneous breaking demonstrated by nonvanishing excitonic and photonic order parameters. In the steady state,  the nature of the condensate changes from an exciton to a polariton and finally to a photon dominated ground-state when the density of excitations increases. Thereby a finite expectation value of the photonic field operator is intrinsically connected with a finite imaginary part of the excitonic order parameter function and, from a physical perspective, with photon loss. Having assumed an electron/hole band symmetric case and charge neutrality, the difference between the self-consistently determined quantity $\mu$ (which takes over the role of a true chemical potential of the system in thermal equilibrium only) and the sum of the chemical potentials of the electron and hole baths $\mu_B$ can be used in order to quantify nonequilibrium effects. For small-to-intermediate excitation densities and large detuning (semimetallic situation) Fermi-surface and Pauli- blocking effects are important and the condensate is reminiscent of the BCS-type excitonic insulator phase, whereas for small detuning (semiconducting situation) the condensate typifies a Bose-Einstein condensate of preformed electron-hole pairs (Excitons). Note that if we would have increased the Coulomb interaction at fixed excitation density we could realize a BCS-BEC crossover in the excitonic condensate due to the growing Hartree shift  between valence and conduction bands. In any case the fully renormalized band dispersions were obtained from the coherent part of the single-particle spectral function and show the opening of the band gaps and significant differences between large and small detuning situations, such as a strong band backfolding and a pronounced band flattening of the valence (conduction) band top (bottom) in the former and latter case, respectively.   As soon as we enter the regime where the photons and therefore nonequilibrium effects play an important role, our results will noticeably depend on the parameters $\gamma$ and $\kappa$ parametrizing respectively the couplings to the electron/hole and photon reservoirs. In this context we have shown that the present steady-state approach cannot be reduced to the case of a closed electron-hole-photon system simply by setting $\gamma$ and $\kappa$ to zero;  instead one gets a description of  thermal equilibrium in the limit of large $\gamma$. On the other hand, the photon leakage/loss strengthens at larger values of $\kappa$. 

It might make sense to emphasize once more the key findings of the nonequilibrium effects of the steady state and to compare our results with previous results for 
the thermal equilibrium situation from Ref.~\cite{PBF16}:
(i) At small-to-moderate  excitation densities $n_{exc}$ the e-h-p subsystem is close to  thermal equilibrium. In particular, for the largest used value 
$\gamma=1$ of the coupling of cavity electrons and holes to their respective baths, the results of Figs.~\ref{fig1}--\ref{fig4} agree very well with those from    
Ref.~\cite{PBF16}. But also for smaller $\gamma$ (and the largest detuning value $d=3.5$) the e-h-p subsystem and the electronic baths stay in a common (quasi-) equilibrium.  Note that  the linear slope 
 with  $\mu = \mu_B$  in Fig.~\ref{fig1}  stands alone for an increase of electrons and holes, whereas the following flattening of $\mu(\mu_B)$ is a 
 Pauli-blocking effect after all quasiparticle states are already occupied by electrons and holes. Any further increase of $n_{exc}$ or $\mu_B$ is solely governed by an
 increase of cavity photons.  
(ii) Increasing $n_{exc}$ further the number of cavity photons increases.  Thereby, for a sufficient large number of cavity photons 
they become affected by their coupling $\kappa$ to the external free photons, which leads to  a loss  of cavity photons. 
This loss is intrinsically connected to the appearance of 
finite imaginary parts of the excitonic order parameters  in Fig.~\ref{fig3} or Fig.~\ref{fig6}, in particular for larger values of $\kappa$. 
For the case of small detuning $d=-0.5$ in Fig. \ref{fig3} or Fig.~\ref{fig6} this effect is more pronounced already at 
small $n_{exc}$ since in the  semiconducting case photons are also present already at smaller excitation densities $n_{exc}$. 
The coupling of  cavity photons to free space photons affects the properties of the system from the very beginning. 
To summarize, nonlinear effects become important whenever a sufficiently large number of cavity photons is present.  Thereby, the
coupling $\kappa$ to external photons  plays an important role but also the detuning of the system and less important 
 the value of the coupling $\gamma$. 
 
The limitations of the present theoretical approach are: (i) The  initial density matrix $\rho_0$ was assumed to be factorizable into a part $\rho_{\rm S}$ for the subsystem $\mathcal H_{\rm S}$ and into a reservoir density $\rho_{\rm R}$, $\rho_0= \rho_{\rm S} \rho_{R}$. Thereby $\rho_{\rm S}$  was assumed to describe thermal equilibrium for $\mathcal H_{\rm S}$, and $\rho_{\rm R}$ should be infinitely large so that it is not changed by renormalization effects. (ii) The interactions $\mathcal H_1$ of the subsystem $\mathcal H_{\rm S}$ was assumed to be `small' and was treated in the renormalization equations in perturbation theory. The renormalization  was only done in small  
steps $\Delta \lambda$, so that extreme high renormalization processes are taken into account in the fully renormalized quantities. 
Therefore, the renormalization method is usually valid for parameters of $\mathcal H_1$ which are of the same order as those of $\mathcal H_0$, i.e.~far beyond usual perturbation theory. 
(iii) Finally, the influence of the reservoirs were taken into account in perturbation theory up to second order in the interaction $\mathcal H_{\rm SR}$  between the
subsystem $\mathcal H_{\rm S}$ and the reservoirs $\mathcal H_{\rm R}$.

Although we exclusively focused on the exciton-polariton problem in this contribution, the extended PRM,  bridging equilibrium and steady state descriptions,  can be used to tackle other strongly open/driven quantum model systems with strong correlations,   which opens opens a new avenue for exploring many-body effects in non-equilibrium situations, i.e., for ultracold atoms in coupled to radiation fields. Work along this line is in progress.
\acknowledgements
The authors would like to thank D. Pagel, D. Semkat, and B. Zenker for valuable discussions. V.-N. P. was funded by Vietnam National Foundation for Science and Technology Development (NAFOSTED) under Grant No.103.01-2017.68. HF is grateful to the Los Alamos National Laboratory for hospitality and support.

\newpage

 \begin{appendix}

\section{Renormalization of $\mathcal H_{\lambda}$}
\label{A}
The renormalization equations for the $\lambda$-dependent parameters of $\mathcal H_\lambda$ will be derived  
from Eq.~\eqref{38} which transforms $\mathcal H_\lambda$ to $\mathcal H_{\lambda- \Delta \lambda}$.  
For sufficiently small renormalization steps $\Delta \lambda$, when the expansion 
in $g$ and $U$ can be limited to ${\cal O}(g^2)$ and ${\cal O}(U^2)$, we have
\begin{eqnarray}
  \label{A1} 
\mathcal H_{\rm S, \lambda -\Delta \lambda} &=&   \mathcal H_{0,\lambda}  +  \mathcal H_{c, \lambda} 
+  \mathcal H_{1,\lambda} \nonumber \\
&+&   [X_{\lambda, \Delta \lambda}, \mathcal H_{0,\lambda} +
{\mathcal H}_{c, \lambda} + \mathcal H_{1, \lambda}]   \nonumber \\
&+&   \cdots \, ,
\end{eqnarray}
where the representation \eqref{52a} for $\mathcal H_{\rm S, \lambda}$ has been used.
Renormalization contributions 
arise from the three commutators on the right hand side 
 which must be evaluated explicitly.  Contributions of order $\mathcal O (X_{\lambda, \Delta \lambda}^2$)  and higher
will be neglected. From the first commutator 
$ [X_{\lambda, \Delta \lambda}, \mathcal H_{0,\lambda}]$ one finds  
renormalization contributions  to $\Delta_{\vec k, \lambda}$ and $\Gamma_\lambda$. They read 
 according to Sec.~\ref{III.B}:  
\begin{eqnarray}
\label{A2}
 \delta  {\Delta}_{\vec k, \lambda}^{(0)} &=&   - \frac{g}{\sqrt N} A_{\vec k 0}(\lambda, \Delta \lambda) \,  
\omega_{0,\lambda} \langle \psi_{0}\rangle  \\
&&  - \frac{U}{N} \sum_{\vec k_1} B_{\vec k_1 \vec k, -\vec k_1, -\vec k }(\lambda, \Delta \lambda) 
(\varepsilon^e_{\vec k_1, \lambda} + 
\varepsilon^h_{-\vec k_1, \lambda}) \, d_{\vec k_1}\,, \nonumber\\
\label{A3}
\delta {\Gamma}_{\lambda}^{(0)}  &=&  \frac{g}{N} \sum_{\vec k} A_{\vec k 0}(\lambda, \Delta \lambda) 
(\varepsilon^e_{\vec k,\lambda} + \varepsilon^h_{-\vec k, \lambda}) \,  d_{\vec k} \, , 
\end{eqnarray}
where we have used expressions \eqref{57}-\eqref{59} for the generator $X_{\lambda, \Delta \lambda}$
\begin{eqnarray}
\label{A3a}
X_{\lambda, \Delta \lambda} &=& X^{g}_{\lambda,\Delta\lambda} + X^{U}_{\lambda,\Delta\lambda}
= - X_{\lambda, \Delta \lambda}^\dag
\end{eqnarray}
with
\begin{eqnarray}
\label{A3b}
 X^{g}_{\lambda,\Delta\lambda} &=&-\frac{g}{\sqrt{N}}\sum_{\vec{k}\vec{q}}
A_{{\bf k{\bf q}}}(\lambda,\Delta\lambda)
\big[:e_{\mathbf{k+q}}^{\dagger}h_{-\mathbf{k}}^{\dagger}\psi_{\mathbf{q}}:  -\textrm{H.c.}\big]\,,\nonumber\\
 && \\
\label{A3c}
 X^{U}_{\lambda,\Delta\lambda} &=&  -\frac{U}{N}
 \sum_{\vec{k}_{1}\vec{k}_{2}\vec{k}_{3}} B_{\vec k_1 \vec k_2; \vec k_3,\vec k_1 + \vec k_3 -\vec k_2 }(\lambda,\Delta\lambda) \nonumber \\
 && \times  :e_{\vec{k}_{1}}^{\dagger}e_{\vec{k}_{2}} \,  h_{\vec{k}_{3}}^{\dagger}h_{\vec{k}_{1}+\vec{k}_{3}-\vec{k}_{2}}:  \, .
\end{eqnarray}
Note that  both parts $X^g_{\lambda, \Delta \lambda}$ and $X^U_{\lambda, \Delta \lambda}$  
contribute to $\delta  {\Delta}_{\vec k, \lambda}^{(0)}$,
whereas  only  $X^g_{\lambda, \Delta \lambda}$
contributes to $ \delta {\Gamma}_{\lambda}^{(0)}$. For the second 
commutator $ [X_{\lambda, \Delta \lambda}, \mathcal H_{c,\lambda}]$ 
 one finds:
 \begin{align}
\label{A4}
& [X^{g}_{\lambda, \Delta \lambda}, {\mathcal H}_{{c}, \lambda}] = \frac{g}{\sqrt N} \sum_{\vec k} \big( A_{\vec k 0}(\lambda, \Delta \lambda)
{\Delta}_{\vec k, \lambda}\nonumber \\
&\hspace*{3cm}\times ( 1- n^e_{\vec k} - n^h_{-\vec k}) \, \psi^\dag_0  + {\rm H.c.} \big) \nonumber \\
&\hspace*{2cm}+ \frac{g}{\sqrt N} \sum_{\vec k} \big( A_{\vec k 0}(\lambda, \Delta \lambda) {\Delta}_{\vec k, \lambda}
\langle \psi^\dag_0 \rangle \nonumber \\
&\hspace*{3cm}\times ( 1- e^\dag_{\vec k} e_{\vec k} 
- h^\dag_{-\vec k} h_{-\vec k})  + {\rm H.c.} \big) \nonumber \\
&\hspace*{2cm} - g {\Gamma}_\lambda \sum_{\vec k} \big( A_{\vec k0}(\lambda, \Delta \lambda) \, 
e^\dag_{\vec k} h^\dag_{-\vec k} + {\rm H.c.} \big)\,,
\end{align}
\begin{align}
\label{A5}
[X^{U}_{\lambda, \Delta \lambda}, {\mathcal H}_{{c}, \lambda}] &=   - \frac{U}{N} \sum_{\vec k \vec k_1} \Big[
B_{\vec k_1,  \vec k, -\vec k_1, -\vec k}(\lambda, \Delta \lambda)\, { \Delta}_{\vec k, \lambda} 
\nonumber \\
& \quad\quad\quad \times \big(  d_{\vec k_1}^*(1- e^\dag_{\vec k} e_{\vec k} - h^\dag_{-\vec k} h_{-\vec k}) \,  
\nonumber \\
& \qquad + (1- n^e_{\vec k} - n^h_{-\vec k}) \,  e^\dag_{\vec k_1} h^\dag_{-\vec k_1} \big)
 + {\rm H.c.}   
\Big]  
  \end{align} 
(omitting irrelevant constants), where again only renormalization contributions of the  operator structure of $\mathcal H_{{\rm S}, \lambda}$
are retained.  We point out that the contributions  in Eqs.~\eqref{A4} and \eqref{A5} which renormalize
$\varepsilon^e_{\vec k}$  and $\varepsilon^h_{-\vec k}$ 
are of second order in the order parameters; they should be small and will be neglected. 
 The remaining contributions renormalize $\Gamma_\lambda$ and $\Delta_{\vec k, \lambda}$:
 \begin{eqnarray}
 \label{A6a}
\delta\Gamma_\lambda^{(c)} &=& \frac{g}{N} \sum_{\vec k} 
 A_{\vec k 0}(\lambda, \Delta \lambda) {\Delta}_{\vec k, \lambda}
( 1- n^e_{\vec k} - n^h_{-\vec k}), \\
 \label{A6ab}
\delta \Delta_{\vec k, \lambda}^{(c)} &=&  - g 
{\Gamma}_\lambda  A_{\vec k0}(\lambda, \Delta \lambda) \ \nonumber \\
&&-  \frac{U}{N} \sum_{ \vec k_1} 
B_{\vec k,  \vec k_1, -\vec k, -\vec k_1}(\lambda, \Delta \lambda)\, { \Delta}_{\vec k_1, \lambda} 
\nonumber \\
&&\;\times  
(1- n^e_{\vec k_1} - n^h_{-\vec k_1}) \, .
 \end{eqnarray}
 Next we look at the last commutator   
$ [X_{\lambda, \Delta \lambda}, \mathcal H_{{\rm 1}, \lambda}]$ in  Eq.~\eqref{A1}. 
Neglecting off-diagonal commutators we first obtain,
  \begin{align}
 \label{A7}
& [X^{g}_{\lambda, \Delta \lambda}, \mathcal H_{{g}, \lambda}] = \frac{2 g^2}{N} \sum_{\vec k \vec q} A_{\vec k - \vec q, \vec q}(\lambda, \Delta \lambda)
(n^\psi_{\vec q} + n^h_{\vec q -\vec k})\,  e^\dag_{\vec k} e_{\vec k} \nonumber \\
&\hspace*{2cm} + \frac{2 g^2}{N} \sum_{\vec k \vec q} A_{\vec k \vec q}(\lambda, \Delta \lambda)
(n^\psi_{\vec q} + n^e_{\vec q +\vec k})\,  h^\dag_{-\vec k} h_{-\vec k} \nonumber \\
&\hspace*{2cm}  - \frac{2 g^2}{N} \sum_{\vec k \vec q} A_{\vec k \vec q}(\lambda, \Delta \lambda)
(1- n^e_{\vec k + \vec q} -n^h_{-\vec k}) \nonumber \\
&\hspace*{2cm} \times \big( \psi^\dag_{\vec q} \psi_{\vec q}  
- \delta_{\vec q,0}(\langle \psi^\dag_{0} \rangle  \psi_{0} - 
\langle\psi_{0}\rangle \psi^\dag_{0}  ) \big) \, ,
 \end{align}
where we have introduced the following expectation value for the photon fluctuations: 
\begin{equation}
\label{A8}
 n^\psi_{\vec q} = \langle \psi^\dag_{\vec q} \psi_{\vec q}\rangle - \delta_{\vec q,0}  \langle \psi^\dag_{0} \rangle 
 \langle\psi_{0}\rangle \, .
 \end{equation}
Equation~\eqref{A7} leads to renormalization contributions of $\varepsilon^e_{\vec k, \lambda}$, 
 $\varepsilon^h_{-\vec k, \lambda}$, $\omega_{\vec q, \lambda}$, and $\Gamma_{\lambda}$:
 \begin{align}
 \label{A9a}
& \delta \varepsilon_{\vec k,\lambda}^{e(g)} =  \frac{2 g^2}{N} \sum_{\vec q} A_{\vec k - \vec q, \vec q}(\lambda, \Delta \lambda)
(n^\psi_{\vec q} + n^h_{\vec q -\vec k})\,,  \\
\label{A9b}& \delta \varepsilon_{\vec k_\lambda}^{h(g)} =  \frac{2 g^2}{N} \sum_{\vec q} A_{\vec k \vec q}(\lambda, \Delta \lambda)
(n^\psi_{\vec q} + n^e_{\vec q +\vec k})\,,  \\
\label{A9c}&  \delta \omega_{\vec q,\lambda}^{(g)} =  - \frac{2 g^2}{N} \sum_{\vec k} A_{\vec k \vec q}(\lambda, \Delta \lambda)
(1- n^e_{\vec k + \vec q} -n^h_{-\vec k})\,,  \\
\label{A9d}& \delta \Gamma^{(g)}_\lambda = \frac{2 g^2}{N \sqrt N} \sum_{\vec k} A_{\vec k 0}(\lambda, \Delta \lambda)
(1- n^e_{\vec k} -n^h_{-\vec k})  \langle\psi_{0}\rangle\,.
  \end{align}
  The evaluation of the second commutator $ [X^{U}_{\lambda, \Delta \lambda}, \mathcal H_{{U}, \lambda}]$ 
 to $[X_{\lambda, \Delta \lambda}, \mathcal H_{1, \lambda}]$ is more evolved. Our starting point is 
  \begin{align}
  \label{A10}
     &  [X^{U}_{\lambda, \Delta \lambda}, \mathcal H_{{U}, \lambda}] =   \frac{U^2}{N^2}  \sum_{\vec k_1 \vec k_2\vec q \atop \vec k_1' \vec  k_2' \vec q'}
 \Gamma^{\vec k_1 ,\vec k_1 - \vec q; \vec k_2, \vec k_2 +\vec q }_{\vec k_1', \vec k_1' - \vec q'; 
  \vec k_2', \vec k_2' + \vec q'}
 (\lambda, \Delta \lambda) \nonumber \\
 & \times \big[ :e^\dag_{\vec k_1} e_{\vec k_1 -\vec q} h^\dag_{\vec k_2} h_{\vec k_2 + \vec q}: ,
 h^\dag_{\vec k_2' + \vec q'} h_{\vec k_2'} e^\dag_{\vec k_1'- \vec q'} e_{\vec k_1'}: \big ] \, ,
  \end{align}
 where we have introduced
\begin{align}
\label{A11}
 &\Gamma_{\vec k'_{1}, \vec k'_{1} -\vec q', \vec k'_{2}, \vec k_2 + \vec q'}
 ^{\vec{k}_{1}, \vec{k}_{1} -\vec q, \vec{k}_{2}, \vec k_2 + \vec q}(\lambda,\Delta\lambda)   \\
&= \frac{1}{2} \big[
  B_{\vec k'_{1},\vec k'_{1} -\vec q', \vec k'_{2}, \vec k'_2 + \vec q'}(\lambda,\Delta\lambda) \,
   \Theta_{\vec{k}_{1},\vec{k}_{1} -\vec q, \vec{k}_{2}, \vec k_2 + \vec q,\lambda} \nonumber\\
&+ B_{\vec k_{1},\vec k_{1} -\vec q, \vec k_{2}, \vec k_2 + \vec q}(\lambda,\Delta\lambda) \, \Theta_{\vec{k}'_{1},\vec{k}'_{1} - \vec q', \vec{k}'_{2}, \vec k'_2 + \vec q',\lambda} 
 \big]\,. 
\end{align} 
We first extract the part 
  of the commutator~\eqref{A10} that renormalizes the electronic one-particle energies. It reads 
 \begin{eqnarray}
 \label{A12}
&& \frac{U^2}{N^2}  \sum_{\vec k_1 \vec k_2\vec q} \sum_{\vec k_1' \vec  k_2' \vec q'}
 \Gamma^{\vec k_1 ,\vec k_1 - \vec q; \vec k_2, \vec k_2 +\vec q }_{\vec k_1', \vec k_1' - \vec q'; 
  \vec k_2', \vec k_2' + \vec q'}
 (\lambda, \Delta \lambda) \nonumber \\
&&\  \times \Big[ \delta_{\vec k_2 + \vec q, \vec k_2' +\vec q} 
:e^\dag_{\vec k_1} e_{\vec k_1 -\vec q}: \, 
  \, h^\dag_{\vec k_2 } h_{\vec k_2'} \,
 :e^\dag_{\vec k'_1 - \vec q'} e_{\vec k'_1}:  \nonumber \\
 &&  \ - \delta_{\vec k_2, \vec k_2'}
 :e^\dag_{\vec k_1} e_{\vec k_1 -\vec q}: \, 
  \, h^\dag_{\vec k_2'+ \vec q' } h_{\vec k_2 + \vec q} \,
 :e^\dag_{\vec k'_1 - \vec q'} e_{\vec k'_1}: \nonumber \\
  &&\ + \delta_{\vec k_1- \vec q, \vec k_1' - \vec q'}
  : h^\dag_{\vec k_2' + \vec q'}h_{\vec  k_2'}:\ e^\dag_{\vec k_1} e_{\vec k_1'} \  : h^\dag_{\vec k_2}
  h_{\vec  k_2 + \vec q}: \nonumber \\
&& \ -  \delta_{\vec k_1, \vec k_1'}
  : h^\dag_{\vec k_2' + \vec q'}h_{\vec  k_2'}:\ e^\dag_{\vec k_1'- \vec q'} e_{\vec k_1 -\vec q}\  
  : h^\dag_{\vec k_2}
  h_{\vec  k_2 + \vec q}: 
\Big]
 \nonumber\,.
 \end{eqnarray}
 From this,  by truncation, we extract those contributions which are proportional to 
 $e^\dag_{\vec k} e_{\vec k} $ or $h^\dag_{-\vec k} h_{-\vec k}$ and  arrive at the renormalization 
  contributions to $\varepsilon^e_{\vec k}$ and $\varepsilon^h_{\vec k}$:
  \begin{eqnarray}
  \label{A13}
 &&  \delta \varepsilon_{\vec k,\lambda}^{e(U)} = \frac{U^2}{N^2} \sum_{\vec k_2 \vec q} 
 \Big(B_{\vec k, \vec k -\vec q; \vec k_2, \vec k_2 +\vec q}(\lambda, \Delta \lambda) \nonumber\\
 && \quad\times \big[
 (n^h_{\vec k_2} - n^h_{\vec k_2 +\vec q}) (1- n^e_{\vec k -\vec q})
 + n^h_{\vec k + \vec q} (1- n^h_{\vec k_2}) \nonumber \\[0.2cm]
 && \quad - B_{\vec k + \vec q, \vec k; \vec k_2, \vec k_2 +\vec q}(\lambda, \Delta \lambda) \nonumber \\
 && \quad  \times \big[
 (n^h_{\vec k_2} - n^h_{\vec k_2 +\vec q}) \, n^e_{\vec k +\vec q}
 + n^h_{\vec k + \vec q} (1- n^h_{\vec k_2}) \big]  \Big),  
 \end{eqnarray}
  \begin{eqnarray}
  \label{A14}
 &&  \delta \varepsilon_{\vec k,\lambda}^{h(U)} = \frac{U^2}{N^2} \sum_{\vec k_1 \vec q} 
 \Big(B_{\vec k_1, \vec k_1 -\vec q; \vec k- \vec q, \vec k}(\lambda, \Delta \lambda) \nonumber \\
 && \quad  \times \big[
 (n^e_{\vec k_1} - n^e_{\vec k_1 - \vec q}) (1- n^h_{\vec k -\vec q})
 - n^e_{\vec k_1} (1- n^e_{\vec k_1 -\vec q}) \nonumber \\[0.2cm]
 && \quad - B_{\vec k_1, \vec k_1 -\vec q; \vec k, \vec k +\vec q}(\lambda, \Delta \lambda) \nonumber \\
 && \quad  \times \big[
 (n^e_{\vec k_1} - n^e_{\vec k_1 -\vec q}) \, n^h_{\vec k +\vec q}
 - n^e_{\vec k_1} (1- n^e_{\vec k_1 -\vec q}) \big]  \Big). 
  \end{eqnarray}
In the same way, the  renormalization contributions to $\Delta_{\vec k, \lambda}$ can be extracted from \eqref{A10}. 
Again, by truncation, we collect the parts being proportional to 
 $e^\dag_{\vec k} h^\dag_{-\vec k}$ or  $h_{-\vec k} e_{\vec k}$. 
 Since $X^{U}(\lambda, \Delta \lambda)$ 
 and $\mathcal H_{{U},\lambda}$ are time-ordered expressions, a truncation within 
$X^{U}(\lambda, \Delta \lambda)$ and within $\mathcal H_{{U},\lambda}$ is thereby forbidden. 
One finds 
   \begin{align}
  \label{A15}
& \delta \Delta_{\vec k, \lambda}^{(U)}  = - \frac{U^2}{N^2} \sum_{\vec k_1 \vec q} \Big( \Gamma^{\vec k_{1} + \vec q, \vec k_{1}; -\vec k, -\vec k + \vec q}
 _{\vec{k}_{1} + \vec q, \vec{k}; -\vec{k}_1, -\vec k + \vec q}(\lambda,\Delta\lambda)
  \nonumber \\
   & \hspace*{3.5cm}\times 
   (n^e_{\vec k_1+ \vec q } - n^h_{-\vec k + \vec q}) d_{\vec k_1} \nonumber \\[0.2cm]
  & + \Gamma^{\vec k, \vec k -\vec q; -(\vec k_1 +\vec q), -\vec k_1}
 _{\vec{k}_{1}, \vec{k} -\vec q; -(\vec{k}_1+\vec q),  -\vec k}(\lambda,\Delta\lambda) 
   (n^e_{\vec k- \vec q } - n^h_{-(\vec k + \vec q)}) d_{\vec k_1}  \Big).
  \end{align}
 Summing up, the following renormalization equations between
   the energy parameters of $\mathcal H_\lambda$ and  $\mathcal H_{\lambda -\Delta \lambda}$
   were found:
  \begin{eqnarray}
  \label{A16}
 \varepsilon^e_{\vec k, \lambda -\Delta \lambda} &=&  \varepsilon^e_{\vec k, \lambda} + 
  \delta\varepsilon^{e(g)}_{\vec k, \lambda} +  \delta\varepsilon^{e(U)}_{\vec k, \lambda}  \,,\\
  \label{A17}
   \varepsilon^h_{\vec k, \lambda -\Delta \lambda} &=& \varepsilon^h_{\vec k, \lambda} 
   +  \delta\varepsilon^{h(g)}_{\vec k, \lambda} 
   +  \delta\varepsilon^{h(U)}_{\vec k, \lambda} \,, \\
   \label{A18}
  \omega_{\vec q, \lambda- \Delta \lambda} &=&  \omega_{\vec q, \lambda} +  \delta\omega^{(g)}_{\vec q, \lambda}
 \,, \\
 \label{A19}
  \Delta_{\vec k, \lambda- \Delta \lambda} &=& \Delta_{\vec k, \lambda} + \delta \Delta_{\vec k,\lambda}^{(0)}
  +  \delta\Delta_{\vec k, \lambda}^{(c)} 
  + \delta\Delta_{\vec k, \lambda}^{(U)}  \,, \\
  \label{A20}
 \Gamma_{\lambda -\Delta \lambda} &=& \Gamma_{\lambda} + \delta \Gamma_{\lambda}^{(0)} +
 \delta \Gamma_{\lambda}^{(c)}
+  \delta \Gamma_{\lambda}^{(g)}\,.
  \end{eqnarray}

 \section{Renormalization of electronic operators}
 \label{B}   


Starting from an appropriate ansatz for the single-fermion operators $ e^\dag_{\vec k, \lambda}$ 
and $h^\dag_{-\vec k,\lambda}$, according to Eqs.~\eqref{90} and \eqref{91}, we have
\begin{align}
\label{B3}
e_{\mathbf{k}, \lambda}^{\dagger} &=    x_{\mathbf{k},\lambda}e_{\mathbf{k}}^{\dagger}
+\frac{1}{\sqrt{N}}\sum_{\mathbf{q}} t_{\mathbf{k -q,q},\lambda} h_{\mathbf{q -k}} \, 
:\psi_{\mathbf{q}}^{\dagger}: \nonumber \\
 &\quad +\frac{1}{N}\sum_{\vec{k}_{1}\vec{k}_{2}}\alpha_{\vec{k}_{1}\vec{k}\vec{k}_{2},\lambda} \, e_{\vec{k}_{1}}^{\dagger}
 :h_{\vec{k}_{2}}^{\dagger}h_{\vec{k}_{1}+\vec{k}_{2}-\vec{k}}: \,,   \\
\label{B4}
h_{\mathbf{- k}, \lambda}^{\dagger}  &=  y_{\mathbf{k},\lambda}h_{-\mathbf{k}}^{\dagger}
+\frac{1}{\sqrt{N}}\sum_{\mathbf{q}} u_{\mathbf{k},\vec{q},\lambda} \,  e_{\vec{q}+\vec k} \,
:\psi_{\vec{q}}^{\dagger}:
\nonumber \\
 &\quad +\frac{1}{N}\sum_{\vec{k}_{1}\vec{k}_{2}}\beta_{\vec{k}_{1}\vec{k}_{2},\vec{k}_2-\vec{k}_{1}-\vec{k},\lambda}:e_{\vec{k}_{1}}^{\dagger}e_{\vec{k}_{2}}: h_{\vec{k}_2-\vec{k}_{1}- \vec{k}}^{\dagger} \,.
  \end{align}
 
 In analogy to the renormalization equations for the parameters of $\mathcal H_\lambda$, 
one derives the following set of renormalization equations for the coefficients 
 $t_{\mathbf{k-q,q},\lambda}^ {}$, $\alpha_{\vec{k}_{1}\vec{k}_{2}\vec{k}_{3},\lambda}$, 
$u_{\mathbf{q},-\vec{k},\lambda}^ {}$, and $\beta_{\vec{k}_{1}\vec{k}_{2}\vec{k}_{3},\lambda}$:
\begin{align}
\label{B5}
t_{\mathbf{k-q,q},\lambda-\Delta\lambda}^ {}  &=  t_{\mathbf{k-q,q},\lambda}^ {}+gx_{\mathbf{k},\lambda}^ {}A_{\mathbf{k-q,q}}^ {}(\lambda,\Delta\lambda)\,,\\
\label{B6}
\alpha_{\mathbf{k_{1}kk_{2}},\lambda-\Delta\lambda}  &=  \alpha_{\mathbf{k_{1}kk_{2}},\lambda}-Ux_{\mathbf{k},\lambda}B_{\mathbf{k_{1}kk}_{2}}(\lambda,\Delta\lambda)\,,
\end{align}
\begin{align}
\label{B7}
&u_{\mathbf{k q},\lambda-\Delta\lambda}^ {}  =  u_{\mathbf{kq},\lambda}^ {}-g\,  y_{\mathbf{k},\lambda}^ {}A_{\mathbf{k,q}}^ {}(\lambda,\Delta\lambda)\,,\\
\label{B8}
&\beta_{\mathbf{k_{1}k_{2},k-k_{1}+k_{2}},\lambda-\Delta\lambda}  =  
\beta_{\mathbf{\mathbf{k_{1}k_{2},k-k_{1}+ k_2}},\lambda}   \nonumber\\
&\hspace*{2.5cm} -  Uy_{\mathbf{k},\lambda}B_{\mathbf{k_{1}k_{2},k-k_{1}+ k_2}}(\lambda,\Delta\lambda)
\, .  
\end{align}
 To obtain renormalization equations for $x_{\vec k, \lambda}$ and $y_{\vec k, \lambda}$
we use the anti-commutator relations for fermionic operators, $[e^\dagger_{\mathbf{k}}(\lambda),e_{\mathbf{k}}(\lambda)]_+=1$ and
$[h^\dagger_{-\mathbf{k}}(\lambda),h_{-\mathbf{k}}(\lambda)]_+=1$, which are valid for any $\lambda$.
We arrive at 
\begin{eqnarray}
\label{B9}
|x_{\mathbf{k},\lambda}|^{2}&= & 1-\frac{1}{N}\sum_{\mathbf{q}}
|t_{\mathbf{k-q, q},\lambda}^ {}|^{2}
(n_{\mathbf{k-q}}^{\psi}+n_{\mathbf{-q}}^{h})\nonumber \\
 &&- \frac{1}{N^{2}}\sum_{\vec{k}_{1}\vec{k}_{2}}|\alpha_{\vec{k}_{1}\vec{k}\vec{k}_{2},\lambda}|^{2}
 \Big[n_{\vec{k}_{1}+\vec{k}_{2}-\vec{k}}^{h}(1-n_{\vec{k}_{2}}^{h})\nonumber\\
 &&- n_{\vec{k}_{1}}^{e}(n_{\vec{k}_{1}+\vec{k}_{2}-\vec{k}}^{h}-n_{\vec{k}_{2}}^{h})\Big] 
  \,, \\[0.2cm]
 \label{B10}
|y_{\mathbf{k},\lambda}|^{2}&= & 1-\frac{1}{N}\sum_{\mathbf{q}}|u_{\mathbf{k,q},\lambda}^ {}|^{2}
(n_{\mathbf{q}}^{\psi}+n_{\mathbf{q+k}}^{e})\nonumber \\
 &&- \frac{1}{N^{2}}\sum_{\vec{k}_{1}\vec{k}_{2}} 
 |\beta_{\vec{k}_{1}\vec{k}_{2},\vec{k}_2-\vec{k}_{1}-\vec{k}_{2},\lambda}|^{2}
 \Big[n_{\vec{k}_{1}}^{e}(1-n_{\vec{k}_{2}}^{e})\nonumber\\
 &&+(1-n_{\vec k_2 -\vec{k}_{1}-\vec{k}}^{h})(n_{\vec{k}_{2}}^{e}-n_{\vec{k}_{1}}^{e})\Big] \,.
\end{eqnarray}
Here, a factorization approximation was used. The expectation values $n^e_{\vec k}$, $n^h_{-\vec k}$, and $n^\psi_{\vec q}$ on the right hand side  
 are best chosen as steady-state expectation values and have been defined before in 
Eqs.~\eqref{88.1} and \eqref{88.2}. 
Moreover $n^\psi_{\vec q}= \langle :\psi^\dag_{\vec q}: \, :\psi_{\vec q}: \rangle$.
Equations \eqref{B5}--\eqref{B8} together with Eqs.~\eqref{B9}, \eqref{B10}, 
taken at $\lambda \rightarrow \lambda -\Delta \lambda$, represent
a complete set of renormalization  equations for the $\lambda$-dependent coefficients in 
Eqs.~\eqref{B3} and \eqref{B4}. 
They connect the parameter values at $\lambda$ with those at $\lambda - \Delta \lambda$. 
The initial parameter values at cutoff $\lambda=\Lambda$ are:
\begin{align}
\label{B11}
&\{x_{\vec{k},\Lambda},y_{\vec{k},\Lambda}\}=1\,,\\
&\{t_{\vec{k}\vec{q},\Lambda},u_{\vec{k}\vec{q},\Lambda},\alpha_{\vec{k}_{1}\vec{k}\vec{k}_{2},\lambda},\beta_{\vec{k}_{1}\vec{k}\vec{k}_{2},\Lambda}  \}= 0\,. 
\end{align}
By integrating the full set of renormalization equations between $\Lambda$ 
and $\lambda =0$ one is led to the fully renormalized one-particle operators:
\begin{align}
\label{B12}
\tilde e_{\vec k}^\dag &= \tilde{x}_{\vec k} e_{\vec k}^\dag
+\frac{1}{\sqrt N}   \sum_{\vec q}  \tilde t_{\vec k-\vec q,\vec q}   h_{\vec q -\vec k} \, 
:{\psi}_{\vec q}^{\dagger}: \nonumber\\
&+ \frac{1}{N}\sum_{\vec k_1\vec k_2}
\tilde{\alpha}_{\vec k_1 \vec k \vec k_2} e_{\vec k_1}^\dag
:h_{\vec k_2}^\dag h_{\vec k_1+ \vec k_2-\vec k}: \, , 
\\
\label{B13}
\tilde h_{- \vec k}^\dag  &= \tilde y_{\vec k} h_{-\vec k}^\dag
+\frac{1}{\sqrt{N}} \sum_{\vec q} \tilde u_{\vec k,\vec q} e_{\vec{q}+\vec k} \,
 :\psi_{\vec q}^\dag: \nonumber \\
 &+ \frac{1}{N}\sum_{\vec k_1 \vec k_2}
\tilde \beta_{\vec k_1 \vec k_2,\vec k_2-\vec k_1-\vec k}
 :e_{\vec k_1}^\dag e_{\vec k_2}:  h_{\vec k_2-\vec k_1- \vec k}^\dag \, .
  \end{align}
With~\eqref{B12} and \eqref{B13} one obtains in the limit $t \rightarrow \infty$:
\begin{align}
\label{B14}
 n^e_{\vec k} &= \langle (\tilde e^\dag_{\vec k} \tilde e_{\vec k})(t \rightarrow \infty) \rangle_{\tilde \rho_0} 
=  |\tilde x_{\vec k}|^2 \hat n^e_{\vec k} \nonumber \\
&\ + \frac{1}{N} \sum_{\vec q} |\tilde t_{\vec k -\vec q, \vec q}|^2
\,  (1- \hat n^h_{\vec k-\vec q}) \, \hat n^\psi_{\vec q}\,
   \nonumber \\
& +\frac{1}{N^2} \sum_{ \vec k_1 \vec k_2} |\tilde \alpha_{\vec k_1 \vec k \vec k_2}|^2
\hat n^e_{\vec k_1} \hat{n}^h_{\vec k_2} (1- \hat n^h_{\vec k_1+ \vec k_2-\vec k})\,,
\end{align}
\begin{align}
\label{B15}
 n ^h_{-\vec k} &= \langle (\tilde h^\dag_{-\vec k} \tilde h_{-\vec k})(t \rightarrow \infty) \rangle_{\tilde \rho_0} =  
|\tilde y_{\vec k}|^2 \hat n^h_{-\vec k}   \nonumber \\
& + \frac{1}{N} \sum_{\vec q} |\tilde u_{\vec k, \vec q}|^2 \, 
\hat n^\psi_{\vec q} (1- \hat n^e_{\vec k +\vec q}) 
  \nonumber \\
&  + \frac{1}{N^2} \sum_{ \vec k_1 \vec k_2} |\tilde \beta_{\vec k_1 \vec k_2,  \vec k_2 -\vec k_1 -\vec k}|^2
\hat n^e_{\vec k_1} (1- \hat n^e_{\vec k_2} )  \hat{n}^h_{\vec k_2- \vec k_1 - \vec k}\,,
\end{align}
and similarly 
\begin{align}
\label{B16}
 d^*_{\vec k}&= 
 \tilde x_{\vec k} \tilde y_{\vec k}\, \hat d^*_{\vec k} 
-\frac{1}{N} \sum_{\vec k_1} 
\big( \tilde x_{\vec k}  \tilde \beta_{\vec k_1, \vec k, - \vec k_1}
\hat n^e_{\vec k}   \nonumber \\ 
&\qquad + \tilde y_{\vec k} \tilde{\alpha}_{\vec k_1,  \vec k, -\vec k_1} (\hat n^h_{-\vec k} -1) \big) 
   \, \hat d^*_{\vec k_1}\,,
   \end{align}
where a small term proportional to $\alpha_{\vec k_1, \vec k, -\vec k_1} \beta_{\vec k_1, \vec k, -\vec k_1}$ of ${\cal O}(U^2)$ was neglected. Another small contribution  being proportional to $ \langle (:\psi^\dag_0:)_{t \rightarrow \infty} \rangle_{\tilde \rho_0}$  was neglected as well. 
The quantities $\hat n^e_{\vec k}$,  $\hat n^h_{-\vec k} $, and $\hat n^\psi_{\vec q} $ on the right hand side of Eqs.~\eqref{B14}--\eqref{B16}
are steady-state expectation values however formed with the renormalized density
$\tilde{\rho}_0$ [also compare Eqs.~\eqref{95}-\eqref{97}]:
\begin{eqnarray}
\label{B17a}
 \hat n^e_{\vec k} &=& \langle (e^\dag_{\vec k} e_{\vec k})(t \rightarrow \infty)\rangle_{\tilde{\rho}_0}\,,\\\label{B17b}
 \hat n^h_{-\vec k} &=& \langle (h^\dag_{-\vec k} h_{-\vec k})(t \rightarrow \infty)\rangle_{\tilde{\rho}_0}  \, , \\\label{B17c}
 \hat n^\psi_{\vec q} &=& \langle (:\psi^\dag_{\vec q}: \, :\psi_{\vec q}:)(t \rightarrow \infty) \rangle_{\tilde \rho_0} \, .
\end{eqnarray}
The corresponding order parameter for the formation of excitons is
\begin{equation}
\label{B18}
\hat d^*_{\vec k} =  \langle (e^\dag_{\vec k} h^\dag_{-\vec k})(t \rightarrow \infty)\rangle_{\tilde{\rho}_0} \, .
\end{equation}

\section{Steady state expectation values}
\label{C}
Evaluating the expectation values $\langle \mathcal A(t) \rangle $ for $t \rightarrow \infty$, we use 
relation \eqref{84} and the steady-state condition \eqref{83}:
\begin{eqnarray}
\label{B1}
&& \langle \mathcal A(t)\rangle = \langle \tilde{\mathcal A}(t)\rangle_{\tilde{\rho_0}}\,, \\
&&
\frac{\rm d}{\rm dt}\langle \mathcal A(t \rightarrow \infty) \rangle = \frac{\rm d}{\rm dt}\langle \tilde{\mathcal A}(t \rightarrow \infty) \rangle_{\tilde \rho_0}
= 0 \, . 
\end{eqnarray}
On the right hand sides, the time dependence is governed by $\tilde{\mathcal H}$. $\tilde \rho_0$ denotes the fully transformed initial density, and  
   $\tilde {\mathcal A}$ is  the transformed operator of $\mathcal A$.

\subsection{Electronic quantities}
\label{C.1}
To analyze time-dependent expectation values for large times, the steady-state condition \eqref{83},
  \begin{eqnarray}
\label{C1}
 \frac{\rm d}{\rm dt}  \langle (C^\dag_{n\vec k} C_{m \vec k})(t) \rangle_{\tilde \rho_0} =0       \quad 
 \mbox{for \,} t \rightarrow \infty \,,
\end{eqnarray}
must be fulfilled. Here, the time dependence is governed by the renormalized Hamiltonian 
\begin{equation}
\label{C2}
\tilde{\mathcal H}=  \tilde{\mathcal H}_{\rm S} + \mathcal H_{\rm R} + \mathcal H_{\rm SR}\,,
\end{equation}
and the expectation values are formed with the transformed initial density $\tilde \rho_0$.
According to this ``recipe'', we first derive equations of motions  
using generalized Langevin equations. These dynamical  equations are best found within the 
Mori-Zwanzig projection formalism~\cite{Mo65a,Mo65b,Zw60}  for a set of dynamical variables 
$ \{ \mathcal A_{\nu} = C_{n \vec k}^\dag C_{m \vec k}, b^\dag_{e, \vec p} b_{e, \vec p},
  b^\dag_{h, -\vec p} b_{h, -\vec p}\}$ ($n,m=1,2$):
 \begin{eqnarray}
\label{C5}
 && \frac{\rm d}{\rm dt}   \mathcal A_\nu(t)  = i \sum_{\mu}  \mathcal A_\mu(t) \, \omega_{\mu \nu} \\
&& \qquad -  \int^t_0 dt' \sum_{\mu}  \mathcal A_{\mu}(t-t') \Sigma_{\mu \nu}(t') +  \mathcal F_{\nu} (t) \, , \nonumber
\end{eqnarray}
 where we have introduced a generalized scalar product 
 $(\mathcal A| \mathcal B) = \langle \mathcal A^\dag \mathcal B \rangle_{\tilde{\rho_0}}$ for operator
 variables $\mathcal A$, $\mathcal B$.
The $\omega_{\mu \nu}$ and  $\Sigma_{\mu \nu}(t)$ are generalized frequencies and self-energies, respectively, 
and $\mathcal F_{\nu} (t)$ is the random force:
\begin{eqnarray}
 \label{C6}
  \omega_{\mu \nu} &=&\sum_{\eta} \chi_{\mu \eta}^{-1} \, \big( \mathcal A_{\eta}| {\tilde {\bf L}} \mathcal A_\nu \big)\,, \\
 \label{C7}
 \Sigma_{\mu \nu}(t) &=& \sum_{\eta} \chi_{\mu \eta}^{-1} \, \big( \mathcal A_{\eta}| {\tilde {\bf L}} \mathbf Q \, 
 e^{i{\mathbf Q \mathbf {\tilde L} \mathbf Q}\, t} \, {\mathbf Q \tilde{\bf  L}  \mathcal A}_\nu \big)\,,  \\
 \label{C8}
 \mathcal F_\nu(t) &=&  
i \, e^{i{\mathbf Q \mathbf {\tilde L} \mathbf Q}\, t} \, {\mathbf Q \tilde{\bf  L}  \mathcal A}_\nu \, .
 \end{eqnarray}
 The quantity $\tilde{\mathbf L}$ is the Liouville operator, defined by the commutator of $\tilde{\mathcal H}$ with any operator observable $\mathcal A$, i.e., $\tilde{\mathbf L} \mathcal A = [\tilde{\mathcal H}, \mathcal A]$, and  $\mathbf Q$ is a generalized projector in the operator space 
 which projects perpendicular to the subspace spanned by the set $\{A_{\nu} \}$.  
 Moreover $\chi_{\mu \eta}^{-1}$ is the inverse of the generalized susceptibility matrix 
 $\chi_{\eta' \mu} = (\mathcal A_{\eta'}| \mathcal A_\mu) $: 
  \begin{equation}
  \label{C9}
 \sum_{\mu} \chi_{\eta' \mu} \chi_{\mu \eta}^{-1} = \delta_{\eta' \eta} \,.
  \end{equation}
  Since $\tilde{\mathcal H}$ does not commute with $\tilde \rho_0$ the expectation values 
$\langle \mathcal A_\nu(t)\rangle_{\tilde{\rho}_0}$ are intrinsically time dependent.

 Let us first consider the equations for  the electronic variables $ \mathcal A^{nm}_{\vec k}:= C^\dag_{n \vec k} C_{m\vec k}$.  
 Because they are  dynamical eigenmodes of $\tilde{ \mathcal H}_{\rm S}$ the 
   frequencies $\omega_{\mu \nu }$ and self-energies 
 $\Sigma_{\mu \nu}$ can be easily evaluated in lowest non-vanishing order perturbation theory in the interaction 
 $\mathcal H_{\rm SR}$.
 One finds in Markov approximation
\begin{eqnarray} 
\label{C10}
&& \frac{\rm d }{\rm dt} \mathcal A^{12}_{\vec k}(t) = i (\tilde E_{1 \vec k} - \tilde E_{2 \vec k})\mathcal A^{12}_{\vec k} 
\nonumber \\
&& \quad 
 - \mathcal A^{12}_{\vec k} \Big[ |\xi_{\vec k}|^2 \, \gamma^e_{\vec k}(\tilde E_{2\vec k})
 +  |\eta_{\vec k}|^2 \, \gamma^e_{\vec k}(\tilde E_{1\vec k}) 
 \nonumber \\
&& \quad  
\phantom{ - \mathcal A^{12}_{\vec k}}+   |\xi_{\vec k}|^2 \, \gamma^h_{\vec k}(-\tilde E_{1\vec k})
 +  |\eta_{\vec k}|^2 \, \gamma^h_{\vec k}(-\tilde E_{2\vec k})
 \Big]  \nonumber \\
 && \quad -\xi_{\vec k}  \eta_{\vec k}^* \Big[
 \,  \mathcal A^{11}_{\vec k}   \,   
 \big( - \gamma^e_{\vec k}(\tilde E_{1\vec k}) +   \gamma^h_{\vec k}(-\tilde E_{1\vec k}) \big)\nonumber \\
&& \phantom{+\xi_{\vec k}  \eta_{\vec k}^* \Big(} 
+ \mathcal A^{22}_{\vec k}   \,   
 \big( - \gamma^e_{\vec k}(\tilde E_{2\vec k}) +   \gamma^h_{\vec k}(-\tilde E_{2\vec k}) \big)
\Big] \nonumber \\
 &&\quad - \pi \xi_{\vec k}  \eta_{\vec k}^* \sum_{\vec p} |\Gamma^e_{\vec k \vec p}|^2 
\big( \delta(\omega^e_{\vec p} -\tilde E_{1 \vec k})
+ \delta(\omega^e_{\vec p} -\tilde E_{2 \vec k})\big)  
\, b^\dag_{e \vec p}  b_{e \vec p}
 \nonumber \\
 && \quad +  \pi \xi_{\vec k}   \eta_{\vec k}^* \sum_{\vec p} |\Gamma^h_{\vec k \vec p}|^2
\big( \delta(\omega^h_{-\vec p} + \tilde E_{1 \vec k}) +   \delta(\omega^h_{-\vec p} + \tilde E_{2 \vec k}) \big)
\nonumber \\
&&  \phantom{ \quad +  \pi \xi_{\vec k} \eta_{\vec k}} \times b_{h,- \vec p}  b^\dag_{h, -\vec p}
+ { \mathcal F}^{12}_{\vec k}
 \nonumber \\
 && \nonumber \\
&& \phantom{\frac{\rm d }{\rm dt} \mathcal A^{12}_{\vec k}(t)} = \Big(\frac{\rm d }{\rm dt} \mathcal A^{21}_{\vec k}(t)\Big)^\dag
\, ,
\end{eqnarray}
 \begin{eqnarray} 
\label{C11}
&& \frac{\rm d }{\rm dt} \mathcal A^{11}_{\vec k}(t) =  - 2 \,  \mathcal A^{11}_{\vec k} \big( |\xi_{\vec k}|^2 \, 
\gamma^e_{\vec k}(\tilde E_{1\vec k}) +   |\eta_{\vec k}|^2 \, \gamma^h_{\vec k}(-\tilde E_{1\vec k})
 \big)  \nonumber \\
 && \qquad +  ( \xi_{\vec k}^* \eta_{\vec k} \, \mathcal A^{12}_{\vec k} 
 + \xi_{\vec k} \eta_{\vec k}^*\mathcal A^{21}_{\vec k} ) \,   \big(  \gamma^e_{\vec k}(\tilde E_{1\vec k}) 
 - \gamma^h_{\vec k}(-\tilde E_{1\vec k})  \big) \nonumber \\
 &&\qquad + 2 \pi |\xi_{\vec k}|^2 \sum_{\vec p} |\Gamma^e_{\vec k \vec p}|^2 
 \delta(\omega^e_{\vec p} -\tilde E_{1 \vec k})  \, b^\dag_{e \vec p}  b_{e \vec p}
 \nonumber \\
 && \qquad + 2 \pi |\eta_{\vec k}|^2 \sum_{\vec p} |\Gamma^h_{\vec k \vec p}|^2
 \delta(\omega^h_{-\vec p} + \tilde E_{1 \vec k}) 
 \,b_{h,- \vec p}  b^\dag_{h,- \vec p} 
 \nonumber \\
 && \qquad + {\mathcal F}_{\vec k}^{11}\,,
\end{eqnarray}
\begin{eqnarray} 
\label{C12}
&& \frac{\rm d }{\rm dt}  \mathcal A^{22}_{\vec k}(t) =  - 2 \,  \mathcal A^{22}_{\vec k} 
\big( |\eta_{\vec k}|^2 \, 
\gamma^e_{\vec k}(\tilde E_{2\vec k}) +  |\xi_{\vec k}|^2 \, \gamma^h_{\vec k}(-\tilde E_{2\vec k})
 \big)  \nonumber \\
 && \qquad +  ( \xi_{\vec k}^* \eta_{\vec k} \,  \mathcal A^{12}_{\vec k} + 
  \xi_{\vec k} \eta_{\vec k}^* \,
 \mathcal A^{21}_{\vec k} ) \, 
  \big(  \gamma^e_{\vec k}(\tilde E_{2\vec k}) 
 - \gamma^h_{\vec k}(-\tilde E_{2\vec k})  \big) \nonumber \\
 &&\qquad + 2 \pi  |\eta_{\vec k}|^2 \sum_{\vec p} |\Gamma^e_{\vec k \vec p}|^2 
 \delta(\omega^e_{\vec p} -\tilde E_{2 \vec k})
 \, b^\dag_{e \vec p}  b_{e \vec p}
 \nonumber \\
 && \qquad - 2 \pi |\xi_{\vec k}|^2 \sum_{\vec p} |\Gamma^h_{\vec k \vec p}|^2 
 \delta(\omega^h_{-\vec p} + \tilde E_{2 \vec k})\, 
  b_{h, -\vec p}  b^\dag_{h,- \vec p}
 \nonumber \\
 &&  \qquad + \mathcal F_{\vec k}^{22}\,.
\end{eqnarray}
Note that the last two terms  in 
equations \eqref{C10}--\eqref{C12} are proportional to the  electron occupation number operators 
$b^\dag_{e \vec p} b_{e \vec p}$ and  $b_{h,-\vec p} b^\dag_{h,-\vec p}$
 of the electronic reservoirs.  However, the equations of motion for 
 $b^\dag_{e, \bf p} b_{e, \bf p}$ and 
$b^\dag_{h, -\bf p} b_{h, -\bf p}$ are not needed. The electronic baths are assumed to be large and stay in thermal equilibrium even when they are coupled 
to the e-h-p system. 

 Moreover, the imaginary parts of the self-energies $\Im\gamma^{e,h}(\omega)$  will be neglected, which would lead to frequency shifts. The remaining real parts  $\Re\gamma^{e,h}_{\vec k}(\omega)$ lead to a 
 damping of electrons and holes  as a result of  the coupling to  the electronic reservoirs:  
\begin{eqnarray}
\label{C13}
&& \Re\gamma^{e}_{\vec k}(\omega) = \pi \sum_{\vec p} |\Gamma^{e}_{\vec k \vec p}|^2
\delta(\omega^{e}_{\vec p} -\omega)  \,, \\
&&  \Re\gamma^{h}_{\vec k}(\omega) = \pi \sum_{\vec p} |\Gamma^{h}_{\vec k \vec p}|^2
\delta(\omega^{h}_{-\vec p} -\omega)\,.
\end{eqnarray}
To simplify the further evaluation we assume  that electrons and holes possess the same damping rate, which is 
also supposed not to depend on $\vec k$ and $\omega$, i.e.,
\begin{equation} 
\label{C14}
\Re\gamma^{e,h}_{\vec k}(\omega) = \Re\gamma_{\vec k}(\omega) \approx \gamma \, .
\end{equation}
 Then Eqs.~\eqref{C10}--\eqref{C12} reduce to
\begin{eqnarray} 
\label{C15}
&& \frac{\rm d }{\rm dt} \mathcal A^{12}_{\vec k}(t) = i (\tilde E_{1 \vec k} - \tilde E_{2 \vec k})\mathcal A^{12}_{\vec k} 
- 2 \gamma \, \mathcal A^{12}_{\vec k}  \nonumber \\
  &&\quad - \pi \xi_{\vec k}  \eta_{\vec k}^* \sum_{\vec p} |\Gamma^e_{\vec k \vec p}|^2 
\big( \delta(\omega^e_{\vec p} -\tilde E_{1 \vec k})
+ \delta(\omega^e_{\vec p} -\tilde E_{2 \vec k})\big)  
\, b^\dag_{e \vec p}  b_{e \vec p}
 \nonumber \\
 && \quad +  \pi \xi_{\vec k}   \eta_{\vec k}^* \sum_{\vec p} |\Gamma^h_{\vec k \vec p}|^2
\big( \delta(\omega^h_{-\vec p} + \tilde E_{1 \vec k}) +   \delta(\omega^h_{-\vec p} + \tilde E_{2 \vec k}) \big)
\nonumber \\
&&  \phantom{ \quad +  \pi \xi_{\vec k} \eta_{\vec k}} \times b_{h,- \vec p}  b^\dag_{h, -\vec p}
+ { \mathcal F}^{12}_{\vec k}
 \nonumber \\
 && \nonumber \\
&& \phantom{\frac{\rm d }{\rm dt} \mathcal A^{12}_{\vec k}(t)} = \Big(\frac{\rm d }{\rm dt} \mathcal A^{21}_{\vec k}(t)\Big)^\dag \,,
\end{eqnarray}
 \begin{eqnarray}
\label{C16}
&& \frac{\rm d }{\rm dt}  \mathcal A^{11}_{\vec k}(t) =  - 2 \gamma \,  \mathcal A^{11}_{\vec k}   \nonumber \\
 &&\qquad + 2 \pi |\xi_{\vec k}|^2 \sum_{\vec p} |\Gamma^e_{\vec k \vec p}|^2 
 \delta(\omega^e_{\vec p} -\tilde E_{1 \vec k})  \, b^\dag_{e \vec p}  b_{e \vec p}
 \nonumber \\
 && \qquad + 2 \pi |\eta_{\vec k}|^2 \sum_{\vec p} |\Gamma^h_{\vec k \vec p}|^2
 \delta(\omega^h_{-\vec p} + \tilde E_{1 \vec k}) 
 \, b_{h,- \vec p}  b^\dag_{h,- \vec p} 
 \nonumber \\
 && \qquad + {\mathcal F}_{\vec k}^{11}\,,
\end{eqnarray}
\begin{eqnarray} 
\label{C17}
&& \frac{\rm d }{\rm dt}  \mathcal A^{22}_{\vec k}(t) =  - 2 \gamma  \,  \mathcal A^{22}_{\vec k} 
  \nonumber \\
 &&\qquad + 2 \pi  |\eta_{\vec k}|^2 \sum_{\vec p} |\Gamma^e_{\vec k \vec p}|^2 
 \delta(\omega^e_{\vec p} -\tilde E_{2 \vec k})
 \, b^\dag_{e \vec p}  b_{e \vec p}
 \nonumber \\
 && \qquad - 2 \pi |\xi_{\vec k}|^2 \sum_{\vec p} |\Gamma^h_{\vec k \vec p}|^2 
 \delta(\omega^h_{-\vec p} + \tilde E_{2 \vec k})  \,
  b_{h, -\vec p}  b^\dag_{h,- \vec p}
 \nonumber \\
 &&  \qquad + {\mathcal F}_{\vec k}^{22}\,.
\end{eqnarray}
The equations of motions for  the expectation values, formed with $\tilde \rho_0$,  
 \begin{eqnarray}
 \label{C18}
  A^{nm}_{\vec k}(t) &=& \langle  \mathcal A^{nm}_{\vec k}(t) \rangle_{\tilde \rho_0} 
  =  \langle (C^\dag_{n \vec k} C_{m \vec k})(t)\rangle_{\tilde \rho_0}\,,
  \end{eqnarray}
can  immediately be found from
Eqs.~\eqref{C10}--\eqref{C12} :
\begin{eqnarray} 
\label{C19}
&& \frac{\rm d }{\rm dt} A^{12}_{\vec k}(t) = \big[i (\tilde E_{1 \vec k} - \tilde E_{2 \vec k})  
- 2 \gamma \big] \,  A^{12}_{\vec k}(t)  \\
  && - \pi \xi_{\vec k}  \eta_{\vec k}^* \sum_{\vec p} |\Gamma^e_{\vec k \vec p}|^2 
\big( \delta(\omega^e_{\vec p} -\tilde E_{1 \vec k})
+ \delta(\omega^e_{\vec p} -\tilde E_{2 \vec k})\big)  
\, \langle b^\dag_{e \vec p}  b_{e \vec p} \rangle_{\tilde \rho_0}
 \nonumber \\
 && +  \pi \xi_{\vec k}  \eta_{\vec k}^* \sum_{\vec p} |\Gamma^h_{\vec k \vec p}|^2
\big( \delta(\omega^h_{-\vec p} + \tilde E_{1 \vec k}) +   \delta(\omega^h_{-\vec p} + \tilde E_{2 \vec k}) \big)
\nonumber \\
&&  \phantom{ +  \pi \xi_{\vec k} \eta_{\vec k}} 
\times \langle b_{h,- \vec p}  b^\dag_{h, -\vec p} \rangle_{\tilde \rho_0}
 \nonumber \\
 && \nonumber \\
&& \phantom{\frac{\rm d }{\rm dt}  A^{12}_{\vec k}(t)} = \Big(\frac{\rm d }{\rm dt}  A^{21}_{\vec k}(t)\Big)^\dag \,,\nonumber
\end{eqnarray}
 \begin{eqnarray}
\label{C20}
&& \frac{\rm d }{\rm dt} A^{11}_{\vec k}(t) =  - 2 \gamma \,  A^{11}_{\vec k}(t)   \\
 &&\qquad + 2 \pi |\xi_{\vec k}|^2 \sum_{\vec p} |\Gamma^e_{\vec k \vec p}|^2 
 \delta(\omega^e_{\vec p} -\tilde E_{1 \vec k})  \, \langle b^\dag_{e \vec p}  b_{e \vec p}\rangle_{\tilde \rho_0}
 \nonumber \\
 && \qquad + 2 \pi |\eta_{\vec k}|^2 \sum_{\vec p} |\Gamma^h_{\vec k \vec p}|^2
 \delta(\omega^h_{-\vec p} + \tilde E_{1 \vec k}) 
 \, \langle b_{h,- \vec p}  b^\dag_{h,- \vec p} \rangle_{\tilde \rho_0}\,,\nonumber
\end{eqnarray}
\begin{eqnarray} 
\label{C21}
&& \frac{\rm d }{\rm dt}  A^{22}_{\vec k}(t) =  - 2 \gamma  \,  A^{22}_{\vec k}(t)  \\
 &&\qquad + 2 \pi  |\eta_{\vec k}|^2 \sum_{\vec p} |\Gamma^e_{\vec k \vec p}|^2 
 \delta(\omega^e_{\vec p} -\tilde E_{2 \vec k})
 \, \langle b^\dag_{e \vec p}  b_{e \vec p} \rangle_{\tilde \rho_0}
 \nonumber \\
 && \qquad +2 \pi |\xi_{\vec k}|^2 \sum_{\vec p} |\Gamma^h_{\vec k \vec p}|^2 
 \delta(\omega^h_{-\vec p} + \tilde E_{2 \vec k})  \,
  \langle b_{h, -\vec p}  b^\dag_{h,- \vec p} \rangle_{\tilde \rho_0}\,,\nonumber
\end{eqnarray}
where the random forces $\mathcal F^{nm}_{\vec k}$ do to contribute  since 
the $\langle \mathcal F^{nm}_{\vec k}\rangle_{\tilde \rho_0}$ vanish at least up to second order in 
$\mathcal H_{\rm SR}$.   Moreover, because the expectation values  of the bath variables 
$ \langle b^\dag_{e \vec p}  b^{}_{e \vec p}\rangle_{\tilde \rho_{0}}$ and 
$\langle b^\dag_{h, -\vec p}  b_{h, -\vec p}^{}\rangle_{\tilde \rho_{0}}$ do not depend on time,
we may use Fermi functions for
\begin{align}
\label{C22}
& \langle b^\dag_{e \vec p}  b^{}_{e \vec p}\rangle_{\tilde \rho_{0}} = 
\frac{1}{1+ e^{\beta [ \omega^e_{\vec p} -(\mu_e - \mu/2)]}}  = f_e(\omega^e_{\vec p}) \,,\\
&\langle b^\dag_{h, -\vec p}  b_{h, -\vec p}^{}\rangle_{\tilde \rho_{0}} =
\frac{1}{1+ e^{\beta [ \omega^h_{-\vec p} -(\mu_h - \mu/2)]}} = f_h(\omega^h_{-\vec p}) . 
\end{align}
Exploiting the presence of the $\delta$-functions, Eqs.~\eqref{C19}--\eqref{C21} can be simplified to 
\begin{eqnarray} 
\label{C23}
 \frac{\rm d }{\rm dt} A^{12}_{\vec k}(t) &=& \big[i (\tilde E_{1 \vec k} - \tilde E_{2 \vec k})  
- 2 \gamma \big] \,  A^{12}_{\vec k}(t)  \nonumber \\
  &-& \gamma \, \xi_{\vec k}  \eta_{\vec k}^*  
 \big( f_e(\tilde E_{1 \vec k}) +  f_e(\tilde E_{2 \vec k})\big)     
 \nonumber \\
 &-&  \gamma \, \xi_{\vec k}   \eta_{\vec k}^* 
 \big(f_h(-\tilde E_{1 \vec k}) + f_h(-\tilde E_{2 \vec k}) -2
\big)
\nonumber \\
&& \nonumber \\
&=& \Big(\frac{\rm d }{\rm dt}  A^{21}_{\vec k}(t)\Big)^\dag \,,
\end{eqnarray}
 \begin{eqnarray}
\label{C24}
 \frac{\rm d }{\rm dt} A^{11}_{\vec k}(t) &=&  - 2 \gamma \,  A^{11}_{\vec k}(t)   + 2 \gamma \, |\xi_{\vec k}|^2  
 f_e(\tilde E_{1 \vec k})
 \nonumber \\
 &+& 2 \gamma \, |\eta_{\vec k}|^2 \big(1- f_h(-\tilde E_{1 \vec k}) \big)\,,
 \end{eqnarray}
\begin{eqnarray} 
\label{C25}
 \frac{\rm d }{\rm dt}  A^{22}_{\vec k}(t) &=&  - 2 \gamma  \,  A^{22}_{\vec k}(t) 
  + 2 \gamma \,   |\eta_{\vec k}|^2 f_e(\tilde E_{2 \vec k})
 \nonumber \\
 &+& 2 \gamma \, |\xi_{\vec k}|^2 \big( 1- f_h(-\tilde E_{2 \vec k}) \big)\, ,
\end{eqnarray}
where Eqs.~\eqref{C13} and \eqref{C14} have been used.  

We are now in the position to evaluate the  limit $t\rightarrow \infty$ for the expectation values 
 \begin{align}
 \label{C26}
  A^{nm}_{\vec k} &=\lim_{t \rightarrow \infty} \langle  \mathcal A^{nm}_{\vec k}(t) \rangle_{\tilde \rho_0} 
  = \lim_{t \rightarrow \infty } \langle (C^\dag_{n \vec k} C_{m \vec k})(t)\rangle_{\tilde \rho_0} \,.
  \end{align}
  Using the steady-state condition \eqref{C1} one finds
 \begin{align}
\label{C27}
&A^{12}_{\vec k} =  \gamma \, \xi_{\vec k}  \eta_{\vec k}^*  \, 
\frac{1}{ \big(  i (\tilde E_{1 \vec k} - \tilde E_{2 \vec k}) -  2 \gamma \big) } \times \\
 & \times 
 \Big[ \big( f_e(\tilde E_{1 \vec k}) + f_e(\tilde E_{2\vec k})
 \big) 
  + \big(f_h(-\tilde E_{1 \vec k}) + f_h(-\tilde E_{2 \vec k}) -2 \big) \Big] \, , 
  \end{align}
and (for  $\gamma \neq 0$)
 \begin{eqnarray}
\label{C28}
 A^{11}_{\vec k}&=&    |\xi_{\vec k}|^2 f_e(\tilde E_{1 \vec k}) +
  |\eta_{\vec k}|^2  \big(1 - f_h(-\tilde E_{1 \vec k}) \big)\,,   \\
\label{C29}
A^{22}_{\vec k}&=&  \, |\eta_{\vec k}|^2 \, f_e(\tilde E_{2 \vec k}) +
  |\xi_{\vec k}|^2 \big(1 - f_h(-\tilde E_{2 \vec k}) \big) \,.
\end{eqnarray}

\subsection{Derivation of equations \eqref{124d}--\eqref{124f} }
\label{C.2}
Let us consider the steady-state result for $\hat d_{\vec k}^*$, $\hat n_{\vec k}^{e}$, and $\hat n_{\vec k}^{h}$, Eqs.~\eqref{124a},~\eqref{124b}, and~\eqref{124c}, respectively. Here, $\hat d^{0*}_{\vec k}$, defined by Eq.~\eqref{118} with Eq.~\eqref{124g}, 
takes the form 
\begin{eqnarray} 
\label{C40}
 \hat d^{0*}_{\vec  k}  
  &=&   \frac{1}{2} \xi_{\vec k}^* \eta_{\vec k}^* F_{1\vec k}^+\,.
\end{eqnarray}
Transforming first the last terms (imaginary parts) in Eqs.~\eqref{124b} and~\eqref{124c}, one gets
\begin{eqnarray}
\label{C42}
 \frac{1}{\gamma} \Im [\tilde \Delta_{\vec k} \hat d_{\vec k}^*] &=&
-\frac{\Im}{\gamma} \Big\{
\frac{1}{\tilde \varepsilon^e_{\vec k} +\tilde \varepsilon^h_{\vec k} +2i \gamma} 
\big[ |\tilde \Delta_{\vec k}|^2 (1-\hat n^e_{\vec k} - \hat n^h_{\vec k}) \nonumber \\
&& - 2i\gamma \tilde \Delta_{\vec k}
\hat d^{0*}_{\vec k}\big]  \Big\}
\end{eqnarray}
or
\begin{eqnarray}
\label{C43}
 \frac{1}{\gamma} \Im [\tilde \Delta_{\vec k} \hat d_{\vec k}^*] &=&
-\frac{\Im}{\gamma} \Big\{
\frac{|\tilde \Delta_{\vec k}|^2}{\tilde \varepsilon^e_{\vec k} +\tilde \varepsilon^h_{\vec k} +2i \gamma} 
\big[  (1-\hat n^e_{\vec k} - \hat n^h_{\vec k}) \nonumber \\
&& - \frac{i \, \gamma}{W_{\vec k}} {\rm sgn}(\tilde \varepsilon^e_{\vec k} +\tilde \varepsilon^h_{\vec k})
F_{1\vec k}^+ \big]  \Big\}\,.
\end{eqnarray}
Thus
\begin{eqnarray}
\label{C44}
 \frac{1}{\gamma} \Im [\tilde \Delta_{\vec k} \hat d_{\vec k}^*] &=&
\frac{|\tilde \Delta_{\vec k}|^2}{(\tilde \varepsilon^e_{\vec k} +\tilde \varepsilon^h_{\vec k})^2 +(2 \gamma)^2}
\big[ 2(1-\hat n^e_{\vec k} - \hat n^h_{\vec k})  \nonumber \\
&+& 
\frac{|\tilde \varepsilon^e_{\vec k} +\tilde \varepsilon^h_{\vec k}|}{W_{\vec k}}\, F_{1\vec k}^+
\big]\,.
\end{eqnarray}
Here, we have used Eq.~\eqref{124a} for $\hat d^*_{\vec k}$ with $\hat d^{0*}_{\vec k}$ given by Eq.~\eqref{C40}
and \eqref{74}:
\begin{eqnarray}
\label{C48}
\hat d^*_{\vec k} &=& \frac{\tilde \Delta^*_{\vec k}}{(\tilde \varepsilon^e_{\vec k} +\tilde \varepsilon^h_{\vec k}) +2i  \, \gamma}
\Big[ (\hat n_{\vec k}^e + \hat n_{\vec k}^h -1)   \nonumber \\
&+& 
 i \gamma \, {\rm sgn }(\tilde \varepsilon^e_{\vec k} +\tilde \varepsilon^h_{\vec k})  \frac{F^+_{1 \vec k}}{W_{\vec k}}
\Big]\,.
\end{eqnarray}
Then, from \eqref{C44} together with Eqs.~\eqref{124b} and \eqref{124c}, one finds: 
\begin{eqnarray}
\label{C45}
  \hat n^e_{\vec k} + \hat n^h_{\vec k} -1  &=& 
\frac{|\tilde \varepsilon^e_{\vec k} +\tilde \varepsilon^h_{\vec k}|}{2W_{\vec k}} F_{1\vec k}^+ + \nonumber  \\
&+&  \frac{1}{2} \frac{ F_{2 \vec k}^+ -2}{1+ \displaystyle 
  \frac{4|\tilde \Delta_{\vec k}|^2}{(\tilde \varepsilon^e_{\vec k} +\tilde \varepsilon^h_{\vec k})^2 + (2 \gamma)^2}   }  \,,
\end{eqnarray} 
 and 
 \begin{eqnarray}
 \label{C47}
&& \hat n^e_{\vec k} - \hat n^h_{\vec k} =  \frac{1}{2} F_{1\vec k}^- 
+ \frac{|\tilde \varepsilon^e_{\vec k} +\tilde \varepsilon^h_{\vec k}|}{2W_{\vec k}} F_{2\vec k}^- \,.
\end{eqnarray} 
\subsection{Photonic expectation values }
\label{C.3}
%
%
To calculate the photon condensation parameter  $\langle \psi^\dag_{\vec q=0}\rangle$,  we  use the  {\it ansatz} 
for the $\lambda$-dependent photon operator, 
\begin{eqnarray}
\label{B19}
\psi^\dag_{\vec q, \lambda} = z_{\vec q,\lambda}\psi^\dag_{\vec q}
+ \frac{1}{\sqrt N} \sum_{\vec k}  v_{\vec k \vec q, \lambda} \, :e^\dag_{\vec k + \vec q} h^\dag_{-\vec k}: \, ,
\end{eqnarray}
where again the operator structure was taken over from a small $X_{\lambda, \Delta \lambda}$ expansion. 
Furthermore,  $:e^\dag_{\vec k + \vec q} h^\dag_{-\vec k}: = e^\dag_{\vec k + \vec q} h^\dag_{-\vec k} -
\langle e^\dag_{\vec k + \vec q} h^\dag_{-\vec k} \rangle $. 
In analogy to the preceding section, one easily obtains renormalization equations for the $\lambda$-dependent coefficients $z_{\vec q,\lambda}$ and 
$v_{\vec k \vec q, \lambda}$:
\begin{align}
\label{B20}
 &v_{\vec k \vec q, \lambda- \Delta \lambda}  = v_{\vec k \vec q, \lambda} 
- g z_{\vec q,\lambda} A_{\vec k \vec q}(\lambda, \Delta \lambda)\,,  \\
\label{B21}
&|z_{\vec q,\lambda}|^2 = 1- \frac{1}{N} \sum_{\vec k} |v_{\vec k \vec q, \lambda}|^2 
(1- n^e_{\vec k+ \vec q} - n^h_{-\vec k}) \,.
\end{align}
Deriving the last equation, the commutator relation $[\psi_{\vec q, \lambda}, \psi^\dag_{\vec q, \lambda}]=1$ was used. 
Equation~\eqref{B20} and Eq.~\eqref{B21}, both taken at $\lambda \rightarrow \lambda - \Delta \lambda$, represent a complete set of renormalization equations for the $\lambda$-dependent coefficient in \eqref{B19}. Here, the initial parameter values are 
\begin{equation}
\label{B22}
z_{\vec q, \Lambda} =1\, , \quad  v_{\vec k \vec q, \Lambda } =0\,.
\end{equation}
The integration between $\lambda =\Lambda$ and $\lambda=0$ leads to the fully renormalized photon 
operator 
\begin{eqnarray}
\label{B23}
\tilde \psi^\dag_{\vec q} = \tilde z_{\vec q}\psi^\dag_{\vec q}
+ \frac{1}{\sqrt N} \sum_{\vec k} \tilde v_{\vec k \vec q} \, :e^\dag_{\vec k + \vec q} h^\dag_{-\vec k}: \,.
\end{eqnarray}
Using Eq.~\eqref{84}, one finds in the large-$t$ limit: 
\begin{eqnarray}
\label{B24}
&&\langle \psi^\dag_{\vec q}({t \rightarrow \infty}) \rangle = 
\tilde z_{\vec q} \langle \psi^\dag_{\vec q}(t\rightarrow \infty) \rangle_{\tilde \rho_0} \nonumber \\
&& \qquad + \frac{1}{\sqrt N} \sum_{\vec k} \tilde v_{\vec k \vec q} \, 
\langle (:e^\dag_{\vec k + \vec q} h^\dag_{-\vec k}:) (t\rightarrow \infty)\rangle_{\tilde \rho_0} \nonumber \\
&& \qquad \simeq   \tilde z_{\vec q} \langle \psi^\dag_{\vec q}(t\rightarrow \infty)\rangle_{\tilde \rho_0} ,
\end{eqnarray}
where the second contribution, being proportional to fluctuation operators, was neglected. 
Similarly, 
\begin{eqnarray}
\label{B26}
 && \langle (\psi^\dag_{\vec q} \psi_{\vec q})(t \rightarrow \infty)\rangle =
|\tilde z_{\vec q}|^2 \langle (\psi^\dag_{\vec q} \psi_{\vec q})(t \rightarrow \infty)\rangle_{\tilde \rho_0} 
\nonumber \\
&& \qquad + \frac{1}{N} \sum_{\vec k} |\tilde v_{\vec k \vec q}|^2 \, \hat n^e_{\vec k+ \vec q} \, 
\hat n^h_{-\vec k}\,.
\end{eqnarray}

We then evaluate the remaining quantities $\langle \psi^\dag_{\vec q}(t\rightarrow \infty) \rangle_{\tilde \rho_0}:=\langle \psi^\dag_{\vec q}  \rangle_{\tilde \rho_0}$ and $\langle (\psi^\dag_{\vec q} \psi_{\vec q})(t\rightarrow \infty)\rangle_{\tilde \rho_0} 
 := \langle \psi^\dag_{\vec q} \psi_{\vec q}\rangle_{\tilde \rho_0}$. Our starting point is an equation of motion for  the 
time-dependent photon creation operator  $\psi^\dag_{\vec q}(t )$. Using again the Mori-Zwanzig approach of Sec.~C.1, 
one obtains with Eqs.~\eqref{69} and \eqref{C5}, 
\begin{equation} 
\label{C30}
\frac{\rm d}{\rm dt}  \psi^\dag_{\vec q}(t ) = i \tilde \omega_{\vec q}   \psi^\dag_{\vec q}(t) 
+ i \sqrt N \tilde \Gamma^* \delta_{\vec q, 0} 
- \kappa  \psi^\dag_{\vec q}(t)     + \mathcal F_{\vec q}^\psi  \, , 
\end{equation}
and
\begin{equation} 
\label{C31}
 \frac{\rm d}{\rm dt} \langle \psi^\dag_{\vec q}(t ) \rangle_{\tilde \rho_0}  = i \tilde \omega_{\vec q}  \langle \psi^\dag_{\vec q}(t) \rangle_{\tilde \rho_0}
+ i \sqrt N \tilde \Gamma^* \delta_{\vec q, 0} 
- \kappa  \langle \psi^\dag_{\vec q}(t) \rangle_{\tilde \rho_0},
\end{equation}
where $\kappa$ is the damping rate of
cavity photons into the free space. For  the 
steady state at $t \rightarrow \infty$ one finds from Eq.~\eqref{C31}
 \begin{eqnarray} 
\label{C32}
\langle \psi^\dag_{\vec q} (t \rightarrow \infty)\rangle_{\tilde \rho_0} = 
- \frac{\sqrt N \hat \Gamma^*}{\tilde \omega_0 + i \kappa} \, \delta_{\vec q, 0} \,,
= \langle \psi^\dag_{\vec q} \rangle_{\tilde \rho_0} 
\end{eqnarray}
and, with Eq.~\eqref{B24}, 
\begin{equation}
\label{C33}
\langle \psi_{\vec q}^\dag(t \rightarrow \infty)\rangle = - 
\hat z_{\vec q=0} \,  \frac{\sqrt N \tilde \Gamma^*}{\tilde \omega_0 + i \kappa}  \, \delta_{\vec q, 0} 
= \langle \psi_{\vec q}^\dag\rangle   \, .
\end{equation}

To evaluate the expectation value 
$\langle (\psi^\dag_{\vec q} \psi_{\vec q})({t \rightarrow \infty}) \rangle_{\tilde \rho_0}$,  one best starts from the 
solution \eqref{146} of the equation of motion \eqref{C30}, thereby neglecting the fluctuation force $\mathcal F_{\vec q}^\psi$:
\begin{equation}
\label{C34}
\psi^\dag_{\vec q}(t) = - \frac{i \sqrt N \tilde \Gamma^* }{ i \tilde \omega_0 - \kappa} \delta_{\vec q,0}
 +  \big(  \psi^\dag_{\vec q} +  \frac{ i \sqrt N \tilde \Gamma^* }{ i \tilde \omega_0 - \kappa} \delta_{\vec q,0}
\big) e^{( i \tilde \omega_{\vec q} - \kappa) t } \,.
\end{equation}
For $t \rightarrow \infty$ one is led to
\begin{equation}
\label{C35}
 \langle (\psi^\dag_{\vec q} \psi_{\vec q})({t \rightarrow \infty}) \rangle_{\tilde \rho_0} =
\frac{N |\tilde \Gamma|^2}{\tilde \omega_0^2 + \kappa^2} \delta_{\vec q, 0}=\langle \psi^\dag_{\vec q} \psi_{\vec q}\rangle_{\tilde \rho_0} \,.
 \end{equation}
For  the fluctuation number $\hat n^\psi_{\vec q}$ of cavity photons 
one obtains with  Eq.~\eqref{C32}:
\begin{equation}
\label{C36}
 \hat n^\psi_{\vec q}= \langle (:\psi^\dag_{\vec q}: \, :\psi_{\vec q}:)(t \rightarrow \infty) \rangle_{\tilde \rho_0} =0 \, .
\end{equation}
Thus, the fluctuation number, formed with $\tilde \rho_0$ vanishes. In contrast, for the full quantity  
$n^\psi_{\vec q}= \langle :\psi^\dag_{\vec q}: \, :\psi_{\vec q}:\rangle$, which is formed by the initial density 
$\rho_0$, one finds from  Eq.~\eqref{B26} and \eqref{C33}:
\begin{equation}
\label{C36a}
n^\psi_{\vec q} = \frac{1}{N} \sum_{\vec k} |\tilde v_{\vec k \vec q}|^2 \, \hat n^e_{\vec k+ \vec q} \, 
\hat n^h_{-\vec k}\,.
\end{equation}
That is, the full fluctuation number $n^\psi_{\vec q}$  of cavity photons is determined 
by the coupling of cavity photons  to electronic particle-hole excitations of the e-h-p system.

\end{appendix}


\end{document}